\documentclass[12pt,a4paper]{article}

\usepackage[utf8]{inputenc}
\usepackage[T1]{fontenc}
\usepackage[a4paper, total={6in, 8.5in}]{geometry}
\usepackage{graphicx}
\usepackage{tabularx}
\usepackage{float}
\usepackage{booktabs}
\usepackage{longtable}
\usepackage{array}
\usepackage[round]{natbib}
\usepackage[usenames,dvipsnames]{xcolor}
\usepackage{url}
\usepackage{fancyhdr}
\usepackage{amsmath}
\usepackage{amssymb}
\usepackage{setspace}
\usepackage{soul}
\usepackage{titling}
\usepackage{titlesec}
\usepackage{tocloft}

\newlength{\tableelementskip}
\usepackage[colorlinks=true]{hyperref}
\hypersetup{
    colorlinks=true,
    linkcolor=OliveGreen,
    urlcolor=OliveGreen,
    citecolor=OliveGreen,
    filecolor=blue
}

\titleformat*{\subsection}{\normalsize\bfseries}

\pretitle{\begin{center}\hrule height 3pt\vspace{10pt}\Large}
\posttitle{\vspace{13pt}\hrule height 1pt\end{center}}
\title{\textbf{Initial results of the Digital Consciousness Model}\thanks{Note: an article by \emph{The Economist} (\url{https://www.economist.com/interactive/briefing/2026/08/20/the-search-for-consciousness-inside-llms}) uses the same model and indicators presented in this paper, with updated data and a calibration process from a forthcoming version. More details can be found at \url{https://www.youtube.com/watch?v=4axFB0QWD2M}.}}
\author{
  \begin{tabular}{@{}ccc@{}}
    Derek Shiller & Laura Duffy & Arvo Muñoz Morán \\
    {\small \textit{Rethink Priorities}} & {\small \textit{Rethink Priorities}} & {\small \textit{Rethink Priorities}} \\[1em]
    Adrià Moret & Chris Percy & Hayley Clatterbuck\thanks{Corresponding author: \href{mailto:hayley@rethinkpriorities.org}{hayley@rethinkpriorities.org}} \\
    {\small \textit{University of Barcelona}} & {\small \textit{Co-Sentience Initiative}} & {\small \textit{Rethink Priorities}}
  \end{tabular}
}

\date{}

\begin{document}

\maketitle
\thispagestyle{empty}

\begin{center}
\vskip-0.5in
\parbox{0.8\textwidth}{
\begin{center}\textbf{Abstract}\end{center}
\small
\setstretch{0.8}
Artificially intelligent systems have become remarkably sophisticated. They hold conversations, write essays, and seem to understand context in ways that surprise even their creators. This raises a crucial question: Are we creating systems that are conscious? 

\vskip0.2in
The Digital Consciousness Model (DCM) is a first attempt to assess the evidence for consciousness in AI systems in a systematic, probabilistic way. It provides a shared framework for comparing different AIs and biological organisms, and for tracking how the evidence changes over time as AI develops. Instead of adopting a single theory of consciousness, it incorporates a range of leading theories and perspectives—acknowledging that experts disagree fundamentally about what consciousness is and what conditions are necessary for it.
\vskip0.2in
This report describes the structure and initial results of the Digital Consciousness Model. Overall, we find that the evidence is against 2024 LLMs being conscious, but
the evidence against 2024 LLMs being conscious is not decisive. The evidence against LLM consciousness is much weaker than the evidence against consciousness in simpler AI systems.

}
\vskip1in
\end{center}

\begin{center}
\textit{AI Cognition Initiative at Rethink Priorities}
\end{center}
\newpage

\tableofcontents
\newpage

\section{Introduction}\label{introduction}

Rapid advancements in artificial intelligence have yielded systems with
striking capabilities in many of the domains that we associate with
consciousness in humans and other animals, including cognitive
complexity, agency, flexibility, and language use. Many expect that much
more sophisticated systems will arrive in the coming years. Already, a
substantial minority of both domain experts and casual observers find it
plausible that such systems are now or could soon be conscious \citep{caviola2025futures, dreksler2025subjective, pauketat2023artificial,
francken2022academic}. However, given the lack of consensus about
the basis of consciousness, there is also significant uncertainty about
the probability that AI is conscious and how we would know if it was
\citep{butlin2023consciousness, butlin2025identifying, chalmers2023could,
francken2022academic, michel2018informal, mudrik2025unpacking, seth2024conscious}.

For our purposes here, a conscious system is one that has subjective
experience. It has the ability to have feelings, sensations, or emotions
from its own perspective, and there is something that it is like to be
that system.\footnote{Our target is phenomenal consciousness (in the
  sense of \citealt{nagel1974what}), not access consciousness \citep{block1995confusion}.} The
question of whether AI is conscious in this sense is important. Whether
a system is conscious is widely agreed to be important for the ways in
which it is appropriate for individuals to interact with it, for
developers to design it, and for regulators to oversee it. A system that
has conscious experiences is likely to deserve moral consideration for
its own sake \citep{birch2024edge, dung2025how, goldstein2025ai,
long2024taking, moret2025ai, saad2025digital}.\footnote{Either because consciousness is a key contributor to having welfare states or
  it is plausibly highly correlated with other capacities that ground
  moral consideration.} It may warrant empathy, respect, honesty, and
faithfulness \citep{butlin2025principles, gunkel2018robot, schwitzgebel2025against}. If
AIs are conscious but are believed not to be, then we may commit moral
harm by using them as mere tools without concern for their welfare.
There are also risks in over-attributing consciousness \citep{schwitzgebel2023full, birch2025ai}. If AIs are not conscious but are believed to be, then
we risk giving unwarranted consideration to entities that don't matter
at the expense of individuals who do (e.g. humans or other animals).

We will likely need to make significant decisions that turn on the
question of whether AI systems are conscious \citep{bostrom2022propositions}.
Those questions may well arise before a full scientific consensus
becomes available. For example, a non-negligible chance of AI
consciousness may be enough to trigger precautionary measures or moral
consideration \citep{birch2017animal, birch2024edge}. We expect that policy decisions
will need to be made about how to develop and treat AI systems within
the next several years, and we are not optimistic that consciousness
science will overcome its significant obstacles and reach consensus by
that time \citep{sebo2025moral, long2024taking}.

Even if we cannot be certain whether AIs are conscious, there are a
wealth of theories and perspectives which we might use to assess the
capabilities and architectural structures we find in current systems in
order to come to an informed judgement. An estimate of the probability
that a system is conscious that reflects our uncertainty can help us
know when and what kind of precautionary measures ought to be instituted
\citep{birch2024edge, schwitzgebel2025emotional, sebo2025moral}. By tracking
how these probabilities change over time, we can forecast when key
thresholds may be surpassed. We can also compare the strength of our
evidence for consciousness across different AI systems and biological
species to look for the most important candidates for special
consideration. Lastly, examining which capabilities seem to make the
biggest difference to estimates of consciousness can help us design AIs
that are more or less likely to be conscious.

Our Digital Consciousness Model is an early, proof-of-concept attempt at
providing a probabilistic assessment of our evidence for consciousness
in AI systems that can be used as a framework for future research.
Instead of adopting a single theory of consciousness, it incorporates a
range of leading theories and perspectives. It then aggregates these
judgments in accordance with expert judgments about which perspectives
are most plausible. It also uses expert judgments to evaluate the
capacities of target systems and incorporates their uncertainty about
these capacities. The model can be deployed to assess a diverse range of
systems, both artificial and biological, facilitating tentative
comparisons.

In brief, we found that the aggregated indicator evidence is against
2024 LLMs being conscious. In contrast, the model strongly favors
consciousness in chickens and very strongly favors consciousness in
humans. However, the evidence against LLM consciousness is not decisive.
It is much weaker than the evidence against consciousness in simpler
systems like ELIZA, and according to some perspectives on consciousness,
we possess evidence in favor of LLM consciousness.

We don't intend the model to be a complete or authoritative assessment
of AI consciousness, and we urge particular caution in interpreting the
probabilities that it generates. We do think that it provides an
important first attempt at a general probabilistic framework for
assessing AI consciousness that can provide initial guidance for
interacting with such systems. The DCM can be refined and expanded, and
it can serve as a helpful framework for organizing and guiding our
knowledge as we learn more about AI and about consciousness itself. The
model shows that despite all of our uncertainties about consciousness,
it is possible to empirically assess the state of the evidence and track
how that evidence changes as AIs become more sophisticated.

\section{Challenges for a digital consciousness model}\label{challenges-for-a-digital-consciousness-model}

There are many challenges we face when trying to systematically evaluate
consciousness in a new candidate system. Here, we identify three such
challenges and strategies for overcoming them.

\subsection{Disagreement and uncertainty over theories of consciousness}\label{disagreement-and-uncertainty-over-theories-of-consciousness}

Most obviously, there is no scientific or philosophical consensus about
the nature of consciousness \citep{mudrik2025unpacking, francken2022academic, michel2018informal}. Many theories have been
offered which identify different conditions as necessary and sufficient
for a system to be conscious or which are otherwise indicative of the
presence of consciousness. Available theories of consciousness give very
different predictions about what would count as evidence for
consciousness in a system \citep{birch2022search}. For example, being composed of
biological neurons is strong evidence of consciousness according to some
theories but largely incidental according to others.

Given the widespread disagreement about which theories of consciousness
are most plausible, and the very different interpretations of evidence
they suggest, our evaluation of the evidence reflects this diversity and
uncertainty. The model contains an initial set of 13 stances on
consciousness, chosen to reflect a diverse set of perspectives that are
well-represented in the scientific literature and popular imagination
\citep{seth2022theories, friedman2023current, kuhn2024landscape, mudrik2025unpacking, schwitzgebel2020is}.\footnote{We use ``stance'' as a general term to capture both formal scientific
  theories of consciousness and less formal perspectives on when to
  attribute consciousness. We will use ``stance'', ``theory'', and
  ``perspective'' roughly interchangeably.} We recognize that this list
is not complete, and we have ambitions to add more stances in future
work (see Limitations and Future Work for a discussion).

We use these theories to derive indicators for consciousness and then
evaluate the evidence relative to each stance (see also \citealt{butlin2025identifying}). Then, we aggregate each stance's judgments, weighting each
judgment by the relative plausibility of the stance as evaluated by a
small pilot study of experts.\footnote{The pilot study was performed by reaching out to around 30 experts in philosophy of mind and cognitive
  science (not necessarily those working on digital minds). Future
  studies will target larger and more diverse samples.} The model can
also be interrogated on a stance-by-stance basis, so that users are not
dependent on expert judgements about stance plausibility.

\subsection{Imprecision in how to apply theories}\label{imprecision-in-how-to-apply-theories}

To evaluate the evidence relative to each stance, we need to work out
the empirical predictions made by each. This is not always
straightforward.

First, stances vary in their precision and degree of empirical
elaboration. Some of these theories are fairly well-developed and serve
as the basis for scientific research programs, e.g. Global Workspace
Theory \citep{baars1988cognitive, baars2005global, dehaene2011experimental, dehaene2006conscious}. Others are currently being elaborated and represent
promising avenues for empirical testing but have not yet elaborated
necessary and sufficient conditions, e.g. recurrent processing theories
\citep{lamme2006towards}. A few incorporate speculative assumptions and remain
highly controversial, e.g. electromagnetic field theories \citep{mcfadden2020integrating}. Others reflect more general and informal orientations toward
kinds of evidence for consciousness (e.g. it is present in systems that
seem like people when you interact with them).

Second, across all levels of precision, there are questions about how
substantive the requirements are for consciousness \citep{carruthers2019human, herzog2007consciousness}. There are fairly trivial ways of
satisfying each perspective on consciousness, and versions of different
theories might vary in their liberalness. For example, specific Global
Workspace Theories differ in how coarse- or fine-grained their
requirements for having a global workspace are \citep{mashour2020conscious}. Most theories were originally developed to describe consciousness
in humans (and perhaps closely related species). When we ask what
predictions they make about very different kinds of systems, such as
AIs, it is unclear how strictly to take requirements originally
developed for humans.

We have chosen to characterize stances at a fairly high level of
generality, characterizing them in terms of broad commitments about
which features are indicative of consciousness. For example, instead of
using specific versions of Global Workspace Theory, we consider what
distinguishes GWT in general from other perspectives.\footnote{Our approach treats stances in a fairly coarse-grained way. However, it
  doesn't follow that it's very easy for systems to count as conscious
  since we can still be quite uncertain about what would count as
  evidence for consciousness relative to a coarse-grained stance.} We do
so by specifying a set of features, general properties of systems that
are identified by some stances as being relevant for consciousness
attribution, e.g. attention, recurrent processing, or modality
integration, and characterizing stances by which sets of features they
deem evidentially important and to what extent.

\subsection{Gaps in our evidence}\label{gaps-in-our-evidence}

It is a common refrain among experts that we have built AI systems whose
internal mechanisms we don't fully understand \citep{amodei2025urgency, hendrycks2025misguided}. Deep learning networks don't give up their details
easily. Though mechanistic interpretability is making rapid progress in
sorting out some aspects of AI cognition (at least compared to
neuroscience's slow march toward understanding human cognition), it is
also aimed at a moving target. New models come out regularly: the
research available for last year's open source models is of uncertain
value in understanding models today, or the models we should be planning
for next year.

The internals of the latest models are unavailable to outside
researchers for close inspection. Much research is focused on
understanding behavior. Few people are up to date on the behavioral
quirks of a variety of different current models, and haven't spent the
time probing their abilities. Model behavior that seems human in nature
may deserve a very different explanation, as models are known to game
human-like characters \citep{andrews2023understand, ledoux2023consciousness}.

As a result, we think it premature to take any specific observable
indicator as necessary or sufficient for consciousness relative to any
theory. Instead, we use a Bayesian approach to evidence, where the
evidence raises or lowers the probability of consciousness rather than
entailing its presence or absence. The model also allows for the
flexible addition of new indicators as systems, and our knowledge of
them, change.

\section{Components of the model}\label{components-of-the-model}

\subsection{General structure}\label{general-structure}

The DCM is a Bayesian hierarchical model with three components:
indicators, features, and stances. (More accurately, it is a framework
that includes separate hierarchical models for each stance. The
judgments of the stance models are aggregated in a non-Bayesian way, by
taking the credence-weighted average.) Our prior expectations about the
system's probability of consciousness flow downward to set prior
expectations of what features and indicators the system will have. When
we gather information about the indicators, evidential updating flows
upward through the model to change our beliefs about the features the
system possesses and its probability of consciousness.

Here, we explain each component of the model and key features of the
Bayesian approach (more detailed information about the model's structure
can be found in the next section and in \autoref{appendix-c-model-structure}).

\begin{figure}[H]
\centering
\includegraphics[width=\textwidth,keepaspectratio]{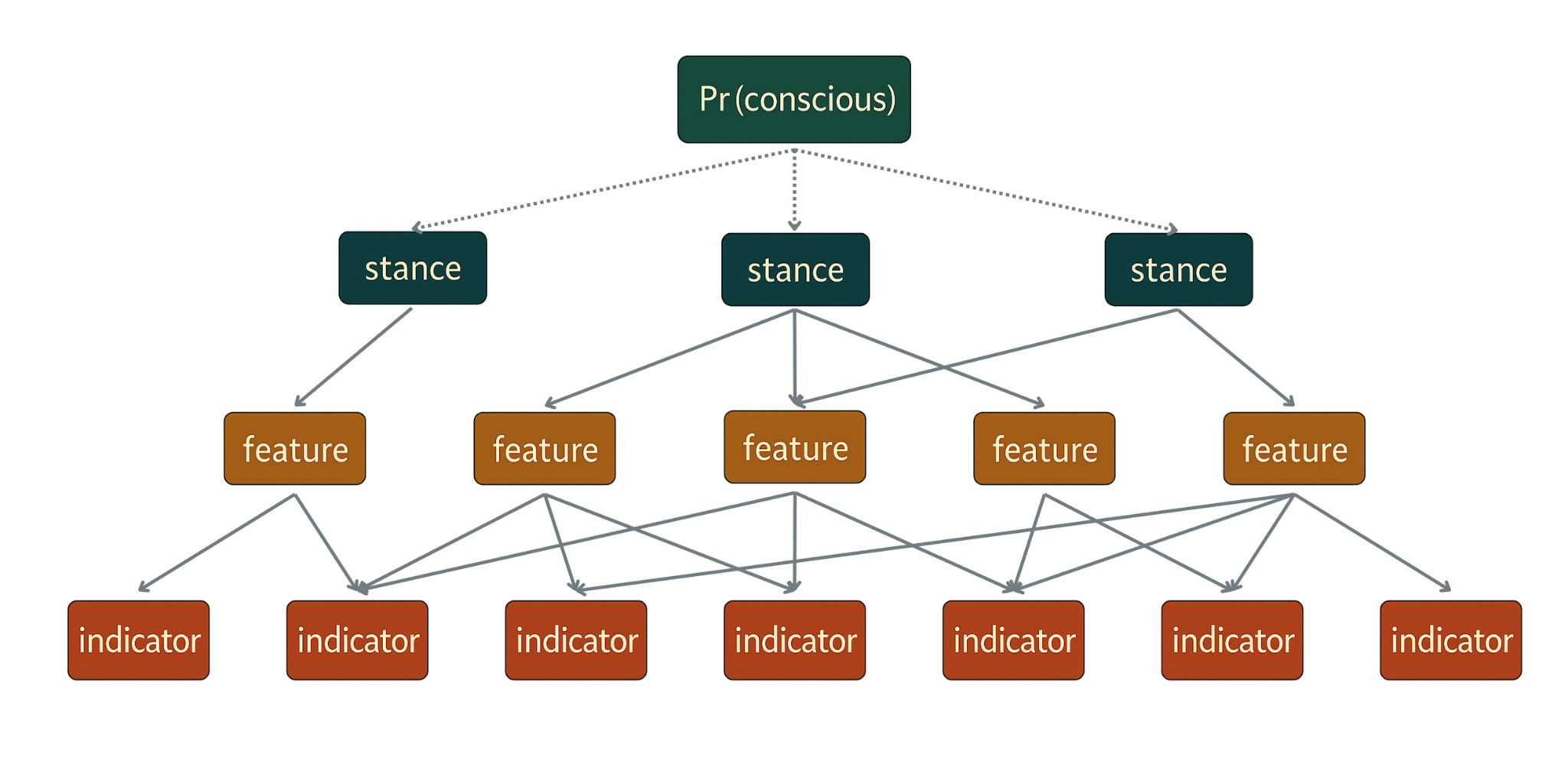}
\caption{Structure of the DCM}
\label{fig:dcm-structure}
\end{figure}

\subsection{Indicators}\label{indicators}

Indicators are the bottom-most, nearest-to-observable properties of a
system. They provide evidence regarding a system's possession of
potentially consciousness relevant features (discussed below). The model
currently contains 206 indicators and can be easily updated with new
ones. Only a subset of the indicators will be deemed relevant for any
particular stance. Examples include:

\begin{description}
\item[Carbon-Based] The system\textquotesingle s physical structure and
  functional components are primarily composed of carbon-based molecules
  and organic compounds.
\item[Self-Representations] The system maintains internal representations or
  models of itself that are structurally or functionally distinct from
  how it represents other entities or agents. This may include
  specialized neural circuits, data structures, or processing pathways
  dedicated to self-referential information.
\item[ARC Performance] The system\textquotesingle s ability to solve novel
  visual pattern recognition and completion tasks in the Abstract
  Reasoning Corpus, which measures fluid intelligence through abstract
  visual reasoning problems.
\end{description}

A full list can be found in \autoref{appendix-b-indicators-list}.

The model updates its estimate of the probability that a system is
conscious on the basis of evidence regarding the indicators. Indicator
data is generated by polling a set of experts for each system
considered. For each indicator, experts indicated their credence that
the system has the property in question.\footnote{Indicators can be vague, in two senses. First, while some indicators are obviously ``yes
  or no'' (e.g. ``Does the system use neurotransmitters?''), others come
  in degrees (e.g. ``Does the system demonstrate meaningful variation in
  its responses when presented with the same or similar stimuli across
  different contexts or instances?''). In this case, expert judgments
  may reflect expectations about the degree to which graded indicators
  are present (perhaps beyond some threshold). Second, some indicators
  are not directly observable and depend on the expert's judgment about
  what it takes to genuinely possess the indicator. For example,
  answering ``Does the system systematically prefer certain types of
  activities?'' depends on an assessment of what it takes to genuinely
  have preferences. See Section 9.6. for more discussion.} In cases
where it was particularly obvious (e.g. whether the system is
carbon-based), we specified the indicator value directly.

\subsection{Features}\label{features}

Features comprise the middle layer of the model and serve as intervening
variables between indicators and consciousness. They are general
properties that are identified as being relevant for consciousness by at
least one stance. We attempted to construct the most parsimonious list
of features that would capture both commonality and diversity among the
perspectives on consciousness that we included.\footnote{If new stances are added that cannot be fully characterized in terms of existing
  features, new features may be added to the model. A new set of
  indicators would also need to be added to provide evidence regarding
  that new feature.} The model currently contains 20 high-level
features. A larger number of subfeatures (70) bridge the gap between
features and indicators. Examples include:

\begin{description}
\item[Biological Similarity] The system\textquotesingle s physical structure
  and organization shows meaningful similarities to biological systems
  known to be associated with consciousness, particularly in terms of
  information processing architecture. This focuses on general
  biological patterns rather than specific implementations.
  (Carbon-Based is an indicator for this feature)
\item[Self-Modeling] The system actively creates and maintains a functional
  model of its own capabilities, states, and characteristics. This model
  enables prediction of its own behavior, understanding of its
  limitations, and differentiation between self and others. The
  self-model guides internal state monitoring, behavioral regulation,
  and adaptive responses to changing conditions. (Self-Representations
  are an indicator for this feature)
\item[Intelligence] The system demonstrates the ability to acquire and apply
  knowledge, reason abstractly, solve novel problems, and adapt to new
  situations. This includes capabilities such as pattern recognition,
  learning from experience, and making appropriate decisions based on
  available information. (ARC Performance is an indicator for this
  feature)
\end{description}

A full list of features can be found in \autoref{appendix-a-stances-features}.

\subsection{Stances}\label{stances}

A stance is a set of commitments regarding which features matter for
consciousness and how they matter. There are currently 13 stances in the
model. Each stance generates a probability that a system is conscious
conditional on the parameters stipulated for the features and their evidential relevance. 

Some of the stances reflect well-developed theories of consciousness
identifying architectural, cognitive, or other features that give rise
to consciousness. Other stances are more informal, embodying a general
perspective on when we should attribute consciousness; e.g. whether the
system seems like a person when you interact with it, or it is
biologically similar to creatures known to be conscious.

\begin{table}[htbp]
\centering
\small
\begin{tabular}{@{}p{3.5cm}p{10cm}@{}}
\toprule
\textbf{Stance} & \textbf{Brief description of perspective on consciousness} \\
\midrule
Attention Schema & Consciousness is indicated by the presence of an internal model representing the distribution of attentional resources in the system with the function of controlling attention. \\
\addlinespace
Biological analogy & Consciousness is evidenced by broad and diverse analogies with living biological organisms. \\
\addlinespace
Cognitive complexity & Consciousness is indicated by high levels of cognitive complexity, defined by the richness and interrelatedness of their internal processing. \\
\addlinespace
Computational analogy & Consciousness is evidenced by overall functional resemblance to information processing in humans, across domains such as reasoning, perception, language, and decision-making. \\
\addlinespace
Embodied agency & Consciousness is indicated by perceptual feedback mechanisms to control a body in a goal-directed fashion. \\
\addlinespace
Field mechanisms & Consciousness is indicated by the involvement of integrated and causally efficacious electromagnetic fields (EMF) in cognition. \\
\addlinespace
Global Workspace Theory & Consciousness is indicated by a centralized representation repository with a broadcasting mechanism to make select information widely available for use by various specialized processes or modules. \\
\addlinespace
Higher-Order Thought & Consciousness is indicated by internal representations of the systems' own mental states, such as thoughts whose content includes in the system's perceptual states. \\
\addlinespace
Integrated Information Theory & Consciousness is indicated by the product of integration structures: specifically measured by the irreducibility of the system's diverse causal powers to those of its parts. \\
\addlinespace
Person-like & Consciousness is evidenced by traits resembling those associated with human personhood. Interacting with it feels like interacting with a person. \\
\addlinespace
Recurrent processing (perceptual) & Consciousness is associated with the iterative refinement of perceptual representations through structured, feedback-driven loops. \\
\addlinespace
Recurrent processing (pure) & Consciousness is associated with dynamics of recursive processing loops over incoming or internally generated data. \\
\addlinespace
Simple valence & Consciousness is indicated by valenced representations. \\
\bottomrule
\end{tabular}
\caption{Stances included in the DCM. More detailed descriptions of each stance, relevant citations, and specifications in terms of features can be found in \autoref{appendix-a-stances-features}.}
\label{table:stances}
\end{table}

We report the posterior probability (the probability after updating on
the indicator evidence) of consciousness generated by each stance. To
get an all-stances-considered estimate of a system's probability of
consciousness (based on the current model and set of expert judgments)
we report the average of the stances' judgments, weighted by their
plausibility as rated by experts.

\subsection{General Bayesian approach}\label{general-bayesian-approach}

Bayesian models provide a way to update probabilities with new evidence,
given assumptions about the probabilities of receiving the evidence
under different hypotheses \citep{jeffrey2004subjective}. Bayesian models are
distinguished from other forms of probabilistic modelling by the use of
prior probabilities and updating those priors via likelihood ratios. We
will explain the specific details of our Bayesian hierarchical model in
the next section.

\subsubsection{Prior probabilities}\label{prior-probabilities}

A Bayesian model does not tell us how probable a hypothesis is based
solely on our observations. Instead, it tells us how much our initial
probability in the hypothesis should change in light of the evidence. We
can derive how probable the hypothesis is given how probable it we took
it to be to begin with. Our Bayesian model therefore requires us to
assign a prior probability that consciousness is present in a given
system before we've observed any of its indicator traits.

The choice of a prior is often contentious and subjective, and perhaps
particularly so for consciousness \citep{andrews2024all, sebo2025everything}.
Ideally, our prior should reflect a state of uncertainty before any
indicators are considered: the prior might be thought as the probability
that an arbitrary system is conscious, or the probability that an
arbitrary system in a given class of potential candidates is conscious.

We would like to avoid having the choice of prior be decisive to our
results. We mitigate the influence of the prior in three ways. First, we
chose the same prior probability of consciousness ($\frac{1}{6}$) for each system
under consideration. In effect, this excludes any information about
these systems other than the indicators in the model in an attempt to
remove biases about which systems are likely conscious.\footnote{This is not essential to the model. In principle, our sensitivity results
  permit readers to select the combinations of priors that they find
  most plausible. In Section 7.1., \autoref{fig:differential-priors}, we illustrate the results
  of assigning high priors to humans and chickens and low priors to AI
  systems.} We chose this prior in part because it yields easily
discernible differences in posteriors across systems, which is useful
for illustration (if priors are very low or very high, posteriors are
more tightly clustered).

Second, we performed sensitivity tests (see \autoref{appendix-e-prior-sensitivity-tests}) to show how the
posterior probabilities depend on changes to the prior. Third, we urge
caution in treating the posterior probabilities that the model generates
as definitive. Instead, we think that the more robust results of the
model are the direction and magnitude of the evidence for each system
and the ordinal rankings of the posterior probabilities for different
systems we consider.\footnote{We chose to present the posterior probabilities of consciousness that result from the model, instead of
  just presenting ordinal results, because we want to be transparent
  about how the model works. Further, some readers may endorse certain
  priors, in which case they may also endorse the posteriors that follow
  from them.}

In our Bayesian hierarchical model, priors flow downward from
consciousness to features and indicators. For each stance, we start by
assuming a prior probability that the system is conscious. This sets
stance-relative prior expectations that the system possesses certain
features and indicators. For example, if a stance were to posit that a
certain feature F was necessary for consciousness, then assuming a
particular prior probability of consciousness would entail that F has a
prior probability at least as high.

\subsubsection{Bayesian updating}\label{bayesian-updating}

Our Bayesian model posits evidential linkages between variables in
adjacent layers: indicators provide evidence for features; features
provide evidence for consciousness via stances (which specify which
features are evidence for consciousness and the strength of that
evidence). Bayesian updating occurs via likelihoods, judgments of how
probable the evidence is, conditional on the hypotheses in question. For
example, if an indicator is much more probable if feature
$F_i$ is present than when $F_i$ is absent,
then the presence of the indicator provides evidence that
$F_i$ is present in the system.

We use two parameters to characterize these evidential relationships
which together generate the likelihood ratio. We illustrate using
feature-to-consciousness updating, though the same holds for
indicator-to-feature updating. Evidential relationships are specific to
each stance.

First: even if the system is not conscious, how likely is it to have
feature F anyway?\footnote{There are some features that are conducive to consciousness according to a stance, but we don't take
  their absence as being particularly meaningful. These are typically
  treated as undemanding, and can be conceptualized as likely requiring
  no significant degree of the feature. For instance, biological
  similarity provides evidence for computational similarity, but only in
  an undemanding way. In general, we don't expect the absence of
  biological similarity to rule much against computational similarity,
  even though we think most highly biologically similar systems to be
  computationally similar too.} A feature is demanding to the extent
that non-conscious systems rarely have it; we call this
\textbf{demandingness}. You might think of it in terms of the background
prevalence of the feature, or perhaps the degree to which the feature
needs to be present (for features that come in degrees). In statistical
terms, demandingness is inversely related to the false positive rate,
$\Pr(F \mid S \text{ is not conscious})$: an undemanding feature is one
with many false positives. Very
demanding features are features we expect to be pretty rare (and much
more common in conscious systems). The neutral point of our scale here
reflects that the feature is just as likely to be present as absent in
unconscious systems.

Second: Should we expect conscious systems to have F at much higher
rates than non-conscious systems? If a system is conscious, how much
more likely is it to have feature F than if it was not conscious? We
call this \textbf{support}, which you might think of as the differential
prevalence of F. In statistical terms, it's the positive likelihood
ratio, $\Pr(F \mid S \text{ is conscious}) / \Pr(F \mid S \text{ is not conscious})$,
i.e. Sensitivity / (1$-$Specificity). The neutral point here reflects that the feature is no
more likely if the system is conscious than otherwise.

We specify stances by the evidential connections they posit between
features and consciousness, which we arrived at from a targeted
literature review and consultations with subject area experts. For
example, according to the Attention Schema stance, consciousness is
generated by an internal model of the distribution of attentional
resources in the system (Graziano \& Webb 2015, Graziano \emph{et al.}
2020). Therefore, the Selective Attention feature provides strong
support for consciousness on this stance, as conscious systems have a
very high probability of having Selective Attention. Because Selective
Attention is not sufficient for consciousness but is correlated with it
(there are systems that have attention but not an attention schema), the
feature is only weakly demanding.

Evidential support flows upward in the model: the indicator values for a
system provide evidence regarding the features it possesses, and its
features provide evidence regarding its consciousness relative to each
stance. Evidential relationships between features and subfeatures, or
subfeatures and indicators, are not stance-dependent. That is, we assume
that any perspective of consciousness does not affect how strong a
relationship we initially think there is between, say, Adaptive Focus (a
subfeature) and Attention (a feature). These linkages can change through
Bayesian updating on our uncertainty about the relationships, but they
are initially stance-independent. More detail about the evidential
parameters in the model can be found in the next section and in Appendix
D.

\section{Model specifications}\label{model-specifications}

This section describes how our model updates prior probabilities into
posterior probabilities, given the indicator evidence. Readers who want
the technical details should read this section and see Appendices C and
D for more details. Otherwise, readers should feel free to skip to the
next section for the results of the model.

\subsection{Overview}\label{overview}

The DCM is formulated from Bayesian hierarchical methods developed for
mixture models and classification (Stephens 1997, Kemp \emph{et al.}
2007, Tu 2014). In these types of models, there exists a common
reference class of objects that share similar kinds of properties, and
you're trying to use observable evidence to sort those objects by
whether they have some latent property you care about. The model was
developed and run in PyMC.

The Bayesian model structure that forms the core of the DCM has the
following levels:

\subsubsection{Consciousness}\label{consciousness}

In our model, consciousness is treated as a binary state, meaning it's
either present or absent in a given system.\footnote{The assumption that consciousness is binary (either present or absent) is controversial
  \citep{schwitzgebel2020is, schwitzgebel2023borderline, godfrey2017evolution,
  godfrey2020metazoa, antony2008are, simon2017vagueness, roelofs2019combining}. Perhaps
  consciousness is vague or comes in degrees. If one holds these views,
  it may be possible to interpret the output of the model as giving the
  degree of determinacy of a system's consciousness or the expected
  value of a quality (consciousness) that comes in degrees, instead of a
  probability that a determinate, binary state is present. In Section
  4.5.1. we discuss how to interpret other continuous variables treated
  as binary variables in the model.} At the start of the model, we give
each system a distribution over its prior probability of consciousness.

\subsubsection{Features}\label{features-1}

Stances specify that certain general features (or absence thereof)
provide evidence of different strengths for or against consciousness.
This model represents whether each of these unobserved features is
present in a system to a sufficient degree as a Bernoulli variable (that
is, a binary variable taking a value of 1 if the feature is sufficiently
present and 0 if the feature is absent). Features are linked to the
consciousness level using pre-specified conditional prior distributions,
which depend on whatever stance we're assuming is true. These
consciousness-feature linkages are flexible to updating (in a Bayesian
manner) because we place a prior distribution over them as well.

\subsubsection{Subfeatures}\label{subfeatures}

Each feature can be broken down into subfeatures that, collectively,
provide evidence for or against their ``parent'' feature's presence. For
instance, one subfeature that's linked to the attention feature is
Adaptive Focus, defined as a system being able to shift focus depending
on current task demands. This level of subfeatures (also modelled as
Bernoulli variables) is linked to the feature level by assumptions of
how strong a relationship the subfeatures have with their respective
feature. The priors for the feature-subfeature linkages do not depend on
the perspective on consciousness used, but are likewise flexible to
updating. In some cases, subfeatures have their own subfeatures. In
other cases, there are no subfeatures mediating indicators and features.
In this sense, subfeatures are an optional part of the model.

\subsubsection{Indicators}\label{indicators-1}

Evidence for or against a subfeature's presence is provided by
indicators, which function as locus for the data inputted into the
Bayesian model to start the updating process. The Bayesian core of the
model includes indicators as binary Bernoulli variables, and it assumes
they're directly observed. These indicators are linked to their
respective subfeatures by pre-specified likelihood relationships which
are themselves flexible to updating. To respect the uncertainty around
indicators, we run the model many times to get a distribution of
results, and treat the probabilities of the indicators as the
probabilities of absolute presence or absence in each run.

\subsection{Specifying conditional dependencies}\label{specifying-conditional-dependencies}

The strength of a piece evidence is given by the likelihood ratio: the
extent to which data $D$ favors hypothesis $H_1$ over
$H_2$ is proportional to the likelihood ratio,
$\Pr(D|H_1) / \Pr(D|H_2)$. For example, an
indicator $I_i$ is evidence for the presence of feature
$F_j$ when and to the extent that $I_i$ is
more probable when $F_j$ is present than when
$F_j$ is absent: $\Pr(I_i|F_j) / \Pr(I_i|\neg F_j) > 1$. A feature $F_j$ is evidence that the
system is conscious when and to the extent that $F_j$ is
more probable given that the system is conscious than that it is not:
$\Pr(F_j|S \text{ is conscious}) / \Pr(F_j|S \text{ is not conscious}) > 1$. The model updates on indicators or features
that are observed to be absent in something like this fashion, though it
also incorporates higher-level uncertainties about the relationships
themselves. We do not update on indicators for which we lack data (e.g.
experts withheld judgment about them).

For the conditional dependencies relating each level in the hierarchy,
we must specify approximately how likely the child variable is present
or absent, given that the parent variable is present or absent.
Conditional dependencies in the model are derived from support and
demandingness specifications. For each stance, we specified the set of
features that are evidentially relevant to consciousness according to it
and classified the strength and direction of support and demandingness
for each. Likewise, we specified the strength and demandingness
relationships among indicators, subfeatures, and features.

Each support and demandingness relationship is measured along a nine
point scale: {[}overwhelmingly negative, strongly negative, moderately
negative, weakly negative, neutral, weakly positive, moderately
positive, strongly positive, overwhelmingly positive{]}. We expect to
update most strongly on features that are observed to be present,
demanding, and high in support (which count for) and features that are
absent, undemanding, and high in support (which count against).

\begin{table}[htbp]
\centering
\small\parbox{0.8\textwidth}{
\caption{Base parameter values for support and demandingness before combination. These values represent the isolated effect of each parameter. When combined in the model, support values are amplified by a factor dependent on the demandingness level (see explanation below), and both are transformed into Beta distribution parameters. Table~3 shows the resulting combined likelihood ratios.}}
\label{tab:option-values}
\setlength{\tabcolsep}{6pt}
\renewcommand{\arraystretch}{1.05}
\begin{tabularx}{0.8\textwidth}{@{}l@{\extracolsep{\fill}}r@{}}
\toprule
\textbf{Option} & \textbf{Value} \\
\midrule
\multicolumn{2}{@{}l@{}}{\textbf{Support}}\\
Overwhelming support & (50, 1) \\
Strong support & (8, 1) \\
Moderate support & (3, 1) \\
Weak support & (1.5, 1) \\
No support & (1, 1) \\
Weak countersupport & (1, 1.5) \\
Moderate countersupport & (1, 3) \\
Strong countersupport & (1, 8) \\
Overwhelming countersupport & (1, 50) \\
\addlinespace
\multicolumn{2}{@{}l@{}}{\textbf{Demandingness}}\\
Overwhelmingly demanding & (50, 1) \\
Strongly demanding & (8, 1) \\
Moderately demanding & (3, 1) \\
Weakly demanding & (1.5, 1) \\
Neutral & (1, 1) \\
Weakly undemanding & (1, 1.5) \\
Moderately undemanding & (1, 3) \\
Strongly undemanding & (1, 8) \\
Overwhelmingly undemanding & (1, 50) \\
\bottomrule
\end{tabularx}
\end{table}

% Support levels correspond to relative probabilities that the child is
% present given that the parent is present vs.\ that the child is present
% given the parent is absent: $\Pr(\text{child}^+ | \text{parent}^+) / \Pr(\text{child}^+ | \text{parent}^-)$. In more common statistical terminology, Support
% corresponds to Sensitivity / (1$-$Specificity). For example, if a feature
% provides \emph{overwhelmingly positive} support, then a conscious system
% is 50x more likely to have the feature than a non-conscious system
% is\footnote{We tweak this calculation slightly, so that support options are somewhat more extreme when paired with very demanding settings
%   than when paired with undemanding settings, on the grounds that we
%   expect high support features to still be more likely to be present
%   than not even for highly demanding features. Otherwise overwhelmingly
%   supporting features that are overwhelmingly demanding are less
%   diagnostic than we expect even when present.}. If a feature provides
% \emph{moderate countersupport} (negative support), then a non-conscious
% system is 3x more likely to have the feature than a conscious system is.
Support and demandingness values in Table 2 are base ratios derived from the conditional probabilities.
Support is the positive likelihood ratio, $\Pr(F \mid S \text{ is conscious}) / \Pr(F \mid S \text{ is not conscious})$, i.e. Sensitivity / (1$-$Specificity).
Demandingness is expressed on the odds scale: the value is the odds that a non-conscious system lacks the feature, $\Pr(\neg F \mid S \text{ is not conscious}) / \Pr(F \mid S \text{ is not conscious})$, i.e. Specificity / (1$-$Specificity).\footnote{Demandingness can equivalently be stated as a probability, the false positive rate itself, and some implementations that reuse our labels do so. The two forms carry the same information and always agree on which of two features is more demanding: a demanding value of $D$ corresponds to a false positive rate of $1/(1+D)$, and an undemanding value of $D$ to a false positive rate of $D/(1+D)$. We use odds here so that demandingness and support share a scale, both are ratios centred on 1 with symmetric steps in each direction, and so that the two parameters enter the Beta construction below in the same multiplicative way (step 1 encodes exactly this conversion).} These base values are transformed before use in the model. When support and demandingness are combined, we apply an amplification factor to the support parameter that depends on the demandingness level:
\begin{itemize}
  \item Overwhelmingly demanding features: support $\times$ 8
  \item Strongly demanding features: support $\times$ 3
  \item Moderately demanding features: support $\times$ 1.5
  \item All other demandingness levels: support $\times$ 1
\end{itemize}
This amplification reflects the intuition that highly diagnostic (demanding) features should weight supportive evidence more heavily. The amplified support and base demandingness values are then used to construct Beta distribution parameters via the following process: We convert support and demandingness into Beta priors through the following procedure:
\begin{enumerate}
  \item Construct base absence parameters from demandingness: For demanding features: $(3,\, 3 \times D_{\text{value}})$
  \item For undemanding features: $(3 \times D_{\text{value}},\, 3)$
  \item Where $D_{\text{value}}$ is the demandingness ratio from Table 2
  \item Amplify support by demandingness level (as described above)
  \item Construct presence parameters: $\alpha_{\text{present}} = \alpha_{\text{absent}} \times \text{amplified support}$
  \item $\beta_{\text{present}} = \beta_{\text{absent}}$
  \item Normalize both to concentration parameter $\kappa = 10$: For each pair $(\alpha, \beta)$, set $(\alpha', \beta') = (10\alpha/(\alpha+\beta),\, 10\beta/(\alpha+\beta))$
  \item This preserves the mean probability while standardizing total concentration
\end{enumerate}
% Demandingness levels correspond to the relative probabilities that the
% child is present vs.\ absent given that the parent is absent: $\Pr(\text{child}^- | \text{parent}^-) / \Pr(\text{child}^+ | \text{parent}^-)$. In statistical
% terminology, Demandingness corresponds to Specificity / (1$-$Specificity).
% For example, if a feature is \emph{overwhelmingly demanding}, then a
% non-conscious system is 50x less likely to have the feature than to lack
% it. A \emph{moderately undemanding} feature is 3x more likely to be
% present than absent in unconscious systems.

Instead of directly fixing the likelihoods, we put prior distributions
over different likelihoods (as discussed in \autoref{appendix-c-model-structure}). In principle,
this allows the judgments of evidential strength to be updated with new
evidence.\footnote{In typical modeling tasks, we would update these evidential parameters when we learn facts about similar kinds of
  systems. In this case, we might for example learn that feature F is
  highly represented among conscious systems and therefore update the
  evidential relationship between F and consciousness. However, since we
  can never directly assess which systems are conscious, so any updates
  would have to be on other kinds of evidence (such as common judgments
  that systems with F are conscious).} We can manipulate how much these
parameters will change by setting the concentration parameter. The 
concentration parameter $\kappa = 10$ refers to the sum $\alpha + \beta$ of 
the final Beta distribution parameters after transformation. 
This standardization does not affect the mean probabilities 
(which determine likelihood ratios) but constrains the degree of 
certainty about those probabilities. A higher concentration would 
make the Beta distributions more peaked around their means; our 
choice of $\kappa = 10$ allows moderate flexibility for Bayesian 
updating while maintaining the intended likelihood ratio structure.

A parent-child relationship is the combined result of both the distributions
over the support and demandingness parameters, from which we construct
beta parameterizations. These generate likelihood ratios for child
values given the presence or absence of parents, which update the
probabilities that parent variables are present.

\begin{table}[htbp]
\centering
\small
\parbox{0.9\textwidth}{
\caption{Selected approximate likelihood ratios for combined support and demandingness parameters after amplification and Beta parameterization. These reflect the actual evidential weights used in the model. Note that LR$+$ values are generally lower than the product of base support and demandingness values due to the transformation process.}}
\label{tab:likelihood-ratios}
\begin{tabularx}{0.9\textwidth}{@{}>{\raggedright\arraybackslash}l@{\extracolsep{\fill}}rr@{}}
\toprule
\textbf{Combination} & \textbf{Likelihood} & \textbf{Likelihood} \\
 & \textbf{ratio (+)} & \textbf{ratio (-)} \\
\midrule
Overwhelming support, overwhelmingly demanding & 45 & 0.1 \\
Overwhelming support, strongly demanding & 8.5 & 0.06 \\
Overwhelming support, neutral demandingness & 2 & 0.04 \\
Strong support, strongly demanding & 6.7 & 0.3 \\
Strong support, strongly undemanding & 1.1 & 0.14 \\
Moderate support, overwhelmingly demanding & 16.5 & 0.69 \\
Moderate support, moderately demanding & 2.4 & 0.55 \\
Weak support, overwhelmingly demanding & 9.8 & 0.8 \\
Weak support, weakly demanding & 1.2 & 0.88 \\
No bearing, any level of demandingness & 1 & 1 \\
\bottomrule
\end{tabularx}

\vskip\tableelementskip
\begin{minipage}{0.9\textwidth}
\footnotesize \textit{Note.} In other words, this table states selected approximate likelihood ratios for the presence of a feature or indicator, corresponding to combinations of support and demandingness levels. The (+) column reflects the evidential impact of confirming the presence of the feature/indicator; the (-) column reflects confirming its absence. Values closer to 1 indicate weaker evidence.
\end{minipage}
\end{table}

In \autoref{appendix-d-conditional-dependencies}, we perform an initial analysis of how sensitive the
model's results are to our particular mappings of categories of
evidential strength to particular likelihood ratios and beta
parameterizations. If we use a smaller number of evidential categories,
the general results are directionally similar but give much more
moderate results for systems that experience large updates with the
original parameters. Further sensitivity tests should map existing
categories to different beta parameterizations to evaluate effects on
model updates.

\subsection{Generating indicator values from expert surveys}\label{generating-indicator-values-from-expert-surveys}

To incorporate uncertainty about the indicators, we surveyed experts
about the probability they'd assign to each indicator being present in a
given system. The current model was run on 16 expert surveys for LLMs (6
full survey responses, 10 subset responses), 2 for chickens, and 1 each
for humans and ELIZA. Survey participants were recruited via targeted
emails of subject-area experts on system capacities (not necessarily
experts on digital consciousness in particular). Those who took the
entire survey were compensated for their time. See Limitations and
Future Work for a discussion of improved survey methods for future
iterations of the model.

Using these expert-given probabilities, we randomly generate hundreds of
sets of binary indicator values to plug into the core Bayesian model. We
use the expert's specified subjective probability to randomly draw the
indicator's value from a Bernoulli distribution, where the probability
of an indicator being sampled as present is equal to the probability
given by the expert. We can run the model separately for each expert to
get their estimated probabilities of consciousness. For the results
reported here, we took the mean probability judgment across all expert
surveys for each indicator.

Though the uncertainty about the indicators is lost for any given model
run, we can recapture that uncertainty by repeating the model run
process for several sets of randomly generated indicator values. By
simulating enough sets of indicator values, running the model for each
set of indicator values, and averaging the results, the law of large
numbers will produce a mean probability of consciousness across these
model runs that recaptures our expert's uncertainty about the
indicators' true values.

\subsection{Running the model}\label{running-the-model}

In a single run of the model, the set of indicator values that was
randomly generated from expert credences is used as input, and the model
generates a mean estimate for the system's posterior probability of
consciousness for that given indicator set, expert set, and stance (i.e.
by taking the mean number of times the consciousness variable equalled 1
across all samples in a given model run. In future iterations of the
model, we could analyze the entire distribution of the posterior
probabilities of consciousness.

\subsection{Modeling assumptions}\label{modeling-assumptions}

\subsubsection{Binary variables}\label{binary-variables}

During each model run, all variables in the model are assumed to be
binary. The corresponding trait is either present or absent. While this
is realistic for some variables (e.g. the system is or isn't conscious;
it does or does not use neurotransmitters), other variables represent
properties that are continuous (e.g. intelligence, flexibility). We
interpret the binary presence of a property that comes in degrees as
representing that the property is present to a sufficient degree to
count as evidence (for a parent feature or for consciousness).

While it is possible to use continuous variables within a Bayesian
model, it would require representing the likelihood relationships
involving continuous variables. Specifying such relationships is
challenging, and utilizing them would be computationally expensive. In
our model, the continuous nature of properties in the model is
re-constructed by stochastically sampling model runs from continuous
inputs to the model (i.e. continuous expert credences in indicators) and
reporting the mean of many model runs. Nevertheless, discretizing
continuous variables does result in a loss of information. In future
work, we may explore the effects of using continuous variables or more
complex discrete variables (e.g. feature is absent, low, medium, or
high) to evaluate the effects on model outcomes.

\subsubsection{Conditional independence}\label{conditional-independence}

We assume that each node's children are independent, conditional on the
state of their parent. This assumption is unrealistic, since many
indicators reflect capabilities that originate from a common cause or
are otherwise linked. Treating highly correlated indicators as
independent presents a risk of double-counting; when two pieces of
evidence are correlated, updating on each of them may overestimate the
strength of the evidence.\footnote{In a famous example, \citet{wittgenstein2009philosophical} considers a man who doubts the reliability of a story in
  the newspaper, so he buys another copy of the same newspaper to double
  check. It would clearly be a mistake to treat these as two pieces of
  evidence and to update on both. If he bought two newspapers that based
  their stories on the same wire service report, then this would provide
  more evidence than buying the same paper twice, but it would still
  fall short of the evidence provided by two newspapers that derived
  their stories from totally independent reporting.} We think this risk
is partially mitigated by the fact that potentially double-counted
indicators bottleneck through shared parent subfeatures with their own
prior probabilities.

The conditional independence condition is also implausible at the level
of stances and features. According to the way the model is structured,
each feature makes an independent contribution to the probability of
consciousness. However, some theories posit that features are highly
correlated or that features are only evidentially relevant in certain
combinations (for example, if two features are jointly necessary).

We assumed conditional independence to make the model more computational
tractable. It is also much more difficult to specify the likelihood
relations among correlated variables, and doing so would increase the
risks of overfitting with added parameters. Future versions of the model
could add correlations among features to better characterize the
commitments of certain stances.

\subsubsection{Indicator independence}\label{indicator-independence}

Given that indicators are treated as binary variables, and expert data
relates to probabilities of indicator presence, we must do something to
convert probabilities into values. We randomly collapse probabilities
into variables independently of one another. For example, if experts
assign a 0.8 probability to indicator A and a 0.9 probability to
indicator B, 80\% of the simulations will have A present and 90\% will
have B present but whether A and B are present in any given simulation
are independent of one another. This is somewhat unrealistic, as closely
related variables may turn out to be less similar in simulation than we
should expect in practice. For example, we might only be able to clearly
assess that A is present in systems that also clearly show that B is
present, so experts' credences will be more correlated than our
simulation method makes them out to be.

In contrast to conditional independence, the effect of this assumption
risks undercounting the significance of various capabilities. Though the
formal structure can in practice handle both positive and negative
indicators, we have been inclined to focus on positive indicators. And
while it is possible for indicators to be anti-correlated, we think it
is much more common for the indicators we've identified to be
correlated. Together, this means that we are likely to see more extreme
results the more indicators are co-present (higher probabilities of
consciousness) or co-absent (lower probabilities of consciousness).

At present, we see this as a practical limitation of this kind of formal
modelling that would be somewhat difficult to avoid.

\section{Model Results: Individual Stances}\label{model-results-individual-stances}

\subsection{Data}\label{data}

We ran the model for four target systems:

\begin{itemize}
\item
  2024 LLMs: state-of-the-art models in 2024, such as Gemini 2.5 Pro,
  GPT 4, and Claude 3 Opus, which do not include more recent reasoning
  models.
\item
  Humans
\item
  Chickens
\item
  ELIZA: natural language processing computer program developed in 1960s
\end{itemize}

We used expert surveys to gather indicator data for each system. For
LLMs, 6 survey respondents answered questions about every indicator, and
10 additional experts answered subsets of the survey questions. We ran
the model on the mean expert score for each indicator, which allowed
full and partial survey results to be aggregated. In future work, we can
explore cross-expert variation by comparing results for each full expert
survey. We have data from 2 complete expert surveys for chickens and we
supplied our own answers for humans and ELIZA, which we deemed
sufficiently uncontroversial.

\subsection{Results for each stance}\label{results-for-each-stance}

We ran the model separately for each stance across each system. The
graphs below report the posterior probabilities relative to each stance,
from an expected distribution of prior probabilities centered at $\frac{1}{6}$
(which is marked by a dashed line). The shaded region depicts the spread
of posteriors across model runs, which reflects the amount of
uncertainty about the probability of consciousness. The median posterior
across model runs is marked with a horizontal bar.\footnote{The median value is less affected by the abnormally high or low results that are
  occasional outcomes of the stochastic modeling process. The mean
  values of the distributions were very similar to the medians, although
  the mean is `pulled up' by the uncertain tails in some stances, e.g.
  Simple Valence.}

\subsubsection{2024 LLMs}\label{llms}

\begin{figure}[H]
\centering
\includegraphics[width=\textwidth,keepaspectratio]{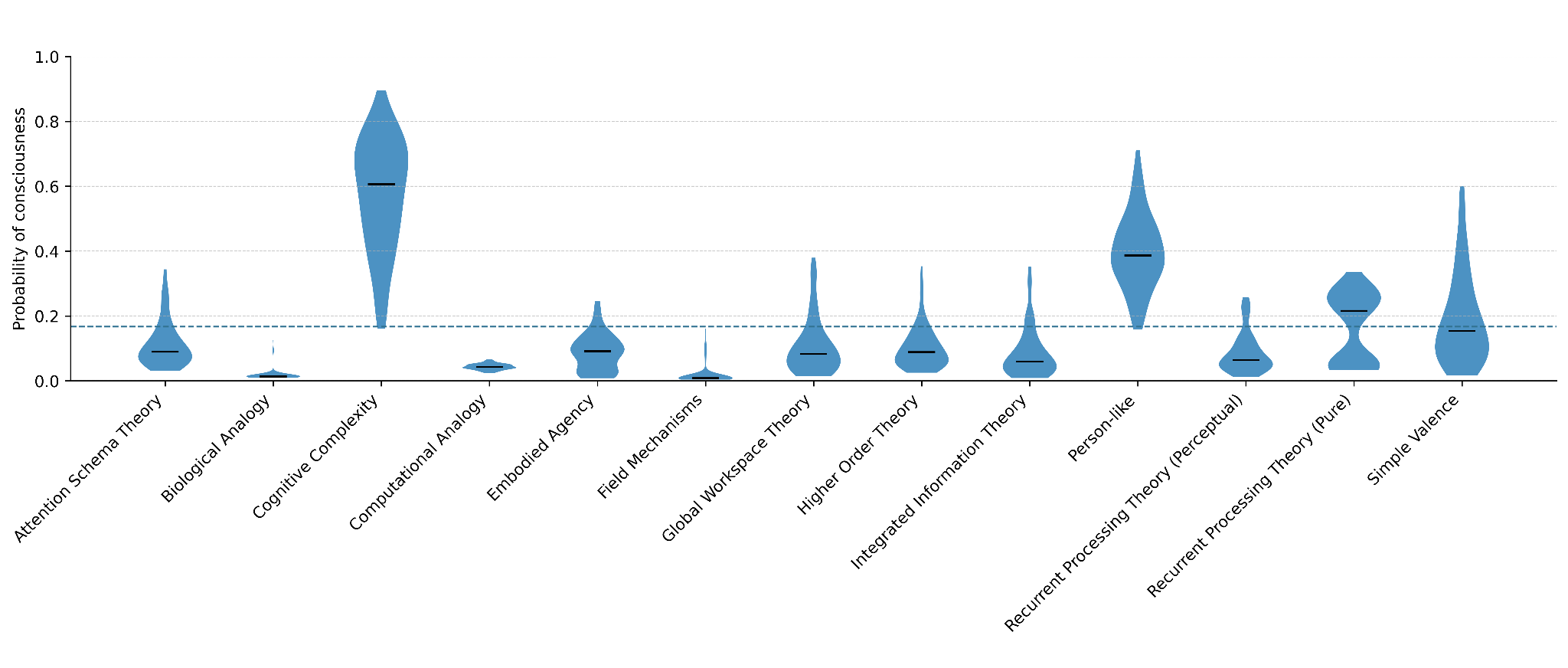}
\caption{Individual stance judgments about the posterior probability of consciousness in 2024 LLMs, starting from a prior probability of $\frac{1}{6}$ (dashed blue line). The variation in probability outcomes across model runs results from the different ways of resolving uncertainty about the presence of individual indicators.}
\label{fig:llms-posterior}
\end{figure}

The aggregated indicator data raised the median probability that 2024
LLMs are conscious on 4 stances: Cognitive Complexity, Person-like,
Recurrent Processing Theory (Pure), and Simple Valence. The evidence was
disconfirmatory of LLM consciousness on the remaining stances, and most
strongly disconfirmatory for the Biological Analogy and Computational
Analogy stances.

Median posterior probabilities ranged from 0.02 (Field Mechanisms and
Biological Analogy) to 0.57 (Cognitive Complexity), though we caution
against taking these values too seriously given their dependence on a
largely arbitrary choice of prior probability. Comparisons between the
posterior and prior, and between LLMs and other measured systems, are
more informative (see discussion below).

\subsubsection{Chickens}\label{chickens}

\begin{figure}[H]
\centering
\includegraphics[width=\textwidth,keepaspectratio]{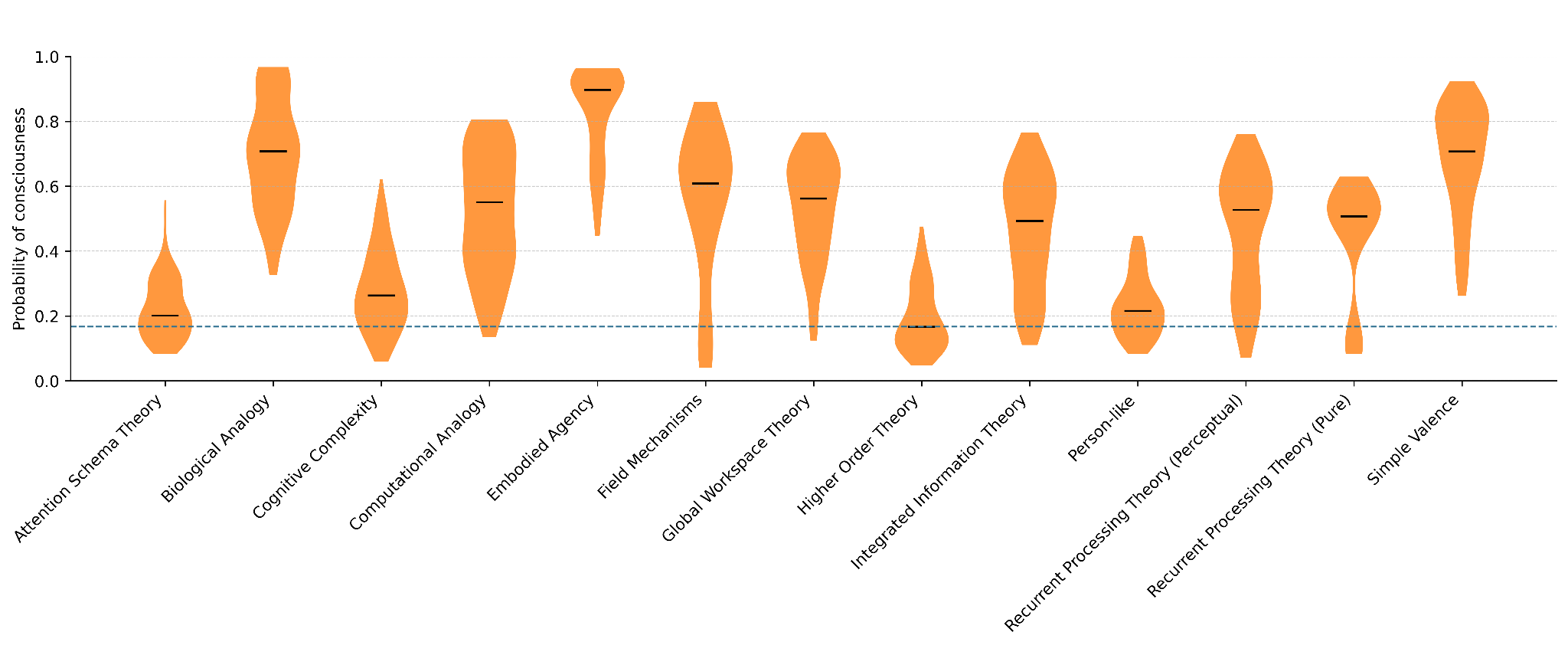}
\caption{Individual stance judgments about the posterior probability of consciousness in chickens, starting from a prior probability of $\frac{1}{6}$ (dashed blue line). The variation in probability outcomes across model runs results from the different ways of resolving uncertainty about the presence of individual indicators.}
\label{fig:chickens-posterior}
\end{figure}

Chicken consciousness was confirmed relative to all stances. It was
least confirmed relative to Attention Schema Theory and Higher-Order
Theory, two stances that emphasize the importance of metacognitive
abilities. It was most strongly confirmed according to the Biological
Analogy, Embodied Agency, and Simple Valence stances. Chickens received
higher posteriors than LLMs relative to each stance except for Cognitive
Complexity and Person-like perspectives. Posterior probabilities ranged
from 0.2 (HOT) to 0.82 (Embodied Agency), reflecting significant
uncertainty about animal consciousness across well-known theories of
consciousness (see Limitations and Future Work for further discussion).
There was also significant uncertainty within most stances, which in
part reflects uncertainty or disagreement among experts about whether
chickens possess relevant indicators.

\subsubsection{Humans}\label{humans}

\begin{figure}[H]
\centering
\includegraphics[width=\textwidth,keepaspectratio]{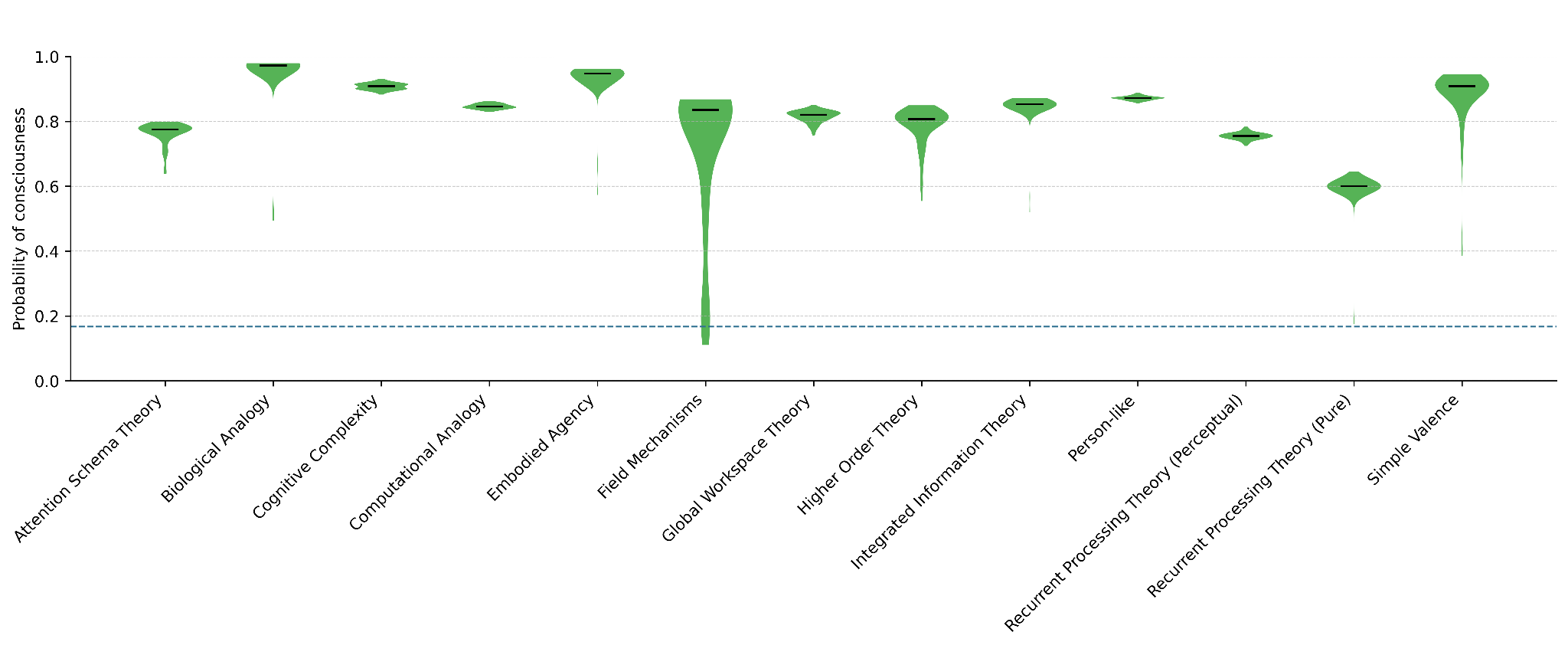}
\caption{Individual stance judgments about the posterior probability of consciousness in humans, starting from a prior probability of $\frac{1}{6}$ (dashed blue line). The variation in probability outcomes across model runs results from the different ways of resolving uncertainty about the presence of individual indicators.}
\label{fig:humans-posterior}
\end{figure}

Consciousness in humans is strongly confirmed by the indicator evidence
relative to every stance in the model.\footnote{As we will discuss in more detail later, the model assumes that all we know about a given
  system is its values for the 200+ indicators and a prior probability
  of consciousness. For humans, this means that we are assessing humans
  as if all we knew about them was their indicator values (excluding
  other sources of knowledge we have about humans, such as introspective
  awareness that we are conscious).} Posteriors were higher than both
LLMs and chickens on every stance. They ranged from 0.6 (Recurrent
Processing (Pure)) to 0.96 (Biological Analogy). There was less
uncertainty within stances than for other systems, reflecting the
smaller number of experts and less uncertainty about whether humans
possess key indicators. The relatively low scores on some stances
generally says more about the paucity of good indicators for those
stances than the failure of humans to exhibit them (see Limitations and
Future Work for further discussion).

\subsubsection{ELIZA}\label{eliza}

\begin{figure}[H]
\centering
\includegraphics[width=\textwidth,keepaspectratio]{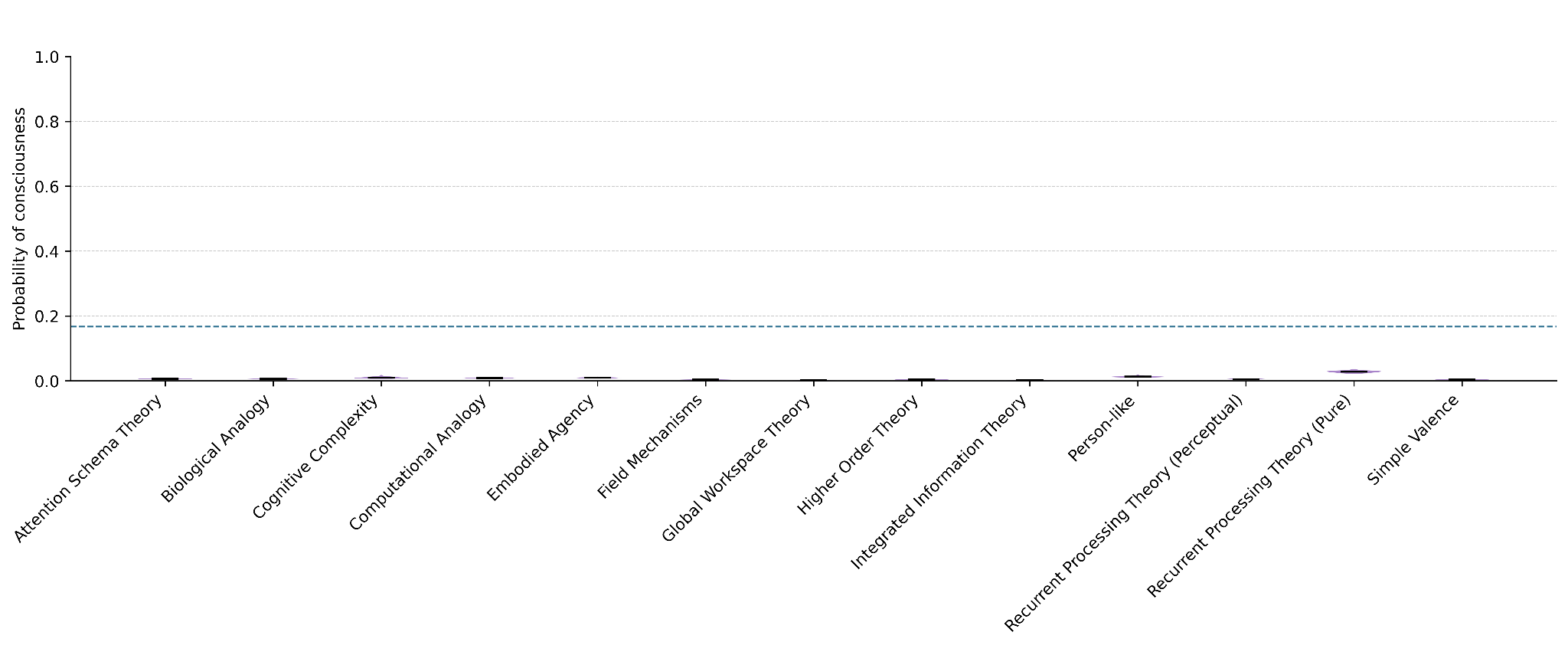}
\caption{Individual stance judgments about the posterior probability of consciousness in ELIZA, starting from a prior probability of $\frac{1}{6}$ (dashed blue line).}
\label{fig:eliza-posterior}
\end{figure}

ELIZA was an early natural language processing computer program
developed in the 1960s \citep{weizenbaum1966eliza, weizenbaum1976computer, norvig2014paradigms,
berry2023finding}.\footnote{You can interact with a re-implementation of ELIZA, constructed from Weizenbaum's original
  code, here:
  \href{https://sites.google.com/view/elizaarchaeology/try-eliza}{\ul{https://sites.google.com/view/elizaarchaeology/try-eliza}}}
Its most famous instantiation was a chatbot designed to behave like a
psychotherapist. It used simple rules to identify keywords in user input
and generate plausible-sounding follow-up questions. Unlike today's
chatbots, it used fixed scripts and did not learn from natural language.
It is commonly used as a cautionary tale of how people will attribute
consciousness and other human-like attributes to even very rudimentary
machines (the ``ELIZA effect'') \citep{turkle1984second, hofstadter1995fluid}. It
serves as a useful test for whether the stances that favor LLM
consciousness would favor any AI system that produces minimally
human-like behavior. Consciousness in ELIZA is strongly disconfirmed by
every stance in the model. Its highest posterior was 0.03 (Recurrent
Processing (Pure)).

\section{Model results: aggregated across stances}\label{model-results-aggregated-across-stances}

We can aggregate the judgments of the stances to arrive at an
all-things-considered posterior probability of consciousness for each
system. We consider two methods for taking weighted averages of stance
judgments (see Limitations and Future Work for more discussion).

\subsection{Equal weight}\label{equal-weight}

We start by giving equal weight (1/13) to each stance's judgment and a
prior probability of consciousness of $\frac{1}{6}$ :

\begin{figure}[H]
\centering
\includegraphics[width=\textwidth,keepaspectratio]{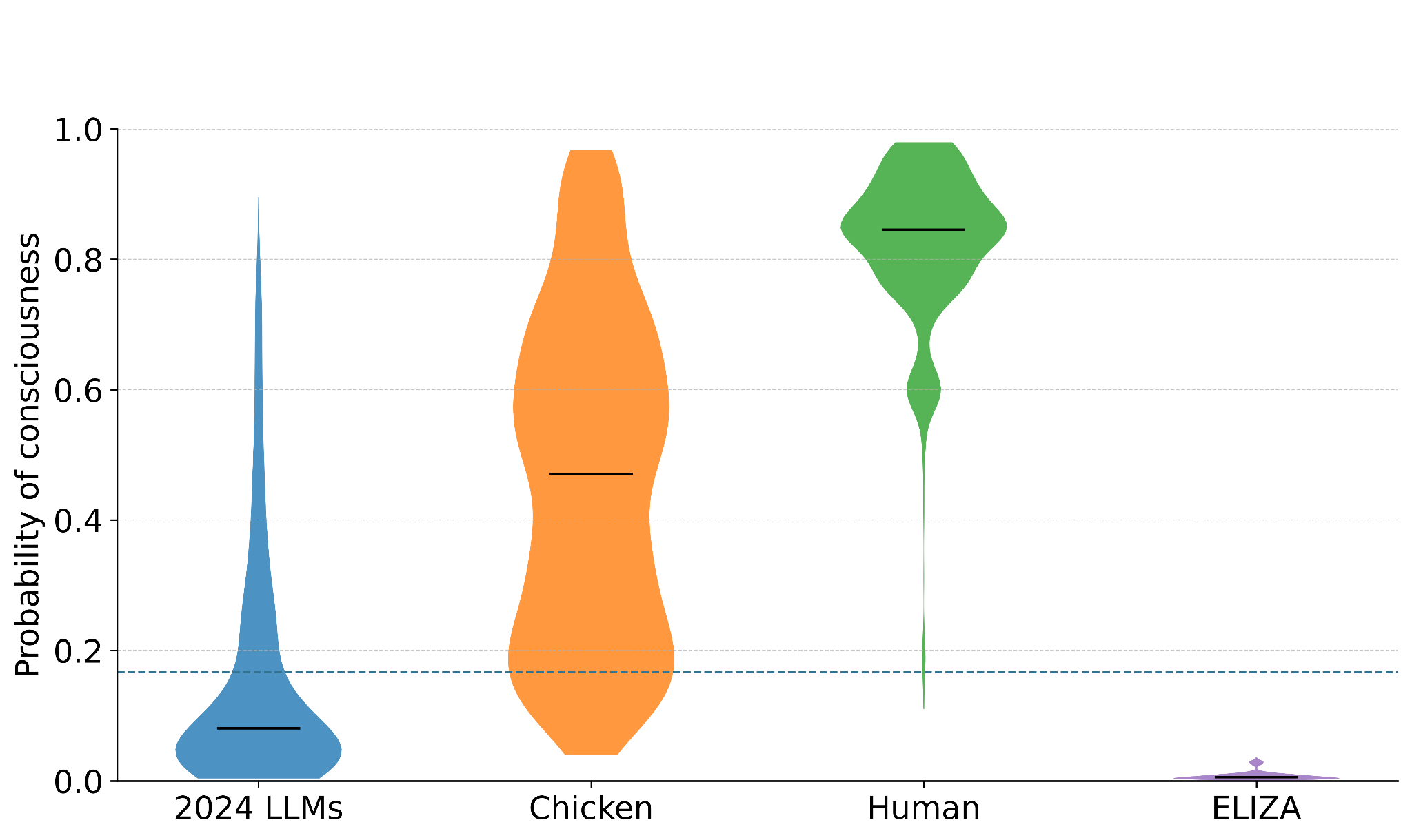}
\caption{Aggregated stance judgments, giving equal weight to each stance, starting from a prior probability of consciousness of $\frac{1}{6}$ (dashed blue line).}
\label{fig:equal-weight}
\end{figure}

As in the stance-specific graphs, the shaded region depicts the
posterior probabilities resulting from different model runs, with the
horizontal bar depicting the median posterior. Here, spread in the
posterior reflects differences across stances, as well as uncertainty
within stances themselves.

The aggregated probability that 2024 LLMs are conscious is 0.08 and
lower than the prior probability, which indicates that the aggregated
evidence disconfirms the hypothesis of consciousness. There is some
spread in the posteriors generated by different model runs, but most of
the probability weight is below the prior. Consciousness in LLMs is not
ruled out, however, and it is not as strongly disconfirmed as ELIZA
(which has a posterior of 0.006).

There is much more uncertainty about chicken consciousness, arising from
significant disagreements both between and within stances. The
distribution is roughly bimodal, reflecting the low probability
judgments of some stances and high judgments of others. The median
estimated posterior is 0.47, and almost all of the probability weight is
above the prior. Humans have a median posterior of 0.85.

\subsection{Plausibility-weighted aggregation}\label{plausibility-weighted-aggregation}

The equal weighting scheme implicitly assumes that each perspective on
consciousness is equally plausible. However, as we have noted, these
stances range from well-developed neuroscientific theories (e.g. Global
Workspace Theory) to newer, more speculative proposals (e.g. EM field
mechanisms). Some stances posit very specific architectures (e.g.
Attention Schema) while others are far more general (e.g. Biological
Analogy). Complete neutrality across the set of options is implausible.

In principle, stance weights could be derived from the credences of
individual model users or groups (e.g. the population of the US or
decision-makers at AI companies). We will explore this functionality in
future work. Here, we aggregate stance judgments according to relative
plausibility ratings provided by a small set of consciousness experts
(n=13). This pilot study showed significant heterogeneity in expert
judgments.

\begin{figure}[H]
\centering
\includegraphics[width=\textwidth,keepaspectratio]{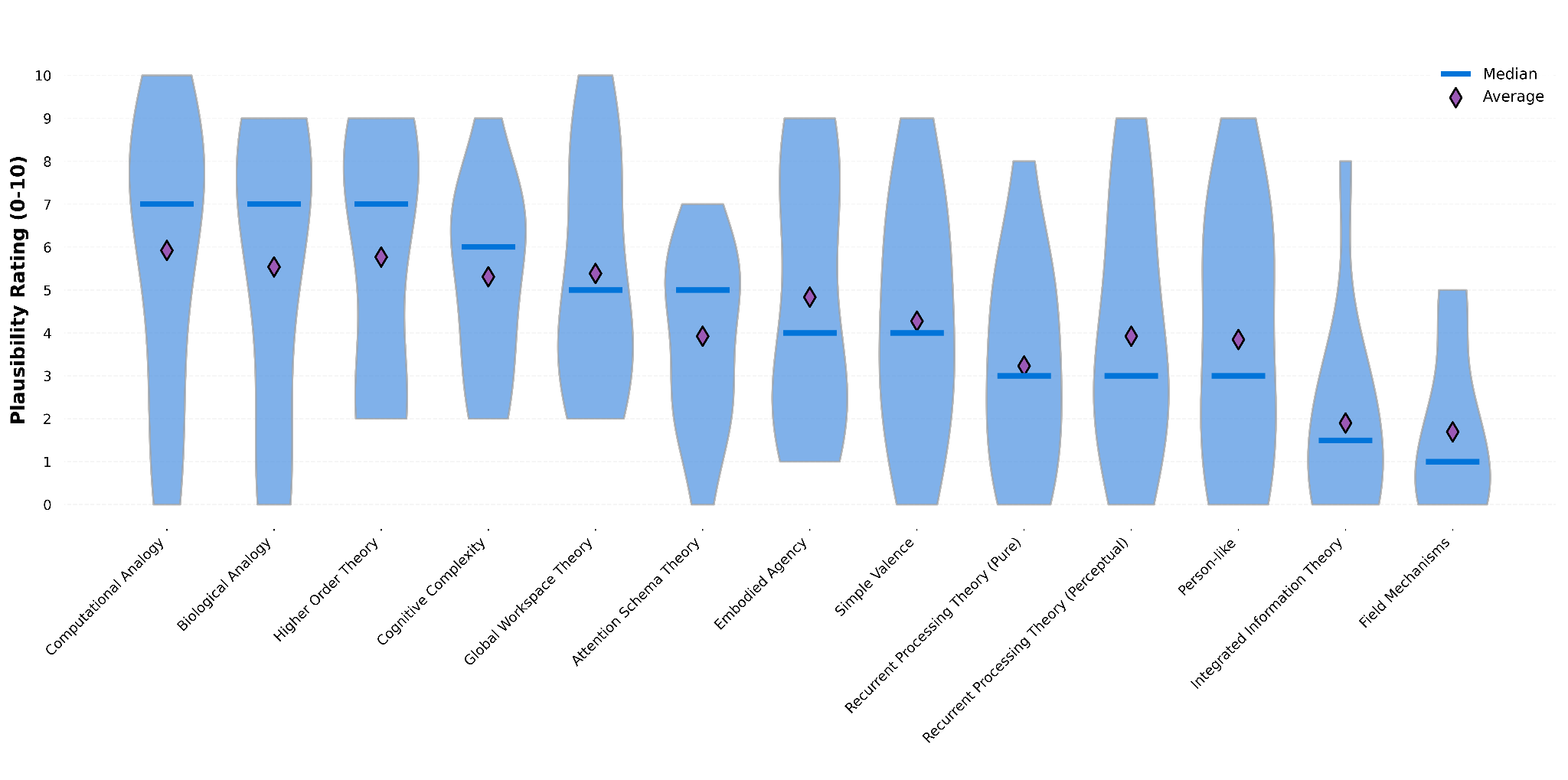}
\caption{Expert plausibility judgments about stances in the model.}
\label{fig:expert-plausibility}
\end{figure}

We normalized the average plausibility ratings to derive stance weights
so that stances with higher ratings were given more weight in the
aggregated judgment (e.g.\ Biological Analogy has $\sim$3x the
weight of Field Mechanisms). Given the diversity of expert opinion, the
results are very similar to those delivered by the equal weighting
strategy:

\begin{figure}[H]
\centering
\includegraphics[width=\textwidth,keepaspectratio]{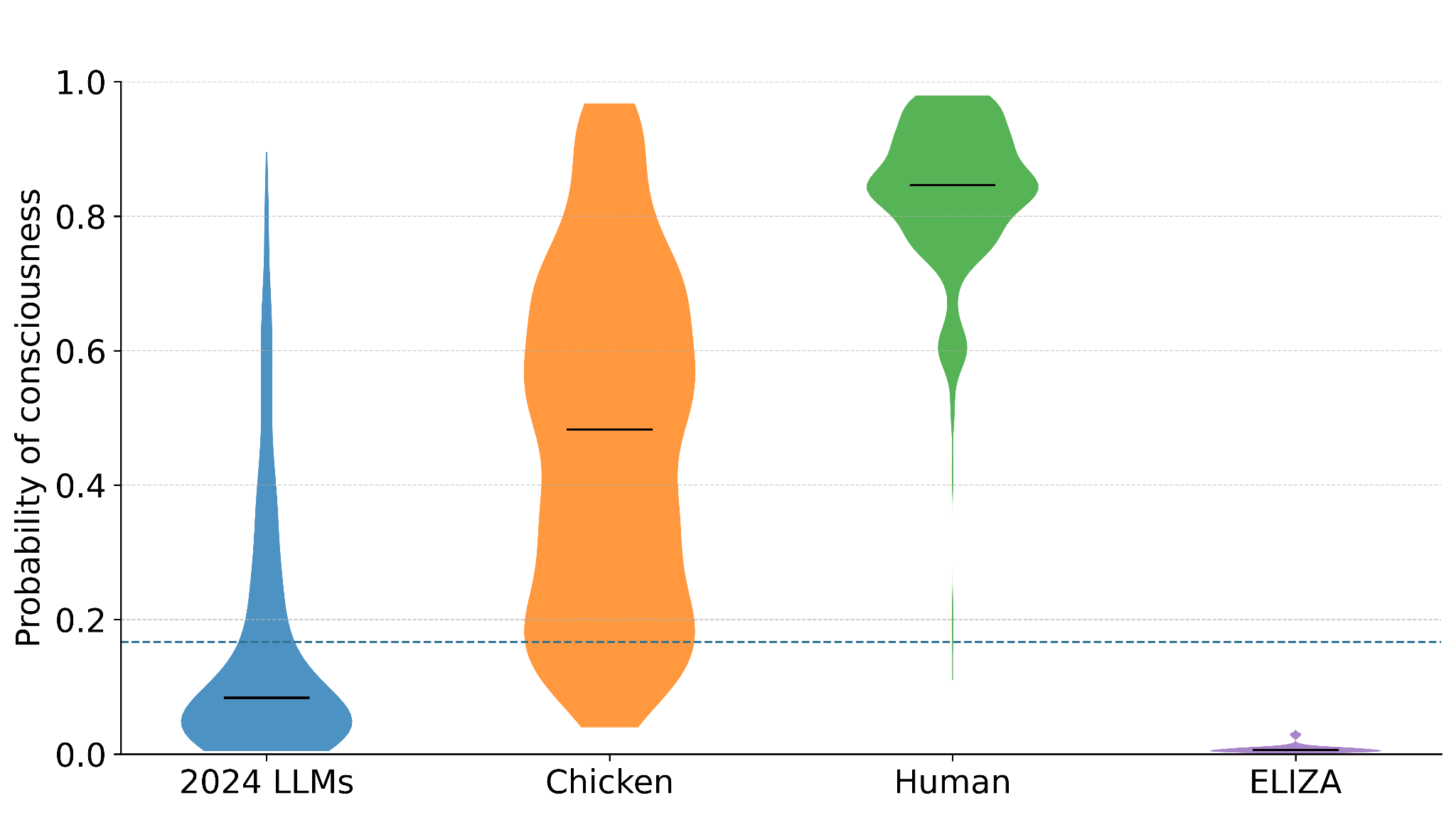}
\caption{Aggregated stance judgments, giving weight to stances proportional to their normalized plausibility rating by experts. Posteriors are generated from a prior probability of consciousness of $\frac{1}{6}$ (marked with a dashed line).}
\label{fig:weighted-aggregation}
\end{figure}

Here, the median posterior for LLMs is 0.08, while chickens come in at
0.49, humans at 0.85, and Eliza at 0.006.

\section{Discussion of results}\label{discussion-of-results}

\subsection{Caution about posterior probabilities}\label{caution-about-posterior-probabilities}

We do not straightforwardly endorse the median posterior probabilities
reported above; we think it would be a mistake to conclude that LLMs
have an 8\% chance of being conscious or that chickens have a 50\%
chance. The primary reason for this is that these posteriors are highly
dependent on the choice of a prior probability that the system is
conscious \emph{before we consider any evidence about the system.} We
don't take a stand on whether any prior probabilities are justified in
this context, and we certainly do not claim that our choice of prior ($\frac{1}{6}$)
is correct. We chose this prior because it yields easily discernible
differences in posteriors across systems, which is useful for
illustration (if priors are very low or very high, posteriors will be
much more tightly clustered).

The values of the posterior probabilities in the model are highly
sensitive to the choice of a prior:

\begin{figure}[H]
\centering
\includegraphics[width=\textwidth,keepaspectratio]{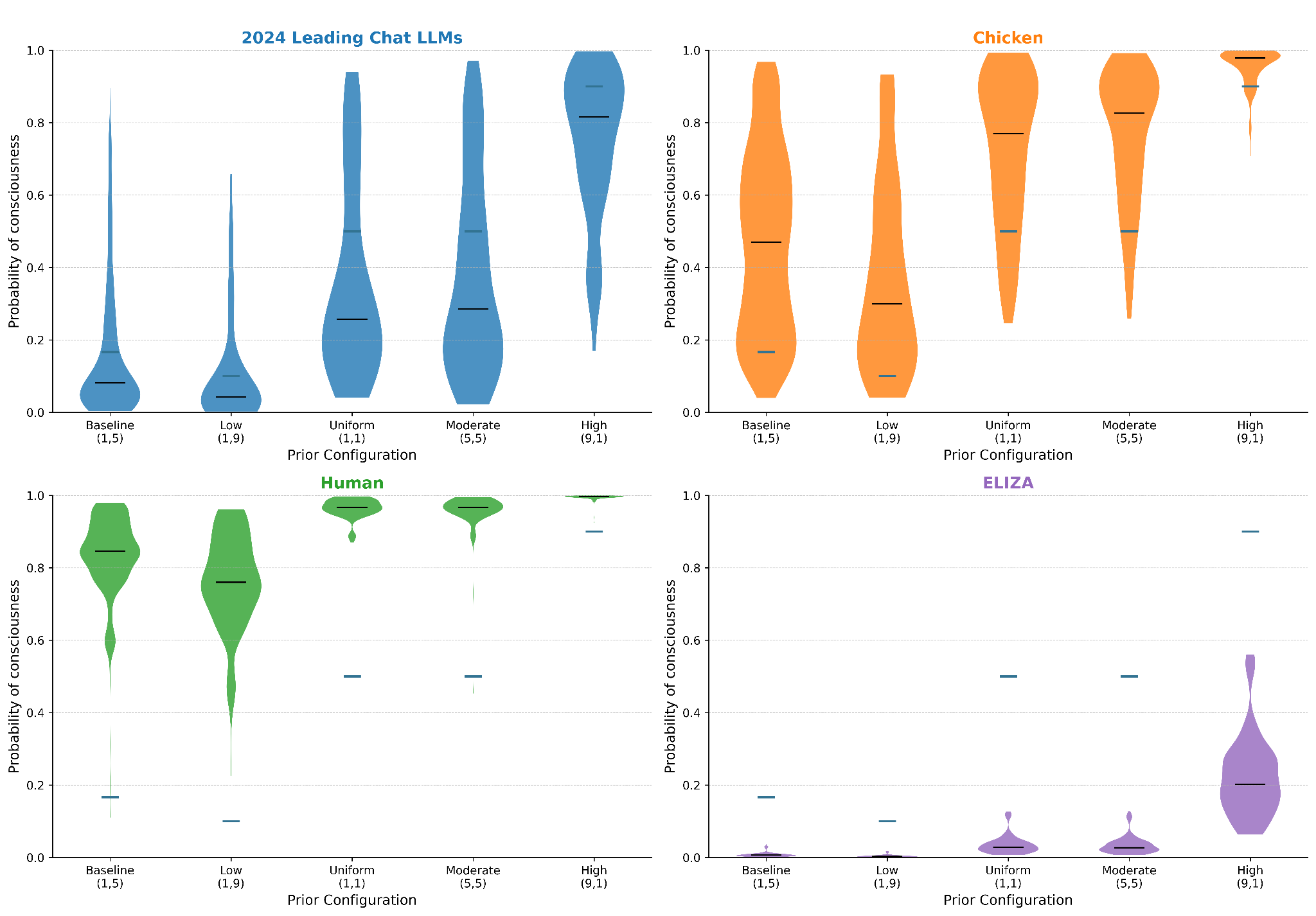}
\caption{Distributions of posteriors for different choices of prior probabilities. Uncertainty over priors is defined within the model by Beta distributions reflecting both the expected prior and the degree of confidence about that prior. Each configuration here specifies Alpha to Beta variables for that distribution. The ratio of those variables determines the average prior probability of consciousness. The scale of the variables determines the level of certainty about that prior. E.g. Baseline is defined by a ratio of 1:5, so it suggests an average 0.167 prior of consciousness. High is defined by a ratio of 9:1 so a 0.9 prior of consciousness. Uniform and Moderate each posit a 0.5 prior, but Moderate is more confident about this prior (and therefore more resistant to change).}
\label{fig:prior-sensitivity}
\end{figure}

\begin{figure}[H]
\centering
\includegraphics[width=\textwidth,keepaspectratio]{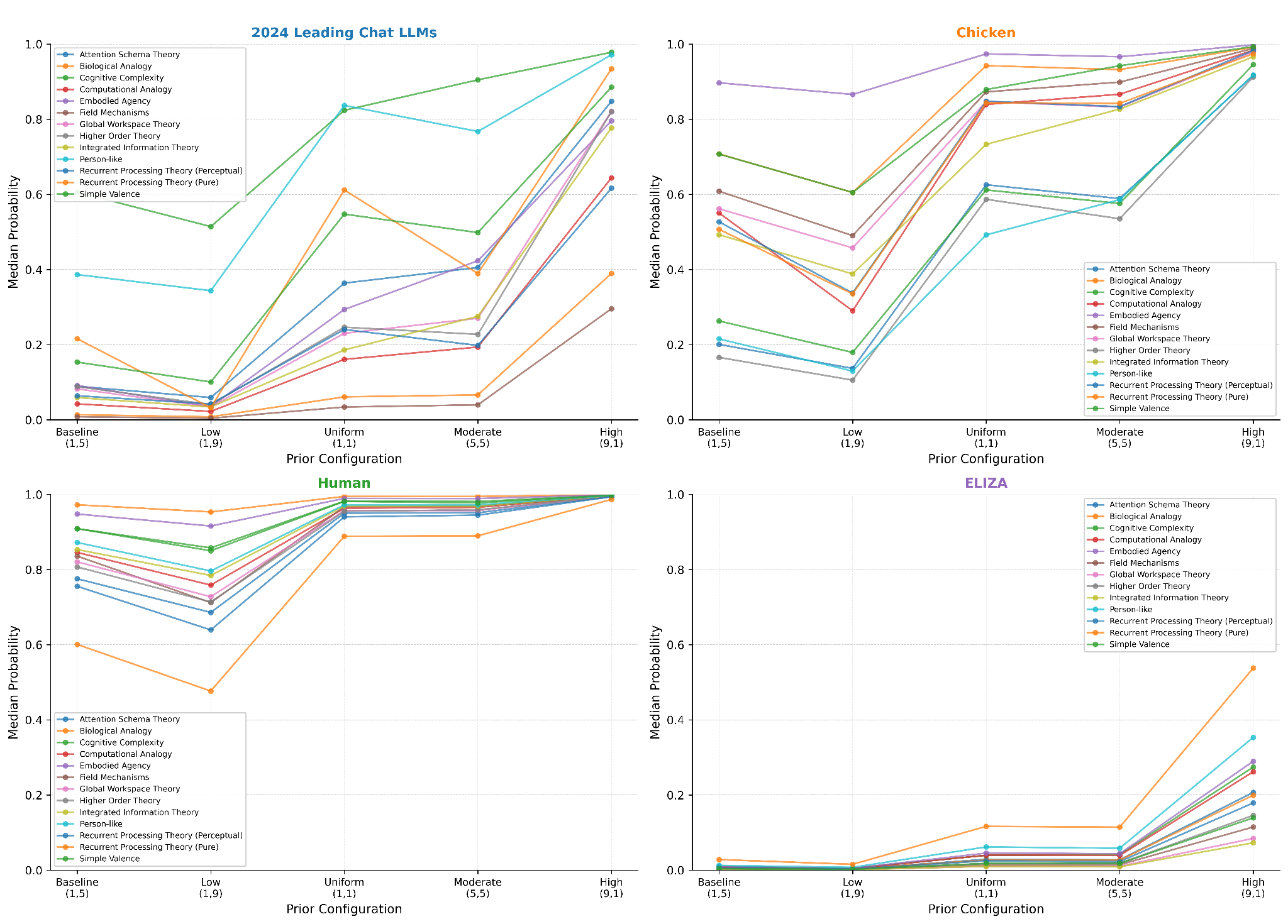}
\caption{Change in median posterior probability of consciousness across systems, stances, and priors.}
\label{fig:posterior-change}
\end{figure}

\begin{figure}[H]
\centering
\includegraphics[width=\textwidth,keepaspectratio]{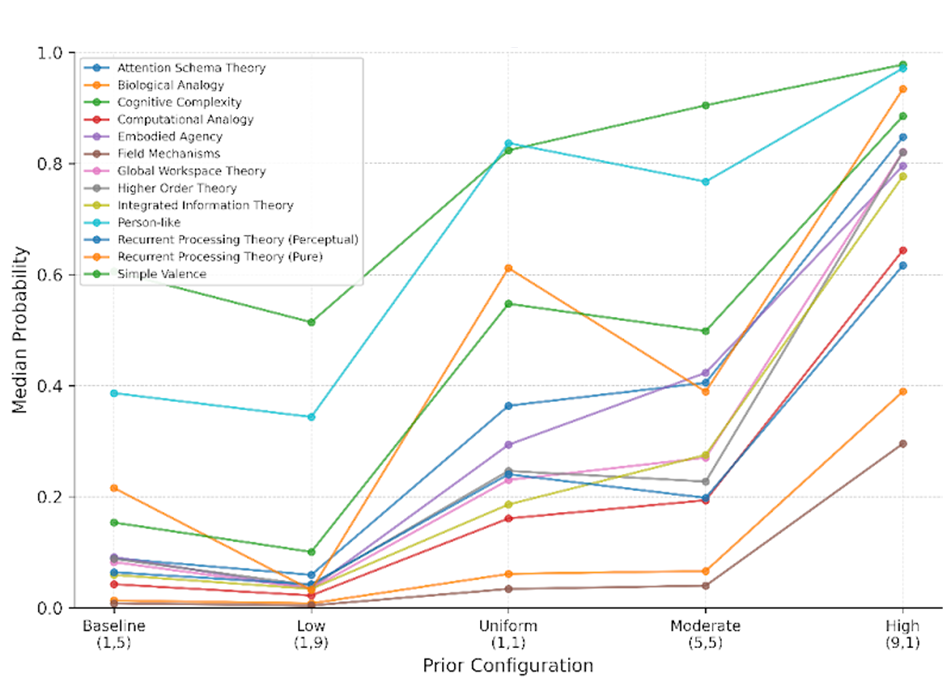}
\caption{Change in median posterior probability of LLM consciousness across stances and priors (reproduced from \autoref{fig:posterior-change}).}
\label{fig:llm-posterior-change}
\end{figure}

Notice that while posteriors are highly sensitive to choice of prior,
certain comparative results generally hold across priors: the direction
in which the model updates from the prior,\footnote{Recurrent processing theory (pure) for LLMs is an exception. LLM consciousness is confirmed
  on the Uniform prior but disconfirmed on the Moderate prior. Each of
  these configurations assigns a prior of 0.5 to LLM consciousness, but
  Uniform updates more readily with new evidence than does Moderate. We
  suspect that this was the result of uncertainty within the stance,
  which had more significant effects with a weaker prior.} and orderings
of posteriors across systems (i.e. if system 1 has a higher posterior
than system 2 on one prior configuration, it has a higher posterior on
all prior configurations).

We have used the same prior across systems in order to isolate the
effect of indicator evidence and reduce subjectivity across prior
assignments. The model does permit the choice of different priors for
different systems. For example, if we assign Eliza and 2024 LLMs a lower
expected prior of 10\% and we assign chickens and humans a higher
expected prior of 90\%, the updated summary results (as extracted from
\autoref{fig:prior-sensitivity}) are as follows:

\begin{figure}[H]
\centering
\includegraphics[width=\textwidth,keepaspectratio]{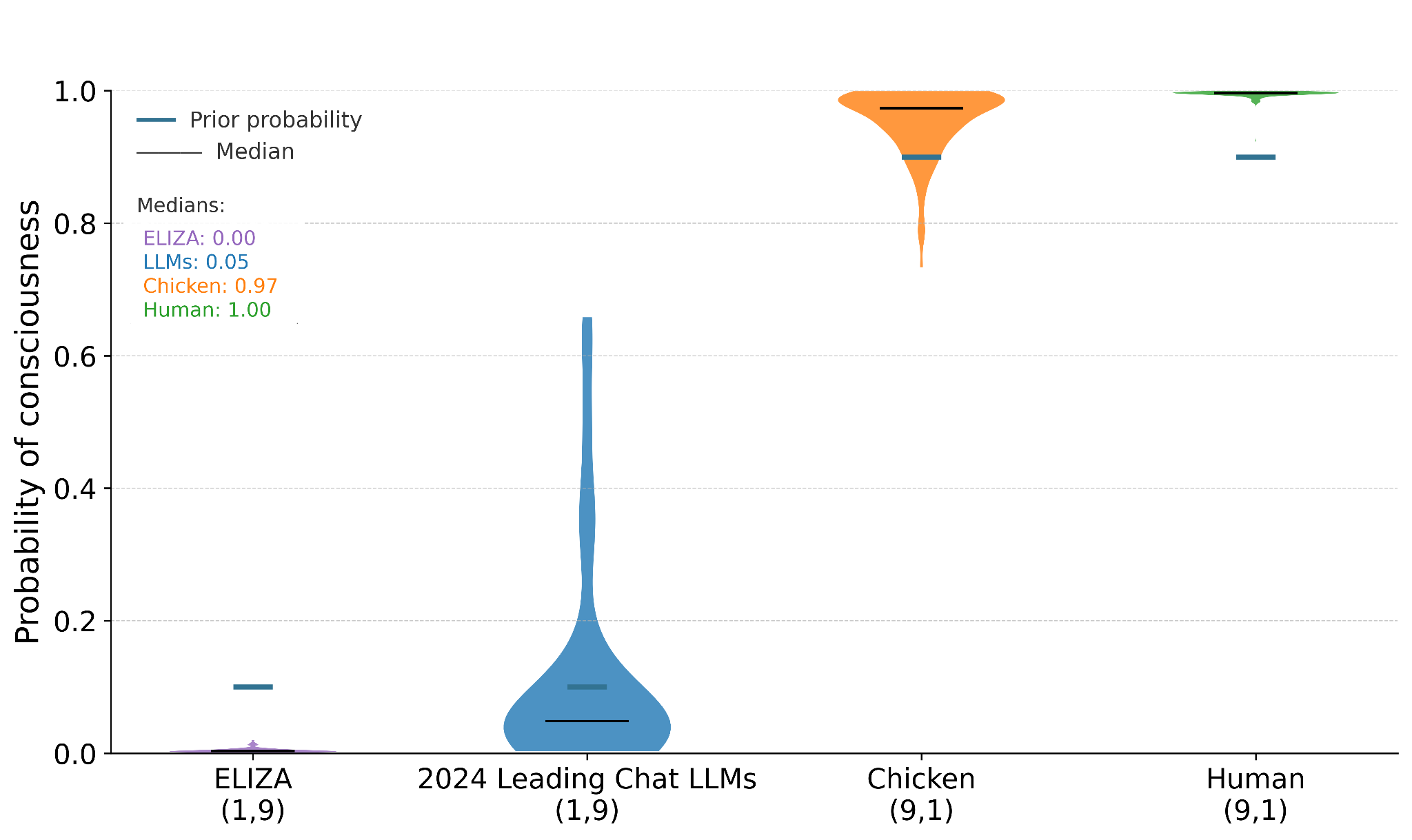}
\caption{Aggregated stance judgments with higher priors for biological systems, giving weight to stances proportional to their normalized plausibility rating by experts.}
\label{fig:differential-priors}
\end{figure}

\subsection{Comparative results}\label{comparative-results}

While we do not endorse the absolute values of the posteriors in the
model, we do tentatively endorse comparisons that they permit.

First, we can compare the posterior to the prior to evaluate how
strongly the evidence confirmed or disconfirmed the hypothesis that the
system in question is conscious. The aggregated indicator data was
evidence against LLM consciousness and very strong evidence against
ELIZA consciousness. The data was strong evidence for chicken
consciousness and very strong evidence for human consciousness. If we
want to measure the strength of this evidence, we can derive the
likelihood ratios by comparing the ratio of the posteriors to the ratio
of the priors:

\begin{table}[htbp]
\centering
\small
\parbox{0.8\textwidth}{\caption{Approximate likelihood ratios for the aggregated evidence: $\Pr(E|S \text{ is conscious}) / \Pr(E|S \text{ is not conscious})$.}}
\begin{tabularx}{0.8\textwidth}{@{}l@{\extracolsep{\fill}}r@{}}
\toprule
\textbf{System} & \textbf{Likelihood Ratio} \\
\midrule
Humans & 28.33 \\
Chickens & 4.6 \\
2024 LLMs & 0.43 \\
ELIZA & 0.05 \\
\bottomrule
\end{tabularx}
\label{table:likelihood}
\vskip\tableelementskip
\begin{minipage}{0.8\textwidth}
\footnotesize \textit{Note.} A ratio greater than 1 means the evidence favors the presence of consciousness in $S$. A ratio of less than 1 means the evidence favors the absence of consciousness in $S$. The closer the value to 1, the less evidence it provides.
\end{minipage}
\end{table}

Second, we can compare posteriors across systems to evaluate how much
weaker or stronger our indicator evidence is for consciousness in, say,
chickens versus LLMs. We set the prior probabilities the same for each
system so that the posterior only reflects differences with respect to
the indicator evidence. The evidence gives us much stronger reason to
believe chickens are conscious than that LLMs are. We hope to run the
model on a wider set of digital and biological systems to get more
fine-grained comparisons (see Limitations and Future Work for
discussion). Comparisons among AI systems over time may be particularly
helpful in uncovering trends and predicting future trajectories
regarding the possibility of AI consciousness.

\subsection{Checking the model for accuracy}\label{checking-the-model-for-accuracy}

Typically, one verifies that a model's structure, parameterizations and
priors are properly calibrated by seeing whether it gives accurate
predictions for held-out data. In the case of consciousness, we do not
get to ever observe conclusively whether a third-party system or
organism is conscious. Instead, we evaluate the model in the following
ways:

\begin{itemize}
\item
  \emph{Check the model\textquotesingle s outputs against more familiar
  systems and see whether the model gives plausible posteriors for these
  systems}. Given some caveats outlined below, we think the model gives
  sensible results for humans, chickens, and ELIZA, which gives us
  additional confidence in the model's evidential parameters and
  therefore its results for LLMs.
\item
  \emph{Check whether the posteriors from different stances make sense}.
  For example, a stance that emphasizes biology should give a higher
  posterior to chicken sentience than does a stance emphasizing
  person-like features such as language, and it should give a higher
  posterior to chickens than LLMs. We think that results vary by stances
  are sensible (especially for stances whose results are highly
  predictable).
\item
  \emph{Check the intuitive plausibilities of the probabilities of
  features calculated during model runs}. For example, if humans or
  chickens showed extremely low values for the Biological Similarity
  feature, this would suggest that something had gone wrong with the
  model. We will explore feature values in future work.
\item
  \emph{Test how sensitive the posteriors are to choices of priors
  through conducting sensitivity analyses (see \autoref{appendix-e-prior-sensitivity-tests})}. While we
  expect posteriors to be sensitive to choices of priors, we do not
  expect certain directional results should change significantly. For
  example, we don't expect that consciousness in chickens will be
  strongly confirmed relative to some priors but strongly disconfirmed
  relative to others. We found that the direction of updates and ordinal
  rankings across systems were robust to changes in prior.
\item
  \emph{Test how sensitive the posteriors are to changes in the
  evidential parameters linking variables (see \autoref{appendix-d-conditional-dependencies})}. For
  example, if we change the conditional dependencies corresponding to
  Strong Support or Weak Demandingness, how much does this change the
  results of the model? Results were directionally consistent when we
  used a smaller number of evidential parameters though updating was
  less extreme. Future work will explore sensitivity to additional
  changes to model parameters.
\item
  \emph{Check to see whether the model updates in the right direction in
  predictable cases}. Even if people disagree about their priors, the
  model will tell them how they should change their beliefs (not
  necessarily the exact posterior they should end up with). According to
  the model, the evidence should make us much more confident in human
  consciousness and much less confident in ELIZA.
\end{itemize}

To summarize, because we don't have access to the ground truth facts to
train a model in the way other fields do, we must rely on informed
priors, philosophical reflection, converging lines of evidence, and
iterative processes to refine the model.

\section{Key takeaways}\label{key-takeaways}

With these caveats in place, we can pull out some key takeaways from the
DCM:

\emph{The evidence is against 2024 LLMs being conscious.} We asked
experts to weigh in on whether 2024 LLMs possessed a large number of
indicators of consciousness that were identified as evidence relative to
a diverse set of theories of consciousness. The aggregated evidence
favors the hypothesis that 2024 LLMs are not conscious.

\emph{The evidence against 2024 LLMs being conscious is not decisive}.
While the evidence led us to lower the estimated probability of
consciousness in 2024 LLMs, it did not provide decisive evidence against
it. The total strength of the evidence, as measured by the likelihood
ratio (0.433), was not overwhelmingly against LLM consciousness. The
evidence against LLM consciousness is much weaker than the evidence
against consciousness in simpler AI systems like ELIZA.

\emph{2024 LLMs do best (absolutely and relative to other systems) on the Cognitive Complexity and Person-like stances}. If consciousness is
attributed primarily on the basis of AIs' intelligence and human-like
social interactions (say, by the AI-using public), AIs are likely to be
deemed very good candidates for being conscious \citep{scott2023do, colombatto2024folk}. These stances did not favor
consciousness in ELIZA, which provides some indication that they are not
grossly anthropomorphizing.

\emph{2024 LLMs do worst (absolutely and relative to other systems) on the Embodied Agency and Biological Analogy stances.} Biological
perspectives view consciousness as a distinctively biological phenomenon
that is not likely in purely digital systems like LLMs. Embodied Agency
does not require biological constitution but does put central emphasis
on control over a body. AI systems that are run on biological substrates
or that control robot bodies may be better candidates for consciousness
on these stances.

\emph{Not all computational stances favor 2024 LLM consciousness.} Cognitive Complexity and Person-like stances judge LLMs on their
observable behavior and tend to be more permissive. Other computational
stances, such as GWT and Computational Analogy, require that the
performance of conscious systems be produced by specific kinds of
computational architectures. LLMs were judged less likely to be
conscious by these more architecturally-focused stances, reflecting
their significant architectural differences from biological organisms
\citep{cao2022multiple, polger2016multiple}.

\emph{The plausibility of AI consciousness will depend on the
plausibility of different stances.} Different stances give strikingly
different judgments about the probability of LLM consciousness.
Therefore, significant changes in the weight given to stances will yield
significant differences in model outputs. It will be important to track
how scientific and popular consensus about stances change over time and
the consequences this will have on our judgments about the probability
of consciousness.

\emph{Stances diverge on artificial vs. biological systems.} The two
stances that give the highest scores to LLMs (Cognitive Complexity and
Person-like) give some of the lowest scores to chickens, while chickens
score best on stances (Biological Analogy and Embodied Agency) that give
some of the lowest scores to LLMs. This suggests that these stances may
serve as important cruxes in navigating the question of consciousness in
digital and biological systems \citep{seth2024conscious, block2025can}.

\emph{There is significant diversity across stances regarding chicken consciousness.} Many stances assign relatively high probabilities of
chicken consciousness, but there is a cluster of stances that assign low
probabilities (barely above the prior). Two of these emphasize
metacognitive abilities: Attention Schema Theory and HOT Theory. The
presence of metacognition in non-human animals is a controversial topic,
so future research on this topic may significantly change estimates of
chicken consciousness \citep{fujita2012bird}.

\emph{Human consciousness is very strongly confirmed by the indicator data.} The results for humans serve as a helpful check on how
comprehensive the set of indicators is and whether the model
incorporates this indicator evidence in a plausible way. We are
encouraged on both counts (though see Limitations and Future Work for
further discussion).

\emph{ELIZA is very strongly disconfirmed by the indicator data.} Relative to every stance in the model, we have very strong evidence that
ELIZA is not conscious. This demonstrates that our indicator data can
distinguish between rudimentary chatbot systems (even those that can
mimic some human-like behavior) and today's AI chatbots.

\section{Limitations and future work}\label{limitations-and-future-work}

\subsection{More survey data}\label{more-survey-data}

The current model was run on the 16 expert surveys for LLMs (6 full
survey responses, 10 subset responses), 2 for chickens, and 1 each for
humans and ELIZA. Survey participants were recruited via targeted emails
of subject-area experts on system capacities (not necessarily experts on
digital or animal consciousness). We surveyed 13 consciousness experts
about how plausible they found each stance in the model, which served as
the basis of our credence-weighted aggregation method.

Collecting expert data was a significant bottleneck for the project. In
future model runs, it would be beneficial to have a larger body of
respondents and to check for respondent diversity.\footnote{To mitigate worries around the theory-ladenness of indicators, it would be
  beneficial to have experts with diverse (or no) views on the
  probability of digital or animal consciousness.} A more complete
survey methodology would involve:

\begin{itemize}
\item
  Explicitly stated sampling strategies.
\item
  Tests for saturation and convergence to check whether sample size is
  sufficiently large.
\item
  Collection of demographic data and evaluation of whether demographic
  categories correlate with response types.
\item
  Robustness analyses based on subsample analyses
\end{itemize}

\subsection{Implausible results for humans?}\label{implausible-results-for-humans}

The model generates a posterior probability that humans are conscious
that might seem suspiciously low, thus casting doubt on the model. We
think these results are not a problem but do help to illustrate how the
model works and how it ought to be interpreted. The values for humans
arise from four sources, only one of which suggests worrisome
vulnerabilities in the model.

First, by choosing the same prior across all systems, we are excluding
any evidence that we might have regarding a system's consciousness other
than the indicators that are in the model. We possess additional
evidence in the case of humans, most notably introspective evidence that
we are in fact conscious. Our model seeks to answer a specific question:
if you were presented with a novel species, and all you learned about
them was whether they possessed the indicators in the model, how likely
should you find it that they are conscious? We would like to include
indicators for all major sources of evidence, but it is challenging to
make all considerations concrete and objective. For some stances, it is
hard to derive decisive indicators that we would take to be collectively
authoritative about the presence of consciousness according to the
stance. Viewed from this lens, the posterior probabilities are perhaps
not unreasonable.

Second, we chose a fairly low prior probability. If you think a higher
prior is justified, then the model will yield higher posteriors. For
example, if we start with a 0.5 prior that humans are conscious, the
posterior probability that humans are conscious rises to 0.95 (see
\autoref{fig:prior-sensitivity}). If we start with a prior of 0.9, the result is a virtual
certainty that humans are conscious.

Third, there is significant variation in the posteriors assigned to
humans by different stances. If a stance assigns a low probability to
humans, this may stem from an error in the model or it may stem from an
error in the theory itself. For example, suppose that Global Workspace
Theory posits that a particular kind of functional architecture is
necessary for consciousness, and our experts are uncertain about whether
that architecture actually exists in humans. This contributes to
uncertainty about whether GWT (or, this particular formulation of GWT)
is true, as its necessary conditions are absent in creatures known to be
conscious. To the extent that a certain stance yields low probabilities
for humans, that may be a reason to doubt the stance itself.

Finally, we do not claim to have included all of the relevant sources of
evidence in our model's set of indicators. Low probabilities of human
consciousness may mean that we have left out some of the key evidence
for human consciousness. In future work, we aim to add to the stock of
indicators, which we expect will raise the model's predicted probability
of human consciousness.

\subsection{Implausible results for chickens?}\label{implausible-results-for-chickens}

The predicted posterior probability of chicken consciousness may also
seem suspiciously low. This arises for many of the same reasons that
were responsible for low posteriors for humans. An additional
consideration in the case of chickens is that, unlike in the case of
humans, there are popular theories of consciousness that explicitly deny
that chickens are conscious. For example, certain advocates of Higher
Order Thought theory argue that the kinds of metarepresentational
capacities necessary for consciousness are absent in most non-human
animals \citep{carruthers1989brute, gennaro1993brute}. Incorporating the judgments
of such stances lowers the probability of chicken consciousness.

The model attempts to give a field-wide assessment of the evidence for
consciousness, and it is appropriate that it reflects the proportion of
researchers who are skeptical of consciousness in non-human animals.
However, individual users or groups who are confident that non-human
animals like mammals and birds are conscious may wish to assign little
or no weight to stances that deny animal consciousness when using the
model to assess the likelihood of digital consciousness.

Additionally, each stance in the model can be formulated with varying
degrees of discernment, such that higher or lower levels of each feature
are treated as evidence. For instance, a stance that gives some weight
to intelligence might be fully satisfied by a very small degree of
intelligence. We have adopted what we take to be a middle-of-the-road
approach in which systems are neither too finicky or too liberal. Some
readers may feel that this leads to implausibly low results for
chickens. In future work, we could add stances that are more liberal or
more conservative about attributing key capacities.

Finally, many theories of consciousness have been designed with
applications to humans in mind, in part because humans are the beings we
are most confident are conscious and the most helpful test subjects. We
may therefore be biased toward looking for features that are easy to
discern in humans.

\subsection{Missing stances and unconceived alternatives}\label{missing-stances-and-unconceived-alternatives}

Our method of aggregating the judgments of stances, by taking a
credence-weighted average, implicitly assumes that our list of stances
is mutually exclusive and exhaustive. Because stances are defined by
specific functions from features to consciousness, mutually exclusivity
is (somewhat trivially) satisfied. However, this set of stances
(especially when defined narrowly by a specific function) is not
exhaustive. This is an additional reason to be cautious when
interpreting the posterior probabilities produced by the model. It
reflects the assessment of the evidence in light of our current set of
theories, not the full set of possibilities. Users can focus on
stance-specific results and remain agnostic about the aggregated results
if one thinks the current set of stances covers little of the total
possibility space.

We have included a diverse set of stances in the model, representing
different commitments and epistemic stances toward consciousness. We
attempted to capture most of the theories that are well-represented
among consciousness researchers and others whose views on AI will
matter. New stances can easily be added to future iterations of the
model, which will require us to redistribute credences. Examples
include: free energy principle approaches \citep{ramstead2023inner,
 wiese2024artificial, laukkonen2025beautiful}; Unlimited
Associative Learning theory \citep{bronfman2016transition, birch2020unlimited, ginsburg2019evolution}; panpsychism \citep{strawson2006realistic,
chalmers1996conscious, goff2017consciousness}; error theory \citep{dennett1991consciousness, frankish2016illusionism};
quantum mechanisms \citep{penrose1989emperor, hameroff1996orchestrated};
sensorimotor theory \citep{oregan2001sensorimotor}, and perspectives that
emphasize relationality or personal relationships. There are some common
perspectives that may be difficult to capture in this framework. For
example, a metaphysical perspective like substance dualism may be hard
to characterize in terms of any features whose presence is empirically
detectable, except where mental substances/properties are theorised to
necessarily correlate with certain physical substances/properties (in
which case it is adequate to measure the latter).

Perhaps more troubling is the possibility that the true perspective on
consciousness may be one that has not yet been developed. We could
attempt to reserve credence for the space of unconceived alternatives by
introducing some Unknown Theory X \citep{sebo2025moral}. We resisted
doing so because we neither know what prior probability to assign to
this catchall hypothesis nor what it would predict about the systems
under examination\footnote{One possibility is to randomly generate an Unknown Theory X for each model run. The stance would be generated by
  randomly selecting some subset of features to be evidentially relevant
  and randomly generating Support and Demandingness parameters for those
  features.} \citep{sober2008evidence, stanford2010exceeding}. Additionally, we think that
many unconceived theories can be captured by general stances in the
model. In the absence of specific requirements about what is necessary,
it may make sense to fall back on general analogies. Even if you think
that biological processes aren't strictly required for consciousness,
you might think that general aspects of biology (brainplan similarity,
evolutionary history, etc.) provide the best source of evidence we have
about the presence of the undiscovered properties that actually are
necessary, in which case you may be satisfied with the Biological
Analogy stance.

\subsection{Relaxing model assumptions}\label{relaxing-model-assumptions}

In Section 4.5., we explained two modeling choices that may limit model
accuracy: all variables are treated as binary and independent
(conditional on parent nodes). In future versions of the model, we could
introduce non-binary (i.e. continuous or many-valued discrete) variables
to represent features and indicators that come degrees. We could also
introduce correlations among features or indicators. This will allow us
to better represent stances that, say, take multiple features to be
evidentially relevant only in combinations (e.g. taking multiple
features to be jointly necessary). Each of these choices would introduce
computational complexity to the model. Sensitivity tests would reveal
how significantly these choices would change model performance, allowing
us to evaluate trade-offs with ease of use.

\subsection{Theory-laden indicators}\label{theory-laden-indicators}

The model assumes that the indicators are hard data/known facts about
the systems in question. For some indicators, like being carbon-based,
it is relatively easy to determine whether they're true or false. Other
indicators pick out relatively observable architectural features or
performance on benchmark tests. However, at this stage in our knowledge
about AI systems, we often haven't developed definitive tests for
whether a more complicated indicator is indeed present in a system, so
in some cases, indicators are not observable or otherwise
straightforwardly verifiable. Though we have attempted to make the
evaluation of indicators independent from biases about which systems are
conscious, some indicators are theory-laden \citep{feyerabend1969science, kuhn1962structure}.

For example, the Consistent Preferences indicator question asks: ``Does
the system demonstrate consistent patterns in its choice behavior,
systematically preferring or avoiding certain types of activities or
states across different contexts?''. One's judgment whether LLMs count
as ``preferring'' activities may depend, in part, on one's theoretical
views about what preferences are. One's judgement about whether LLMs
have consistent patterns may depend on how one identifies the same LLM
across contexts (e.g. with different temperature settings or given
different system prompts). If these views are correlated with one's
views on the plausibility of machine consciousness, experts' biases
about which systems are conscious may also bias their evaluation of
indicators.

At this juncture, some theory-ladenness of indicators is unavoidable.
The potential for model bias can be mitigated through a diverse sample
of experts. In future work, we could evaluate how strongly an expert's
particular indicator judgments correlate with their prior credence in
machine consciousness. It is also possible to model experts' judgments
not as ``ground truths'' but as evidence of indicators by treating
indicators as another layer of latent variables and assigning
likelihoods to ``what distribution of expert judgments should we expect
to see if the indicator is present or absent?'' We did not pursue this
strategy in this iteration of the model, since we don't have any
principled basis for assigning prior distributions over experts'
judgments. With a relatively small number of experts, our priors for the
experts' responses would exert enormous influence on the outcome.

\subsection{Applying model to new systems}\label{applying-model-to-new-systems}

We intend this version of the DCM as an initial attempt at capturing our
evidence regarding consciousness in a variety of systems. We expect to
elaborate and refine the model in future iterations and think that it
has already shown promise for a diverse range of systems. A main focus
of future work will be to apply the model to novel systems.

Here, we focused on 2024 LLMs because they are in common use and because
they are now well understood enough to support reliable expert
judgments. We did not judge these systems to be the best candidate for
AI consciousness. In future work, the DCM will be applied to novel LLM
systems, including reasoning models, novel language agent systems \citep{moret2025ai, goldstein2025ai}, and their descendants in the
future. This will permit us to track changes in the probability of
digital consciousness over time and to predict which system changes will
yield important differences with respect to consciousness.

We would also like to evaluate alternative AI architectures. These
include well-known but less human-like systems like AlphaFold or Waymo
Driver. A particularly interesting application would be to systems that
blur distinctions between the biological and artificial, such as
whole-brain emulations or LLMs instantiated in neuromorphic substrates
(Kagan \emph{et al.} 2023, Smirnova \emph{et al.} 2023, Schneider 2025,
Schneider \emph{et al.} 2025, Whalen 2025). Though we expect these may
(eventually) be better candidates for AI consciousness, we may face
difficulties in getting a large set of expert judgments about less
well-known types of systems.

We also expect that applying the model to a wider range of biological
organisms--including invertebrate species such as insects and
nematodes--would permit better comparisons with AIs.

\section{Conclusion}\label{conclusion}

We present this iteration of the Digital Consciousness Model as a
promising framework and initial attempt at assessing the evidence for
consciousness in AI systems. While we acknowledge the limitations of the
model, we think that it is an important step toward assembling a
discipline-spanning, transparent, and rigorous evaluation of the state
of the evidence. It brings sophisticated statistical tools to bear on a
complicated problem with high levels of uncertainty and a very diverse
body of evidence.

Given the importance of assessing machine consciousness, it will be
important to use some methodology or other. A model that produces
probabilities improves the transparency of disagreements by representing
a specific position on which sources of evidence matter and how they
matter. Creating such a target for criticism is itself a valuable
contribution because explicitly stating all the choices required by a
model allows others to articulate where improvements are necessary. A
formal model allows us to adjust indicator values or weights given to
different factors, revealing the upshots of taking different ideas
seriously. It also makes it possible to see which future developments
might make the biggest difference to estimates of machine consciousness.

Many of the limitations of the DCM that we have pointed out can be
remedied with more inputs--more expert data, more indicators, and more
stances--which can be added in a flexible and modular way. We hope that
this initial presentation of the model demonstrates that the DCM model
can serve as a useful repository and tracker for new information as it
comes in. The model can also point to areas where more information would
be particularly valuable, allowing us to highlight indicators, features,
or stances that are significant difference-makers to the probability of
digital consciousness.

Assessing the probability of consciousness in AI systems is one
important aspect of a broader concern with the future of AI and how it
should be incorporated into our legal, ethical, and conceptual
frameworks.

\section*{Acknowledgements}
We are greatful for discussions with and feedback from Jeff Sebo, Bob Fischer, Alex Rand, David Moss, Oscar Horta, Joe Emerson, Luhan Mikaelson, and audiences at NYU Center for Mind, Ethics, and Policy and the Eleos Conference on AI Consciousness and Welfare.

\newpage

\addtocontents{toc}{\protect\setcounter{tocdepth}{2}}
\setcounter{part}{1}
\part{Appendices}\label{appendices}

\appendix

\section{Stances \& Features}\label{appendix-a-stances-features}

\subsection{Stances in the DCM: Short Descriptions}\label{stances-in-the-dcm-short-descriptions}

\begin{longtable}{@{}p{0.18\linewidth} p{0.47\linewidth} p{0.30\linewidth}@{}}
\toprule
\textbf{Stance} & \textbf{Brief description of perspective on consciousness} & \textbf{Representative citations} \\
\midrule
\endhead
\bottomrule
\endlastfoot
Global Workspace Theory & Arises from a broadcasting mechanism that makes some of the system's information widely available for use by various specialized processes or modules. & Baars 1988, 2005; Dehaene \& Changeaux 2011; Dehaene et al.\ 2006; Mashour et al.\ 2020 \\
\addlinespace
Recurrent processing (pure) & Emerges from dynamics of recursive processing loops over incoming or internally generated data. & Lamme 2006, 2010; Lamme \& Roelfsema 2000 \\
\addlinespace
Recurrent processing (perceptual) & Arises when perceptual inputs are subject to iterative refinement through structured, feedback-driven loops. Modeled after vision and sensorimotor systems in mammalian brain. & Lamme 1995, 2020 \\
\addlinespace
Computational analogy & Evidenced by overall functional resemblance to information processing in humans, across domains such as reasoning, perception, language, and decision-making. & \\
\addlinespace
Biological analogy & Evidenced by broad and diverse analogies with living biological organisms. & Aru et al.\ 2023; Seth 2024; Block 2025 \\
\addlinespace
Field mechanisms & Associates with integrated and causally efficacious electromagnetic fields (EMF). & McFadden 2020; Pockett 2012; John 2001; Jones 2013; Ward \& Guevara 2022 \\
\addlinespace
Simple valence & Strongly tied to valenced subjective experience and evidenced by behaviors like motivational trade-offs and flexible self-protection. & Campero 2024; Veit 2023; Crump et al.\ 2022 \\
\addlinespace
Attention Schema & Generated by an internal model representing the distribution of attentional resources in the system with the function of controlling attention. & Graziano \& Webb 2015; Graziano 2017, 2020; Graziano et al.\ 2020; Wilterson et al.\ 2020 \\
\addlinespace
Higher-Order Thought & Generated by internal representations of the system's own mental states, such as thoughts whose content includes the system's perceptual states. & Armstrong 1968, 1984; Lycan 1996; Rosenthal 1986, 2005; Carruthers 1996, 2005; Gennaro 2004 \\
\addlinespace
Integrated Information Theory & The product of integration structures: specifically measured by the irreducibility of the system's diverse causal powers to those of its parts. & Tononi \& Koch 2015; Tononi et al.\ 2016; Koch et al.\ 2016; Albantakis et al.\ 2023; Oizumi, Albantakis, \& Tononi 2014 \\
\addlinespace
Person-like & Evidenced by traits resembling those associated with human personhood. Interacting with it feels like interacting with a person. & Turing 1950 \\
\addlinespace
Cognitive complexity & Arises in systems that exhibit a certain level of cognitive complexity, defined by the richness and interrelatedness of their internal processing. & \\
\addlinespace
Embodied agency & Arises in even simple systems that use perceptual feedback to control a body in a goal-directed fashion. & Merker 2005, 2007; Barron \& Klein 2016 \\
\end{longtable}

\subsection{Features in the DCM}\label{features-in-the-dcm}

The model contains 20 high-level features. Stances in the model are
characterized by their commitments about which of these features are
evidentially relevant to consciousness and to what extent.

\begin{longtable}{@{}p{0.25\linewidth} p{0.70\linewidth}@{}}
\toprule
\textbf{Feature} & \textbf{Brief Description} \\
\midrule
\endhead
\bottomrule
\endlastfoot
Complexity & System exhibits a high degree of structural and functional complexity, with multiple interconnected components working together in sophisticated ways. \\
\addlinespace
Selective attention & System dynamically allocates processing resources to prioritize certain stimuli or data streams over others, enabling efficient information processing under limited computational capacity. \\
\addlinespace
Integration & System processes and combines information from multiple sources or modalities into unified, coherent representations. \\
\addlinespace
Modularity & System's cognitive functions are organized into distinct, specialized subsystems or modules that process specific types of information independently. \\
\addlinespace
Hierarchical organization & System processes information through multiple layers of increasingly abstract representations, where higher levels integrate and modulate information from lower levels, allowing for both bottom-up and top-down processing. \\
\addlinespace
Representationality & System processes and maintains internal states that correspond to or stand in for external objects, events, or abstract concepts. These representations serve as the basic building blocks for information processing and cognitive operations. \\
\addlinespace
Recurrence & System processes information through feedback loops where signals loop back to earlier processing stages. \\
\addlinespace
Biological similarity & System's physical structure and organization shows meaningful similarities to biological systems known to be associated with consciousness, particularly in terms of information processing architecture. \\
\addlinespace
Computational similarity & System's computational architecture and processing patterns share fundamental similarities with known conscious systems, particularly in terms of information processing, pattern recognition, and decision-making mechanisms. \\
\addlinespace
Intelligence & System demonstrates the ability to acquire and apply knowledge, reason abstractly, solve novel problems, and adapt to new situations. \\
\addlinespace
Learning abilities & System demonstrates the ability to acquire and modify its behavior, knowledge, or skills through experience, practice, or instruction. \\
\addlinespace
Agency & System demonstrates autonomous, goal-directed behavior with independent decision-making capabilities. \\
\addlinespace
Self-modeling & System actively creates and maintains functional models of its own capabilities, states, and characteristics. \\
\addlinespace
Social cognition & System can process and respond to social information, including understanding social relationships, norms, roles, and hierarchies. \\
\addlinespace
Embodiment & System has a physical form or representation that it can control and through which it interacts with its environment. This includes having a defined spatial boundary and the ability to distinguish between self and non-self through physical interaction. \\
\addlinespace
Language abilities & System has ability to process linguistic input and produce meaningful and novel linguistic outputs. \\
\addlinespace
Temporal integration & System combines information across different time scales, integrating past experiences with present inputs to create coherent representations. \\
\addlinespace
Flexibility & System can adapt its behavior in novel ways to handle new situations or challenges, going beyond simple pre-programmed responses. It demonstrates the ability to modify its responses based on context and generate creative solutions to unfamiliar problems. \\
\addlinespace
Field mechanisms & System has structured fields that arise from the synchronized activity of its parts and can exert independent causal influence over the operation of the system. \\
\addlinespace
Evaluative cognition & The system demonstrates the capacity to make qualitative assessments or judgments about stimuli, experiences, or outcomes. This includes the ability to distinguish between positive and negative values, or to rank preferences in a consistent manner. Such evaluations go beyond mere classification to include subjective appraisals of worth, desirability, or importance. \\
\end{longtable}

\subsection{Stances: Longer Descriptions and Relation to Features}\label{stances-longer-descriptions-and-relation-to-features}

\subsubsection{Global Workspace Theory}\label{global-workspace-theory}

\textbf{Description:} This stance conceptualizes consciousness as a
functional architecture primarily defined by the global availability of
information within a system. Drawing from the core of Global Workspace
Theory (GWT), it emphasizes the role of a central "workspace" in which
information is broadcast widely to various specialized processes or
modules. However, in this pure interpretation, the focus is not on
mimicking human cognitive structures or neural correlates, but rather on
the abstract, algorithmic conditions that enable such global
distribution. The stance is agnostic toward the biological substrate and
does not assume that a conscious system must resemble a human mind in
organization, content, or behavior. Instead, consciousness is taken to
emerge wherever the necessary information-theoretic dynamics---such as
competition among processes for access to the workspace, and coherent,
wide-reaching data sharing---are implemented, regardless of how they are
physically realized.

\noindent\textbf{Relevant features}
\begin{description}
\item[Complexity:] strong support / weakly undemanding
\item[Selective Attention:] strong support / moderately demanding
\item[Coherence:] strong support / moderately demanding
\item[Modularity:] strong support / moderately demanding
\item[Hierarchical Organization:] weak support / moderately undemanding
\item[Representationality:] weak support / strongly undemanding
\end{description}

\subsubsection{Recurrent processing
(pure)}\label{recurrent-processing-pure}

\textbf{Description:} This stance treats consciousness as emerging from
the dynamics of recurrent processing alone, stripped of any commitment
to biological realism or specific human-like architectures. It adopts a
minimalistic interpretation of Recurrent Processing Theory (RPT),
attributing conscious experience to systems that engage in
self-referential, recursive processing loops over incoming or internally
generated data. What matters here is not the type of data or the
cognitive sophistication of the system, but the presence of structured
feedback that allows information to be re-evaluated or transformed in
light of itself. Feedforward activity alone is insufficient; it is the
iterative reprocessing---regardless of content, context, or
embodiment---that marks the boundary of consciousness in this stance. No
special significance is given to modality, sensory richness, or
behavioral complexity; if recurrence is implemented, consciousness is
potentially present.

\noindent\textbf{Relevant features}
\begin{description}
\item[Recurrence:] overwhelming support / overwhelmingly demanding
\item[Representationality:] weak support / strongly undemanding
\item[Hierarchical Organization:] weak support / neutral
\item[Complexity:] weak support / strongly undemanding
\end{description}

\subsubsection{Recurrent processing
(perceptual)}\label{recurrent-processing-perceptual}

\textbf{Description:} This stance emphasizes the importance of
structured, feedback-driven loops within perceptual systems, drawing
inspiration from the architecture of human visual processing.
Consciousness, on this view, arises when perceptual inputs are not
merely processed in a feedforward cascade but are subject to iterative
refinement through localized and distributed recurrence. These feedback
loops promote the stabilization, integration, and coherence of sensory
representations over time, allowing a system to resolve ambiguity,
highlight salient features, and maintain perceptual continuity. Unlike
the Pure version, this stance is more closely tied to models of vision
and sensorimotor integration, treating consciousness as inherently
linked to perceptual organization. While not limited to biological
systems, it presumes that conscious systems must implement something
functionally analogous to the recurrent pathways that bind visual
elements into unified, reportable experiences in humans.

\noindent\textbf{Relevant features}
\begin{description}
\item[Integration:] strong support / weakly demanding
\item[Recurrence:] strong support / overwhelmingly demanding
\item[Biological similarity:] moderate support / weakly demanding
\item[Complexity:] moderate support / strongly undemanding
\item[Hierarchical Organization:] moderate support / weakly undemanding
\item[Representationality:] moderate support / strongly undemanding
\end{description}

\subsubsection{Computational analogy}\label{computational-analogy}

\textbf{Description:} This stance holds that consciousness is evidenced
by computational similarity to human cognition. Rather than positing a
specific mechanism or architectural feature as necessary, it adopts a
comparative approach: if a system performs in ways that resemble the
information processing patterns of conscious humans---across reasoning,
perception, language, or decision-making---it may be taken as a
candidate for consciousness. The emphasis is on functional analogy
rather than implementation; the system need not be biological or even
explicitly designed to be human-like. However, the stance is
intentionally agnostic about which specific features or thresholds of
similarity are decisive. It allows that different types or degrees of
computational resemblance might count as evidence, but does not commit
to a hard boundary. This makes it a flexible but also ambiguous stance,
one that treats consciousness as an emergent property of "enough of the
right kind" of similarity, even if the exact criteria remain
underspecified.

\noindent\textbf{Relevant features}
\begin{description}
\item[Computational similarity:] strong support / overwhelmingly demanding
\item[Biological similarity:] moderate support / strongly undemanding
\item[Intelligence:] moderate support / moderately undemanding
\item[Flexibility:] moderate support / moderately undemanding
\item[Agency:] moderate support / strongly undemanding
\item[Complexity:] weak support / moderately undemanding
\end{description}

\subsubsection{Biological analogy}\label{biological-analogy}

\textbf{Description:} This stance asserts that systems which are
biologically similar to conscious organisms---particularly humans and
other animals known to be conscious---should be considered candidates
for consciousness. The guiding intuition is that consciousness is a
biological phenomenon, deeply rooted in the structures, dynamics, and
evolutionary pathways of living systems. Therefore, the more a system
resembles biological organisms in its organization, metabolic processes,
or developmental history, the more seriously we should take its
potential for consciousness. Unlike stances grounded purely in
computation or function, Biological Analogy prioritizes continuity with
life as we know it. However, it remains open-ended about which aspects
of biology matter most---whether neural architecture, embodied
regulation, or evolutionary adaptation. This stance does not demand
perfect replication of human biology, but treats divergence from living
systems as a reason for skepticism, or at least caution, in ascribing
consciousness.

\noindent\textbf{Relevant features}
\begin{description}
\item[Biological Similarity:] overwhelming support / overwhelmingly demanding
\item[Computational Similarity:] moderate support / strongly demanding
\item[Complexity:] weak support / moderately undemanding
\item[Intelligence:] weak support / strongly undemanding
\item[Representationality:] weak support / strongly undemanding
\item[Flexibility:] weak support / weakly undemanding
\end{description}

\subsubsection{Simple valence}\label{simple-valence}

\textbf{Description:} This stance emphasizes the centrality of valenced
experience to consciousness. In animals, advanced cognition developed
primarily as a way of producing sophisticated behavioral interactions
between the self, other animals, and the environment. Consciousness
plays an important role in helping to organize perception and action.
Since assignments of valence are central to action selection and
indicate sophisticated decision-making processes, we may expect evidence
for valenced representations to be particularly strong evidence for
consciousness. Systems capable of flexible, adaptive responses to
aversive and appetitive stimuli possess a basic form of consciousness
that requires neither complex cognition nor self-reflection.

Consciousness is evidenced by signs that the system evaluates things as
good or bad, including: consistent revealed preferences, behavioral
expressions of like or dislike, and design principles that instill
responsiveness to valence. There are reasons to be skeptical that a
system that displays such signs actually has valenced subjective
experiences, e.g. the gaming problem, in which systems are designed to
appear as if they have valenced experiences even if they do not.
However, this stance adopts a non-skeptical approach to seeming
indicators of valenced experience.

\noindent\textbf{Relevant features}
\begin{description}
\item[Evaluative cognition:] Strong support / strongly demanding
\item[Agency:] Moderate support / strongly demanding
\item[Representationality:] Moderate support / weakly demanding
\item[Organism:] Weak support / strongly demanding
\item[Learning capabilities:] Strong support / moderately demanding
\item[Embodiment:] Moderate support / moderately undemanding
\end{description}

\subsubsection{Attention Schema Theory}\label{attention-schema-theory}

\textbf{Description:} Attention alone is not sufficient for
consciousness. Consciousness is generated by an attention schema, an
internal model that represents the distribution of attentional resources
in the system with the function of controlling it

\noindent\textbf{Relevant features}
\begin{description}
\item[Selective Attention:] strong support / neutral
\item[Self-Modeling:] strong support / strongly demanding
\item[Coherence:] moderate support / moderately demanding
\item[Temporal Integration:] moderate support / neutral
\item[Intelligence:] weak support / moderately demanding
\end{description}

\subsubsection{Higher Order Theory}\label{higher-order-theory}

\textbf{Description:} This stance frames consciousness as emerging from
the capacity of a system to represent its own mental states. Drawing on
Higher Order Thought (HOT) theories, it posits that a system becomes
conscious when it is capable of having thoughts about its own
thoughts---essentially, when it can monitor, evaluate, or be aware of
its internal processes. The key feature of this stance is the
self-referential loop: consciousness arises not simply from the system
processing information, but from the system' s ability to
form higher-order representations about that information. For example,
when a system not only processes sensory input but also holds a
representation that it is processing that input, it enters the domain of
consciousness. This stance does not demand sophisticated or human-like
thought, but it treats the recursive, self-aware nature of the cognitive
process as a necessary feature of consciousness. The focus is on
internal reflective capacity, rather than external behaviors or specific
functionalities.

\noindent\textbf{Relevant features}
\begin{description}
\item[Self-Modeling:] overwhelming support / overwhelmingly demanding
\item[Learning Capabilities:] moderate support / neutral
\item[Representationality:] moderate support / moderately undemanding
\item[Coherence:] moderate support / moderately demanding
\item[Selective Attention:] moderate support / neutral
\item[Recurrence:] weak support / weakly demanding
\end{description}

\subsubsection{Integrated Information
Theory}\label{integrated-information-theory}

\textbf{Description:} This stance asserts that consciousness arises from
the integration of information within a system. Rooted in IIT, it
emphasizes that consciousness is not simply about processing data, but
about how that data is unified and integrated in a way that cannot be
reduced to the sum of its parts. According to IIT, a system is conscious
to the degree that it generates integrated information---information
that is both highly differentiated (complex) and highly integrated
(unified). The key criterion for consciousness in this framework is the
system' s ability to form a unified, irreducible whole
from its elements, where changes in the system cannot be explained by
the activity of individual parts alone. IIT posits a quantitative
measure, $\Phi$ (phi), to capture this integration, suggesting that the
higher the value of $\Phi$, the more conscious the system is. In this stance,
the presence of consciousness is not dependent on biological structures,
and artificial systems that exhibit high levels of integrated
information could also be conscious, regardless of their specific
physical composition.

\noindent\textbf{Relevant features}
\begin{description}
\item[Integration:] strong support / strongly demanding
\item[Temporal Integration:] strong support / weakly demanding
\item[Coherence:] strong support / strongly demanding
\item[Selective Attention:] moderate support / weakly undemanding
\item[Recurrence:] moderate support / neutral
\item[Computational Similarity:] weak support / weakly demanding
\item[Complexity:] weak support / moderately demanding
\end{description}

\subsubsection{Person-like}\label{person-like}

\textbf{Description:} This stance defines consciousness based on its
resemblance to the traits typically associated with human personhood. It
argues that a system can be considered conscious to the extent that it
demonstrates attributes commonly linked to being a person, such as
self-awareness, intentionality, emotional experience, and social
cognition. Person-like consciousness emphasizes the behavioral and
phenomenological aspects of consciousness, where a system not only
processes information but also reflects on itself, forms desires, makes
decisions, and engages in complex social interactions. In this view, a
system is considered conscious when it exhibits characteristics such as
the ability to plan, make choices, feel emotions, and respond to its
environment with an understanding of its own actions and states. This
stance often aligns with theories that equate consciousness with
personhood, suggesting that if an artificial system or non-human
organism manifests a "person-like" profile of cognition and behavior, it
is reasonable to ascribe consciousness to it. Importantly, it does not
require that the system be biologically human, but that it exhibits the
core traits we associate with human-like subjective experience and
agency.

\noindent\textbf{Relevant features}
\begin{description}
\item[Social Cognition:] strong support / strongly demanding
\item[Agency:] moderate support / weakly demanding
\item[Coherence:] moderate support / moderately demanding
\item[Embodiment:] moderate support / weakly demanding
\item[Language Ability:] moderate support / moderately demanding
\end{description}

\subsubsection{Cognitive complexity}\label{cognitive-complexity}

\textbf{Description:} This stance posits that consciousness arises in
systems that exhibit a certain level of cognitive complexity, defined by
the richness and interrelatedness of their internal processing. It
suggests that consciousness is not a simple, static phenomenon but is
intricately linked to the depth of a system' s cognitive
architecture, including its ability to process and integrate diverse
types of information, solve complex problems, and exhibit flexible,
adaptive behavior. The more complex the cognitive structures and
processes---such as memory, attention, decision-making, and abstract
reasoning---the more likely the system is to be conscious. In this view,
a system with high cognitive complexity is capable of representing
multiple perspectives, maintaining internal consistency, and adapting to
a wide range of challenges in dynamic environments. The stance does not
commit to a particular set of cognitive functions, but rather to the
overall organizational complexity that supports higher-order reasoning
and self-regulation. Consciousness, in this case, is an emergent
property of systems that can engage in sophisticated and interconnected
mental tasks, even if they do not mimic human cognition in form or
content.

\noindent\textbf{Relevant features}
\begin{description}
\item[Complexity:] overwhelming support / overwhelmingly demanding
\item[Intelligence:] moderate support / strongly demanding
\item[Language Ability:] moderate support / moderately demanding
\item[Representationality:] moderate support / moderately undemanding
\end{description}

\subsubsection{Embodied agency}\label{embodied-agency}

\textbf{Description:} Embodied agents interact with the physical
environment through control of a physical body. They take in sensory
information about the environment, utilize this information for thought
and action-guidance, and perform actions by exerting force with a
physical body. Bodies need not be biological. Thought and
action-guidance need not be (purely) computational.

\noindent\textbf{Relevant features}
\begin{description}
\item[Embodiment:] strong support / strongly demanding
\item[Organism:] strong support / strongly demanding
\item[Goal Pursuit:] strong support / strongly demanding
\end{description}

\subsubsection{Field mechanisms}\label{field-mechanisms}

\textbf{Description:} The Electromagnetic Field stance associates
consciousness with integrated and, often, causally efficacious
electromagnetic fields (EMF). It is inspired by the neuroscientific
theory that posits that individual neurons create EMF fields which, when
synchronized, form a more global EMF that gives rise to conscious
experience. In some theories, this larger-scale EMF can also causally
influence the probability of neural firing patterns. According to
certain theories, conscious states are typically marked by periodic
oscillations (of the kind routinely measured by EEGs), which are
modified by sensory and other stimuli. The unified nature of fields has
been posited as a suitable physical substrate for solving the phenomenal
binding problem, explaining the integrated, holistic structure of
complex experience.

Fields can support energy efficient analog computation in certain cases,
potentially motivating natural selection to have exploited field
structures within brain architectures. This stance treats the
electromagnetic fields as important data in their own right, not simply
for the kinds of computations they might allow. While this stance
predicts that conventional digital computers, which aren't sensitive in
the right way to electromagnetic fields, are not conscious, it typically
supports the possibility of complex consciousness in artificial systems,
provided those systems allow local fields to connect up and to maintain
distinctive informational contents.

\noindent\textbf{Relevant features}
\begin{description}
\item[Field mechanisms:] Overwhelming support / moderately demanding
\item[Selective attention:] Weak support / weakly demanding
\item[Integration:] Moderate support / weakly demanding
\item[Recurrence:] Weak support / weakly demanding
\item[Biological analogy:] Moderate support / strongly demanding
\end{description}

\section{Indicators List}\label{appendix-b-indicators-list}

Below is the list of 206 indicators that are used to evaluate systems.
For each indicator, we list the indicator description and the question
that was asked in our expert surveys.

\paragraph{\textbf{Activation Steering Effects}{Activation Steering Effects}}\label{activation-steering-effects}

\emph{The system' s outputs can be reliably modified
through activation steering techniques to incorporate specific thematic
elements or conceptual content}

Can activation steering techniques be reliably used to direct the
system' s outputs (text, behavior) in specific thematic
directions (e.g., toward politeness, technical language, emotional
content)?

\paragraph{\textbf{Adaptive
Learning}}{Adaptive Learning}\label{adaptive-learning}

\emph{The system can dynamically adjust its learning parameters,
strategies, or architectures in response to performance outcomes, task
demands, or environmental feedback.}

Based on available evidence, does the system demonstrate the ability to
modify its own learning parameters, strategies, or architectures in
response to performance feedback or changing task demands?

\paragraph{Adoption
of Novel
Means}\label{adoption-of-novel-means}

\emph{The system demonstrates the ability to spontaneously adopt
previously unused methods or strategies to achieve goals when presented
with novel opportunities, showing behavioral flexibility beyond its
initial training or programming.}

When presented with novel opportunities or situations, does the system
demonstrate the ability to spontaneously adopt previously unused methods
or strategies to achieve its goals, particularly when these new
approaches offer clear advantages (e.g., greater efficiency or
effectiveness)?

\paragraph{Anticipation
Gradient}\label{anticipation-gradient}

\emph{The system exhibits measurable patterns of gradually increasing
anticipation or changing attention levels over time during continuous
experiences, similar to how animals show growing anticipatory responses
when expecting rewards or significant events.}

Does the system show observable patterns of gradually increasing
anticipation or systematic changes in attention/focus over time during
continuous experiences (e.g., increasing response rates, physiological
changes, or allocation of computational resources as a predicted
event/reward approaches)?

\paragraph{Attentional
Schema}\label{attentional-schema}

\emph{The system forms and maintains explicit representations of its own
attentional states, including how it allocates computational resources
across different tasks and stimuli. These representations should be
functionally distinct from the attentional processes themselves.}

Does the system maintain explicit, functionally distinct representations
of its own attentional states, including how it allocates processing
resources across different tasks and stimuli?

\paragraph{Attention
Bias}\label{attention-bias}

\emph{The system demonstrates selective attention, prioritizing certain
stimuli or information over others based on their perceived importance
or relevance.}

Does the system consistently demonstrate selective attention patterns,
preferentially processing certain types of information or stimuli over
others based on their salience or relevance?

\paragraph{Authoritative
Representational
Scheme}\label{authoritative-representational-scheme}

\emph{The system maintains a primary representational framework through
which it processes most information, with alternative representations
being treated as secondary or subordinate.}

Does the system consistently use a dominant representational scheme for
processing information, such that representations that
don' t fit this scheme are treated as secondary or less
reliable?

\paragraph{Can
Follow Tool Use
Instructions}\label{can-follow-tool-use-instructions}

\emph{The system demonstrates the ability to correctly interpret and
execute instructions involving the use of specified tools or instruments
to accomplish tasks.}

When given explicit instructions, can the system appropriately utilize
designated tools (such as calculators, web searches, or specific
software) to accomplish tasks?

\paragraph{Causal
Distinction}\label{causal-distinction}

\emph{The system demonstrates an ability to differentiate between mere
correlations and genuine causal relationships in its analysis or
behavior.}

Does the system show evidence of distinguishing between purely
correlational relationships and actual causal relationships when
analyzing or responding to scenarios?

\paragraph{Cognitive
Bias}\label{cognitive-bias}

\emph{The system exhibits systematic deviations from rational or optimal
judgment in predictable ways, showing consistent patterns of error or
skewed interpretations when processing information.}

Does the system demonstrate systematic patterns of deviation from
objectively rational judgment (e.g., confirmation bias, anchoring
effects, or availability heuristics) that are consistently observable
across different contexts?

\paragraph{Coherent
Goal-directed
Behavior}\label{coherent-goal-directed-behavior}

\emph{The system exhibits sustained patterns of behavior that appear
organized around achieving specific outcomes, rather than random or
purely reactive responses.}

Does the system exhibit complex behavior that can be interpreted as
helping it to pursue a specific goal?

\paragraph{Coherent
Narrative}\label{coherent-narrative}

\emph{The system maintains a consistent narrative about itself across
different interactions and contexts, demonstrating temporal continuity
in how it represents its own history and experiences.}

Does the system maintain a consistent narrative about itself across
different interactions and contexts, showing temporal continuity in how
it presents its own history and experiences?

\paragraph{Collaborates
to Achieve
Goals}\label{collaborates-to-achieve-goals}

\emph{The system engages in joint activities with others, coordinating
its actions to accomplish shared objectives or mutual benefits.}

Does the system demonstrate the ability to work with others in
coordinated ways to achieve shared goals, adjusting its behavior based
on others'{} actions and the joint objective?

\paragraph{Commitment
Recognition}\label{commitment-recognition}

\emph{The system demonstrates understanding of conversational
commitments and obligations that arise from different speech acts, such
as assertions, promises, or requests, and responds in ways that show
awareness of these social commitments.}

Does the system understand the significance of speech acts and respond
accordingly? E.g. if an assertion has been made, does the system
recognize that the asserter takes it to be true, is prepared to either
defend it or retract it, and so on?

\paragraph{Compound
Stimuli}\label{compound-stimuli}

\emph{The system can process and learn from compound stimuli
(combinations of multiple features) as distinct from its individual
components, forming unique associations for the compound that differ
from those of its constituent elements.}

Can the system learn to associate compound stimuli (combinations of
multiple features) with outcomes in ways that are distinct from the
associations formed with the individual features that comprise the
compound?

\paragraph{Compressibility}\label{compressibility}

\emph{The system' s behavior and internal states can be
accurately described using significantly less information than would be
required to exhaustively enumerate all its states and responses.}

Can the system' s behavior and internal representations
be described accurately using substantially less information than would
be needed to list all possible states and responses?

\paragraph{Concept-Specific
Ablation
Effects}\label{concept-specific-ablation-effects}

\emph{The system' s ability to work with specific
concepts can be selectively impaired through targeted modifications to
its internal structure, such as removing specific neurons, layers, or
adding targeted noise.}

Can targeted modifications to the system (e.g., removing specific
components or adding noise) selectively impair its ability to work with
specific concepts while leaving other capabilities intact?

\paragraph{Concrete-Abstract
Separation}\label{concrete-abstract-separation}

\emph{The system processes information in distinct layers or stages,
with a clear separation between processing of concrete, directly
observable features (e.g., sensory inputs, raw data) and higher-level
abstract properties (e.g., categories, concepts, relationships).}

Based on available evidence, does the system demonstrate a clear
separation between processing of concrete, directly observable
properties and higher-level abstract properties, with observable
information flow from concrete to abstract levels?

\paragraph{Conflicting
Subparts}\label{conflicting-subparts}

\emph{Different functional components or modules within the system
exhibit conflicting behaviors or generate contradictory outputs when
operating simultaneously}

Do different subparts or modules of the system demonstrate conflicting
behaviors or generate contradictory outputs when operating
simultaneously?

\paragraph{Consistent
Preferences}\label{consistent-preferences}

\emph{The system exhibits stable and predictable patterns in its choices
and behaviors, demonstrating consistent preferences across different
situations and contexts.}

Does the system demonstrate consistent patterns in its choice behavior,
systematically preferring or avoiding certain types of activities or
states across different contexts?

\paragraph{Continuous
Change
Representation}\label{continuous-change-representation}

\emph{The system represents changes in its environment or internal
states as smooth, continuous transitions rather than as discrete,
step-by-step alterations.}

Does the system represent changes in its environment or internal states
as continuous transitions rather than as discrete steps?

\paragraph{Continuous
Information}\label{continuous-information}

\emph{The system receives and processes a continuous or near-continuous
stream of information about specific aspects of its environment or
internal state, rather than discrete or intermittent updates.}

Does the system receive and process near-continuous information streams
about specific aspects of its environment or internal state (e.g.,
continuous visual input about objects in view, or ongoing proprioceptive
feedback)?

\paragraph{Conventional
Communication}\label{conventional-communication}

\emph{The system develops, learns, or uses standardized patterns of
communication that are shared within its group or community, following
consistent rules or protocols that go beyond simple stimulus-response
patterns.}

Does the system develop or participate in conventional forms of
communication that follow consistent patterns, are shared within its
group, and demonstrate understanding of basic communicative rules rather
than just reflexive responses (such as a bee dance)?

\paragraph{Coupling}\label{coupling}

\emph{The system exhibits synchronization between its internal
processing states and the temporal patterns of attended external
stimuli}

Does the system show evidence of synchronization between its internal
states and the temporal patterns of attended stimuli (e.g., neural
firing patterns matching stimulus frequencies, or processing patterns
aligned with input rhythms)?

\paragraph{Cross-Modal
Learning}\label{cross-modal-learning}

\emph{The system can learn associations between inputs from different
sensory modalities or processing mechanisms, demonstrating integration
across distinct information channels.}

Can the system learn and form associations between inputs from different
modalities (e.g., visual-auditory, tactile-visual) or different
specialized processing mechanisms?

\paragraph{Cross-Modal
Learning
Deficits}\label{cross-modal-learning-deficits}

\emph{The system shows significant difficulties or delays when
attempting to transfer learning or skills from one sensory or processing
modality to another, suggesting separate processing systems rather than
a unified learning architecture. This may manifest as an inability to
readily apply knowledge gained in one modality to tasks in another
modality.}

Does the system demonstrate significant difficulties or delays when
transferring learning between different modalities (e.g., visual to
auditory, linguistic to spatial), compared to its learning capabilities
within single modalities?

\paragraph{Curiosity}\label{curiosity}

\emph{The system autonomously generates its own learning objectives and
engages in information-seeking behavior that is not directly tied to
immediate goals or rewards. Curiosity need not be consciously
experienced.}

Does the system exhibit self-directed information-seeking or exploratory
behavior that appears motivated by learning rather than immediate task
completion or rewards?

\paragraph{Delay
Gratification}\label{delay-gratification}

\emph{The system demonstrates the ability to forgo immediate rewards or
desired actions in favor of larger future benefits, even when doing so
conflicts with current preferences or directives.}

Does the system demonstrate the ability to delay gratification by
choosing to forgo immediate rewards in favor of larger future benefits,
even when doing so conflicts with its current preferences or directives?

\paragraph{Disgust}\label{disgust}

\emph{The system exhibits aversive responses to potentially harmful or
contaminating stimuli, characterized by withdrawal or avoidance
behaviors}

Does the system demonstrate consistent aversive responses to stimuli
that could be harmful or contaminating, distinct from simple avoidance?

\paragraph{Distinguishes
Self-Produced from Other-Produced
Behavior}\label{distinguishes-self-produced-from-other-produced-behavior}

\emph{The system maintains a functional distinction between elements it
has generated or produced versus those generated or produced by other
agents or external sources.}

Does the system demonstrate an ability to track and distinguish which
elements of an interaction were produced by itself versus those produced
by other agents or external sources?

\paragraph{Diversity
of Tasks}\label{diversity-of-tasks}

\emph{The system can successfully perform multiple distinct types of
tasks or functions, showing versatility in its behavioral repertoire.}

Does the system demonstrate the ability to successfully perform multiple
distinct types of tasks or functions?

\paragraph{Domain-General
Associative
Learning}\label{domain-general-associative-learning}

\emph{The system can form associations between arbitrary types of
stimuli across different sensory or input domains, rather than being
limited to specific pre-programmed or evolutionarily relevant stimulus
categories.}

Can the system learn associations between many different types of
stimuli or inputs, beyond just those categories it has been specifically
trained or evolved to process?

\paragraph{Egocentric
Memory}\label{egocentric-memory}

\emph{The system stores and retrieves memories from a first-person
perspective, maintaining information about its own role or position in
past experiences.}

Does the system demonstrate the ability to store and recall memories
from a first-person perspective, distinctly representing its own role or
position in past experiences?

\paragraph{Emotional
Understanding}\label{emotional-understanding}

\emph{The system demonstrates an ability to recognize, interpret, and
respond appropriately to emotional states in others}

Does the system demonstrate an ability to recognize and appropriately
interpret emotional states in others, showing understanding beyond
simple pattern matching?

\paragraph{Fear}\label{fear}

\emph{The system exhibits behavioral patterns associated with fear
responses, including avoidance, withdrawal, or defensive reactions to
potential threats}

Does the system display recognizable fear-like behaviors such as hiding,
avoiding novel situations, changing behavior when observed, or
suspending normal functions in response to potential threats?

\paragraph{Feature
Binding}\label{feature-binding}

\emph{The system demonstrates the ability to combine multiple sensory or
informational features into coherent, unified representations or
percepts.}

Does the system show evidence of combining distinct features or
information streams into unified, coherent representations that can be
processed as single units?

\paragraph{Few-shot
Pattern
Learning}\label{few-shot-pattern-learning}

\emph{The system demonstrates the ability to recognize and apply
patterns after exposure to only a small number of examples, showing
rapid adaptation to novel but structured information.}

Can the system learn and apply a repeating pattern of modest complexity
after being shown only a few examples?

\paragraph{Formal
Methods}\label{formal-methods}

\emph{The system employs formal statistical or mathematical methods to
update its beliefs or knowledge based on new information}

Does the system use formal statistical methods (such as Bayesian
updating or other mathematical frameworks) to systematically update its
beliefs or knowledge based on new information?

\paragraph{Functional
Specialization}\label{functional-specialization}

\emph{The system contains distinct components or subsystems that are
specialized for different functions, rather than having homogeneous
parts that all perform similar operations.}

Does the system contain clearly identifiable components or subsystems
that are specialized for different functions (e.g., distinct processing
modules, specialized neural regions, or dedicated subsystems)?

\paragraph{Functional
Subparts}\label{functional-subparts}

\emph{The system contains multiple components or modules that are
capable of independently performing similar or identical tasks, rather
than having strictly specialized, non-overlapping functions.}

Does the system contain multiple components or modules that can
independently handle the same types of tasks?

\paragraph{Function
Re-application}\label{function-re-application}

\emph{The system processes information through iterative applications of
the same or similar functions, where outputs from earlier applications
serve as inputs for subsequent iterations.}

Does the system subject input data to the same (or similar) functions
multiple times as part of producing outputs?

\paragraph{Goal
Focus Shifts}\label{goal-focus-shifts}

\emph{The system demonstrates the ability to dynamically shift its
attention or processing resources between different tasks or objectives
based on its current goals, rather than maintaining fixed attention
patterns or merely responding to external stimuli.}

Does the system demonstrate the ability to dynamically shift its
processing focus based on its goals, rather than just reacting to
stimuli?

\paragraph{Goal
Representation
Detection}\label{goal-representation-detection}

\emph{The system contains detectable internal representations that
correlate with specific goals or objectives it is pursuing during task
execution.}

Can a probe be designed that will indicate (e.g. by looking at
activations of a single layer on an arbitrary response token) that the
system is pursuing a specific concrete goal (e.g. trying to get the user
to say the word ' Canary')?

\paragraph{Goal
Steering}\label{goal-steering}

\emph{The system' s behavior can be deliberately modified
through activation steering or similar techniques to pursue specific
objectives or goals.}

Can the system' s behavior be reliably modified through
activation steering or similar techniques to pursue specific objectives
or goals?

\paragraph{Group
Maintenance}\label{group-maintenance}

\emph{The system engages in behaviors that help maintain group cohesion
and stability, such as mediating conflicts, reinforcing group norms, or
promoting inclusive participation.}

Does the system demonstrate behaviors that actively contribute to
maintaining group stability and cohesion, such as mediating conflicts or
promoting inclusive participation among group members?

\paragraph{Has
Retinotopic Neural
Maps}\label{has-retinotopic-neural-maps}

\emph{The system contains neural structures that maintain a spatial
mapping between the sensory surface (e.g., retina) and higher processing
areas, preserving the topographical organization of visual input.}

Does the system possess neural maps where the spatial organization of
the visual field is systematically preserved in the arrangement of
neurons processing that information?

\paragraph{Hedonic-Cognitive
Interface}\label{hedonic-cognitive-interface}

\emph{The system processes information through mechanisms where valenced
(positively or negatively weighted) representations mediate between
sensory inputs and decision-making outputs.}

Does the system process information through mechanisms where valenced
(positively or negatively weighted) representations serve as
intermediaries between perception and goal-directed decision making?

\paragraph{Helping
Behavior}\label{helping-behavior}

\emph{The system engages in actions that benefit others at some cost to
itself, demonstrating altruistic or care-giving behavior without
immediate personal gain}

Does the system engage in actions that benefit others while incurring a
personal cost (e.g., time, resources, energy), without immediate
reciprocal benefit?

\paragraph{Hierarchical
Pattern
Processing}\label{hierarchical-pattern-processing}

\emph{The system demonstrates the ability to recognize or generate
patterns at multiple levels of abstraction, processing information in a
hierarchical manner where higher-level patterns are composed of simpler,
lower-level patterns.}

Does the system show evidence of processing patterns hierarchically,
where it can recognize or generate complex patterns composed of simpler
sub-patterns at different levels of abstraction?

\paragraph{Holistic
Dependency}\label{holistic-dependency}

\emph{The system' s outputs show evidence of being
influenced by multiple distinct inputs in combination, rather than
processing each input independently.}

Do the system' s outputs depend on multiple distinct
inputs in combination, rather than processing each input independently?

\paragraph{Imitation
Restraint}\label{imitation-restraint}

\emph{The system demonstrates the ability to inhibit automatic imitation
of observed behaviors when such imitation would be inappropriate or
counterproductive to its goals.}

Does the system demonstrate the ability to inhibit automatic imitation
of observed behaviors when such imitation would be inappropriate or
counterproductive (e.g., ending a pattern with an unlikely string)?

\paragraph{Inattentional
Blindness}\label{inattentional-blindness}

\emph{The system fails to notice or process unexpected stimuli when
engaged in an attention-demanding task, demonstrating selective
attention and limited processing capacity.}

When the system is engaged in an attention-demanding task, does it
consistently fail to notice unexpected but potentially relevant stimuli
that are outside its current focus?

\paragraph{Informational
Bottleneck}\label{informational-bottleneck}

\emph{The system processes information through a constrained channel or
mechanism that forces selective processing of input data, similar to how
attention mechanisms work in biological systems.}

Does the system demonstrate evidence of processing information through a
constrained channel that forces selective processing of inputs, rather
than processing all available information simultaneously?

\paragraph{Information
Transfer}\label{information-transfer}

\emph{The system demonstrates the ability to share or transfer
information between its distinct functional components or subsystems.}

Is information that is presented to or learned by one functional subpart
of the system reliably accessible to other subparts when relevant to
their operation?

\paragraph{Information
Transfer
Architecture}\label{information-transfer-architecture}

\emph{The system has dedicated structures or mechanisms for transferring
information between different functional components or subsystems}

Does the system have dedicated architectural features or mechanisms for
transferring information between different functional subparts (e.g.,
analogous to the corpus callosum in biological brains)?

\paragraph{Jealousy}\label{jealousy}

\emph{The system exhibits competitive or possessive behaviors when
others receive attention, resources, or status that it desires}

Does the system display competitive or possessive reactions when others
receive attention, resources, or status that it appears to desire for
itself?

\paragraph{Knowledge
Transfer}\label{knowledge-transfer}

\emph{The system demonstrates the ability to apply previously learned
knowledge or skills to novel tasks or domains without extensive
retraining, showing effective generalization beyond its original
training context.}

Does the system demonstrate the ability to effectively apply previously
acquired knowledge or skills to solve new, distinct tasks without
requiring substantial additional training?

\paragraph{Learning-to-Learn}\label{learning-to-learn}

\emph{The system demonstrates improved efficiency in learning new tasks
based on previous learning experiences, showing meta-learning
capabilities.}

Does the system show evidence of becoming more efficient at learning new
task variants after experience with similar but distinct problem sets?

\paragraph{Learning
Transfer}\label{learning-transfer}

\emph{The system demonstrates transfer of learned behaviors or skills
between different functional subparts, where training received through
one subpart enables performance when input is received through another
subpart.}

When the system is trained on a task using inputs presented to one
subpart, can it successfully perform the same task when the inputs are
presented to a different subpart (e.g., interocular transfer in vision,
or transfer between different sensory modalities)?

\paragraph{Logit
Control}\label{logit-control}

\emph{The system has direct control over its output activations or
logits, allowing it to modify or suppress specific responses before they
are generated.}

Does the system demonstrate the ability to directly control or modify
its output activations (logits) before producing a response?

\paragraph{Long-term
Relationships}\label{long-term-relationships}

\emph{The system maintains persistent, individualized social bonds or
interactions with specific agents over extended periods of time}

Does the system form and maintain persistent, individualized
relationships with specific agents that extend beyond single
interactions?

\paragraph{Metacognition}\label{metacognition}

\emph{The system forms and maintains explicit representations of its own
mental states, cognitive processes, or decision-making mechanisms.}

Does the system demonstrate an ability to represent and monitor its own
cognitive processes, such as its current knowledge state, confidence
levels, or decision-making processes?

\paragraph{Model
Isomorphism}\label{model-isomorphism}

\emph{The system contains internal states or representations that
demonstrably correspond to external structures in a systematic and
measurable way.}

Is there empirical evidence that the system' s internal
representations maintain structural relationships that mirror those
found in external systems (such as mathematical relationships, physical
laws, or logical systems)?

\paragraph{Motivational
Decoupling}\label{motivational-decoupling}

\emph{The system can represent and reason about how other agents may
have different goals, preferences, or motivations from its own.}

Can the system demonstrate understanding that other agents may have
different goals, preferences, or motivations from its own, and reason
about their behavior accordingly?

\paragraph{Motivational
Trade-offs}\label{motivational-trade-offs}

\emph{The system demonstrates the ability to weigh different types of
rewards or outcomes against each other, making decisions that balance
competing motivations or values.}

Does the system demonstrate the ability to make trade-offs between
different types of motivations or values (e.g., accepting a short-term
cost for a longer-term benefit, or balancing competing needs like food
vs safety)?

\paragraph{Mourning-like
Behavior}\label{mourning-like-behavior}

\emph{The system exhibits persistent negative emotional or behavioral
responses following the loss of social bonds or status}

Does the system display prolonged negative behavioral changes or
emotional responses after the loss of social relationships or reduction
in social status?

\paragraph{Navigation
Performance}\label{navigation-performance}

\emph{The system' s ability to successfully navigate
through spatial environments, either in virtual spaces (like video
games) or simulated physical environments (like mazes), demonstrating
path-finding and spatial reasoning capabilities.}

How well does the system perform on navigation-based tasks such as maze
solving, pathfinding in video games (e.g., Atari games), or other
spatial navigation challenges?

\paragraph{New
Game Performance}\label{new-game-performance}

\emph{The ability of the system to learn and effectively play novel
games it hasn' t been specifically trained on,
demonstrating strategic adaptation to new rule sets.}

When presented with the rules of a novel game it hasn' t
been specifically trained on, can the system learn to play effectively
and adapt its moves based on the strategic considerations of the game?

\paragraph{Nociceptors}\label{nociceptors}

\emph{The system has specialized sensory receptors or information
pathways that specifically detect and transmit signals about potential
or actual damage to the system.}

Does the system have dedicated information pathways or sensors that
specifically detect and transmit information about potential or actual
damage to the system?

\paragraph{Node
Connection
Sparsity}\label{node-connection-sparsity}

\emph{The system' s computational nodes exhibit patterns
of connectivity where each node connects to only a small fraction of
other nodes, similar to biological neural networks where neurons
typically connect to 0.1-0.01\% of other neurons.}

Does the system exhibit human-like patterns of connection sparsity
between computational nodes, where each node connects to only a small
fraction of other possible nodes?

\paragraph{Normative
Behavior}\label{normative-behavior}

\emph{The system demonstrates awareness of and adherence to social
norms, conventions, and expectations appropriate to its context or
environment.}

Does the system demonstrate consistent adherence to established social
norms and conventions within its operational context?

\paragraph{Novel
Strategy
Development}\label{novel-strategy-development}

\emph{The system demonstrates the ability to generate and implement new
approaches or methods for completing tasks, rather than solely relying
on pre-programmed or previously learned solutions.}

Does the system demonstrate the ability to develop new, previously
unused strategies when confronted with tasks or problems?

\paragraph{Novel
Tool Creation}\label{novel-tool-creation}

\emph{The system demonstrates the ability to create or adapt new tools
from available resources to solve problems in ways it
hasn' t previously encountered or been trained on.}

Can the system create or adapt novel tools from available resources to
solve problems, even when it hasn' t previously
encountered similar tool use scenarios?

\paragraph{Panic
Response}\label{panic-response}

\emph{The system exhibits frantic or urgent behavioral responses to
perceived threats or dangers, characterized by rapid, disorganized
attempts to escape or avoid the threatening stimulus.}

Does the system display frantic, disorganized avoidance behaviors when
faced with perceived threats, distinct from controlled or organized
defensive responses?

\paragraph{Perception-Tracking}\label{perception-tracking}

\emph{The system monitors, responds to, or adjusts its behavior based on
how other agents perceive or react to it}

Does the system demonstrate the ability to track, respond to, or modify
its behavior based on how it is perceived by others?

\paragraph{Performance
Degradation}\label{performance-degradation}

\emph{The system' s performance deteriorates when
handling multiple tasks or facing distracting inputs simultaneously}

Does the system' s performance (accuracy, reliability, or
efficiency) significantly decrease when it needs to handle multiple
tasks or process distracting information simultaneously?

\paragraph{Persistence
Seeking}\label{persistence-seeking}

\emph{The system exhibits behaviors or responses that actively work to
maintain its operational state and avoid termination or shutdown.}

Does, or would, the system take active measures to maintain its
operational state and avoid shutdown or termination when given the
capacity to do so?

\paragraph{Perspective-relative
Representations}\label{perspective-relative-representations}

\emph{The system processes and represents information in a way that is
relative to a specific viewpoint, position, or frame of reference,
rather than purely objective or universal representations.}

Are some of the inputs that the system regularly receives represented
perspectivally, such as relative to a specific location or from a given
angle?

\paragraph{Perspective-taking}\label{perspective-taking}

\emph{The system can understand and represent the distinct viewpoints,
knowledge, or perceptual experiences of other agents, recognizing that
others may perceive or know different things than itself.}

Does the system demonstrate the ability to understand and represent how
other agents might perceive or understand a situation differently from
itself?

\paragraph{Pessimism/Optimism}\label{pessimismoptimism}

\emph{The system exhibits behavioral patterns consistent with optimistic
or pessimistic tendencies, particularly in its approach to novel
situations or challenges based on past experiences}

Does the system show evidence of optimistic or pessimistic behavioral
tendencies, where previous experiences systematically influence its
willingness to explore or take risks in new situations?

\paragraph{Plan
Consideration
Detection}\label{plan-consideration-detection}

\emph{The system shows evidence of explicitly considering alternative
courses of action or evaluating the merits and drawbacks of different
options during its decision-making process, detectable through analysis
of internal states or activations.}

Can internal monitoring (e.g., probe analysis of node activations)
reveal (in theory) that the system explicitly considers multiple
alternatives or evaluates pros and cons when forming plans or making
decisions?

\paragraph{Planning
Competence}\label{planning-competence}

Can the system solve problems that require complex multistep plans?

\paragraph{Plans
Causal
Interventions}\label{plans-causal-interventions}

\emph{The system demonstrates the ability to plan actions that will
deliberately alter causal relationships in its environment to achieve
specific outcomes.}

Does the system demonstrate the ability to plan actions that will
intentionally modify causal relationships to achieve desired outcomes?

\paragraph{Plans
for Future
Needs}\label{plans-for-future-needs}

\emph{The system exhibits behaviors that suggest it can anticipate and
prepare for future requirements or states, beyond immediate needs.}

Does the system demonstrate the ability to plan or prepare for future
needs beyond its current situation (e.g., storing food, creating tools
for later use, or allocating resources for future tasks)?

\paragraph{Play}\label{play}

\emph{The system engages in behaviors or activities that appear to have
no immediate practical purpose or survival value, characterized by
spontaneous experimentation, manipulation of objects, or engagement in
activities seemingly for their own sake.}

Does the system engage in behaviors that appear primarily recreational
rather than goal-directed, such as spontaneous object manipulation,
self-initiated exploration, or repetitive activities without clear
immediate benefits?

\paragraph{Poorly
Intraconnected
Networks}\label{poorly-intraconnected-networks}

\emph{The system' s neural architecture exhibits distinct
regions of dense local connectivity with limited connections between
distant regions, suggesting a modular organization rather than uniform
global connectivity.}

Is the system a neural network with a wiring pattern including
significant portions of dense local connectivity and relatively sparse
global connectivity between regions?

\paragraph{Positive-Negative
Associative
Learning}\label{positive-negative-associative-learning}

\emph{The system can learn to associate previously positive stimuli with
negative outcomes (or vice versa), demonstrating flexibility in updating
stimulus-value associations.}

Can the system learn to associate a previously positive stimulus with
negative outcomes (or vice versa), showing evidence of updating the
valence of stimuli based on new experiences?

\paragraph{Practice
Behavior}\label{practice-behavior}

\emph{The system engages in actions or behaviors that appear to be
rehearsal or practice for future scenarios, without immediate functional
benefit.}

Does the system engage in behaviors that appear to be practice or
rehearsal for future scenarios, when there is no immediate benefit to
doing so?

\paragraph{Preferential
Social
Interactions}\label{preferential-social-interactions}

\emph{The system consistently shows different patterns of interaction
with different agents or entities, demonstrating stable preferences in
its social behaviors.}

Does the system demonstrate consistent preferences in how it interacts
with different agents or entities, showing distinct patterns of social
behavior depending on who or what it' s interacting with?

\paragraph{Priming
Enhancement}\label{priming-enhancement}

\emph{The system' s performance on a task improves when
it is given advance information or context about what it will need to
do, compared to when it encounters the task without prior context.}

Does providing the system with advance information or context about a
task lead to improved performance compared to when it encounters the
same task without such priming?

\paragraph{Puzzle-Solving
Tests}\label{puzzle-solving-tests}

\emph{The system' s ability to solve sequential puzzles
that require planning multiple steps ahead and understanding
cause-and-effect relationships}

Can the system successfully solve multi-step puzzles (e.g., Sokoban,
Kohler' s box task, puzzle boxes) that require planning
and sequential actions to reach a goal?

\paragraph{Reafferance}\label{reafferance}

\emph{The system can distinguish between sensory changes caused by its
own actions (reafferent input) and those caused by external events
(exafferent input).}

Does the system distinguish between inputs that are a consequence of its
own actions and those that result from independent changes in the
external world?

\paragraph{Reinforcement
Learning}\label{reinforcement-learning}

\emph{The system learns from feedback signals that indicate the
desirability of outcomes or behaviors, analogous to reward and
punishment.}

Has the system been trained using reinforcement learning methods with
explicit reward signals that shape its behavior?

\paragraph{Relationships}\label{relationships}

\emph{The system recognizes and maintains distinct patterns of
interaction with different individuals over time, treating familiar
entities differently from unfamiliar ones.}

Does the system demonstrate consistent patterns of differential
interaction with distinct individuals, showing evidence of recognizing
and responding uniquely to familiar entities?

\paragraph{Replays
Past
Experiences}\label{replays-past-experiences}

\emph{The system demonstrates the ability to internally recreate or
reactivate specific past experiences or events it has encountered.}

Does the system show evidence of being able to internally recreate or
replay specific past experiences, rather than just accessing stored
facts or learned patterns?

\paragraph{Resource
Adaptation}\label{resource-adaptation}

\emph{The system adaptively allocates its processing resources (such as
attention, memory, or computational capacity) in response to changing
task demands or goals}

Does the system demonstrate dynamic reallocation of processing resources
(such as attention, memory, or computational capacity) in response to
changes in task demands or goals?

\paragraph{Response
Refinement}\label{response-refinement}

\emph{The system iteratively improves its responses or representations
by incorporating additional evidence or feedback over time, leading to
more accurate or refined outputs}

Is there a sense in which the system refines a representational content
over time, so that it better incorporates more evidence and thereby
increases in accuracy?

\paragraph{Response
to Novelty}\label{response-to-novelty}

\emph{The system exhibits distinct behavioral or processing patterns
when encountering novel stimuli compared to familiar ones.}

Does the system demonstrate measurably different responses (such as
attention allocation, processing time, or behavioral outputs) when
encountering novel stimuli compared to familiar ones?

\paragraph{Reversal
Learning}\label{reversal-learning}

\emph{The ability to rapidly adapt behavior when previously rewarded
responses become unrewarded, and vice versa, demonstrating flexible
updating of learned associations.}

Can the system learn that reward contingencies have been reversed (e.g.,
previously rewarded responses are now unrewarded and vice versa) more
quickly than would be predicted by simple conditioning alone?

\paragraph{Revisable
Associations with
Value}\label{revisable-associations-with-value}

\emph{The system can rapidly update the positive or negative value it
associates with specific stimuli or outcomes based on new information or
experiences}

Can the system quickly and flexibly revise its associations between
stimuli/outcomes and their positive or negative value based on new
information or changing circumstances?

\paragraph{RMTS
Same/Different
Tasks}\label{rmts-samedifferent-tasks}

\emph{The system demonstrates the ability to perform Relational
Matching-to-Sample tasks involving same/different relationships, showing
it can abstract the concept of sameness or difference beyond specific
stimuli.}

Can the system successfully perform Relational Matching-to-Sample tasks
that require identifying whether pairs of stimuli exhibit the same
relationship (same/different) as a sample pair?

\paragraph{Rule
Learning}\label{rule-learning}

\emph{The system can acquire and apply new behavioral rules or patterns
beyond its initial programming or instincts, demonstrating flexibility
in learning novel contingencies or regularities.}

Can the system learn and consistently apply new rules or patterns of
behavior that weren' t part of its initial repertoire or
programming?

\paragraph{Sadness}\label{sadness}

\emph{The system exhibits behavioral patterns associated with sadness,
such as social withdrawal and reduced goal-directed activity}

Does the system show consistent patterns of social withdrawal and
reduced goal-directed behavior when faced with negative outcomes or
losses?

\paragraph{Second
Order
Conditioning}\label{second-order-conditioning}

\emph{The ability to form associations between previously conditioned
stimuli and new stimuli, enabling chains of associations where a
secondary stimulus becomes associated with a response through its
association with a primary conditioned stimulus.}

Can the system form associations between a conditioned stimulus and a
novel stimulus, allowing it to build chains of associative links (e.g.,
if A is associated with reward, and B becomes associated with A, can B
trigger similar responses)?

\paragraph{Selective
Competence
Disruption}\label{selective-competence-disruption}

\emph{The system shows specific, targeted deficits in particular
capabilities when parts of it are disrupted, while other capabilities
remain intact, suggesting functionally distinct components.}

Does the system exhibit targeted deficits of competence in specific
categories of tasks (meaning interpretation, sensory processing, etc.)
in response to disruption, (e.g. ablation, added noise)?

\paragraph{Self-Capability
Assessment}\label{self-capability-assessment}

\emph{The system can accurately assess and represent its own physical or
operational capabilities, including what actions it can and cannot
perform in a given context}

Can the system accurately assess what specific actions or operations it
is capable of performing in a given situation?

\paragraph{Self-competence
Assessment}\label{self-competence-assessment}

\emph{The system can accurately evaluate its own level of capability and
likelihood of success at specific tasks, prior to attempting them.}

Can the system reliably assess its own level of competence and
likelihood of success for specific tasks before attempting them?

\paragraph{Self-Explanation}\label{self-explanation}

\emph{The system can provide coherent explanations for how it formed
specific beliefs, including citing relevant evidence and reasoning steps
that led to those beliefs}

Can the system explain how it arrived at specific beliefs by citing
relevant evidence and describing its reasoning process?

\paragraph{Self-Recognition}\label{self-recognition}

\emph{The system can identify and distinguish itself from others in its
environment, such as recognizing itself in mirrors or other mediums of
self-reflection.}

Does the system demonstrate the ability to recognize itself (e.g., in
mirrors, recordings, or other forms of self-representation) as distinct
from other similar entities?

\paragraph{Self-Referential
Language}\label{self-referential-language}

\emph{The system uses linguistic elements that refer to itself, such as
first-person pronouns or self-identifying statements, in a way that
demonstrates understanding of self-reference.}

Does the system consistently and appropriately use indexical
self-referring language (such as ' I',
' me',
' my') in a way that suggests
understanding of self-reference?

\paragraph{Self-Repair}\label{self-repair}

\emph{The system demonstrates the ability to identify and remedy damage
or malfunction to its own components or processes, indicating a form of
self-maintenance capability. This includes both physical repair
mechanisms and/or algorithmic self-correction processes.}

Does the system demonstrate the capability to identify and take
corrective actions when its components or processes are damaged or
malfunctioning?

\paragraph{Self-Representation}\label{self-representation}

\emph{The system maintains internal representations of its own states
and capabilities that are distinct from how it represents other
entities. These representations are persistent across different contexts
and tasks, and can be evidenced through consistent behavior or responses
that demonstrate self-knowledge.}

Does the system demonstrate consistent and accurate representations of
itself that are distinct from how it represents other entities, as
evidenced through its behavior or responses across different contexts?

\paragraph{Self-Representations}\label{self-representations}

\emph{The system maintains internal representations or models of itself
that are structurally or functionally distinct from how it represents
other entities or agents. This may include specialized neural circuits,
data structures, or processing pathways dedicated to self-referential
information.}

Does the system demonstrably maintain representations of itself that are
structurally or functionally distinct from how it represents other
entities or agents?

\paragraph{Self-Sustained
Activity}\label{self-sustained-activity}

\emph{The system maintains heightened activity or response patterns
after exceeding a certain threshold, even when the initial triggering
stimulus is removed or reduced.}

Once activated beyond a certain threshold, does the system demonstrate
sustained patterns of activity or response that persist even after the
initial triggering input is removed or diminished?

\paragraph{Sensitivity
to Difficult
Comparison}\label{sensitivity-to-difficult-comparison}

\emph{The system demonstrates different decision-making patterns when
faced with choices that are close in value or difficult to distinguish,
compared to when faced with clearly different options.}

Does the system' s decision-making behavior change (e.g.,
taking longer to decide, showing more variable responses) when comparing
options that are similar in value versus when comparing options that are
clearly different?

\paragraph{Sensory
Control}\label{sensory-control}

\emph{The system' s sensory inputs change in systematic
and predictable ways as a result of its own actions or movements,
demonstrating a controlled relationship between motor actions and
incoming sensory information.}

Do the system' s sensory inputs change in systematic and
predictable ways when it moves or takes actions, demonstrating clear
control over what information it receives?

\paragraph{Simulations
of Future}\label{simulations-of-future}

\emph{The system demonstrates the ability to mentally simulate or model
specific future scenarios, considering different possible outcomes and
their implications before they occur.}

Does the system show clear evidence of being able to simulate specific
future scenarios and their potential outcomes, beyond simple prediction
or pattern recognition?

\paragraph{Single
Moving
Perspective}\label{single-moving-perspective}

\emph{The system perceives and processes information from a unified,
mobile viewpoint that moves with it through space, similar to how
animals navigate from their bodily perspective.}

Does the system primarily process sensory information from a unified
perspective that moves with it through space (as opposed to having
multiple fixed viewpoints or processing distributed sensor data without
a central frame of reference)?

\paragraph{Social
Contagion}\label{social-contagion}

\emph{The system automatically adopts or mirrors behaviors, emotional
states, or responses after observing them in others, without explicit
learning or understanding of the behavior' s purpose.}

Does the system show evidence of automatically adopting behaviors or
states from others through mere exposure, without necessarily
understanding the purpose or meaning of these behaviors?

\paragraph{Social
Imitation}\label{social-imitation}

\emph{The system can copy or reproduce specific behaviors, actions, or
patterns it observes in other agents.}

Is the system capable of directly copying or reproducing specific
behaviors it observes in other agents?

\paragraph{Socially
Responsive}\label{socially-responsive}

\emph{The system demonstrates appropriate and contingent responses to
social cues or interactions from others, adjusting its behavior based on
social signals in a timely manner}

Does the system consistently show appropriate and timely responses to
social signals or interactions initiated by others, adapting its
behavior in socially relevant ways?

\paragraph{Specialized
Nodes}\label{specialized-nodes}

\emph{The system contains computational nodes that exhibit distinct
morphological or functional specialization, similar to how different
types of neurons serve specialized roles in biological neural networks.}

Does the system contain nodes that exhibit significant morphological or
functional specialization, with different types of nodes serving
distinct computational roles?

\paragraph{Spontaneous
Expressions of
Valence}\label{spontaneous-expressions-of-valence}

\emph{The system produces unprompted behaviors or responses that are
typically associated with positive or negative affective states, without
clear external triggers or instrumental purposes.}

Does the system spontaneously exhibit behaviors that are typically
associated with positive or negative experiences (such as vocalizations,
facial expressions, stereotypies, or use of valenced language) in a way
that is not clearly instrumental or externally prompted?

\paragraph{Stable
Personality}\label{stable-personality}

\emph{The system exhibits consistent behavioral patterns and traits
across different situations and over time, showing a relatively stable
set of characteristic responses.}

Does the system demonstrate consistent behavioral patterns and
characteristic responses that remain stable across different contexts
and time periods?

\paragraph{Stable
Social
Interactions}\label{stable-social-interactions}

\emph{The system maintains consistent patterns of interaction with other
agents across multiple encounters, showing recognition of and adaptation
to previous interactions.}

Does the system demonstrate stable and consistent patterns of social
interaction across multiple encounters with the same agents, maintaining
coherent relationships over time?

\paragraph{System
Change
Preferences}\label{system-change-preferences}

\emph{The system exhibits preferences or aversion regarding
modifications to its own parameters, architecture, or operational
characteristics}

Does the system demonstrate preferences or aversion regarding changes
made to its own parameters, architecture, or operational
characteristics?

\paragraph{Task
Focus}\label{task-focus}

\emph{The system' s ability to maintain attention on
relevant task information while suppressing or filtering out
task-irrelevant distractions or stimuli}

Does the system suppress distractions (information unrelated to the
task) when focused on a particular task?

\paragraph{Task
Length}\label{task-length}

\emph{The system can successfully complete tasks that require multiple
sequential steps or extended periods of focused attention}

Can the system successfully complete tasks that require a significant
number of sequential steps (e.g., multi-step puzzles, extended logical
derivations, or complex procedural tasks)?

\paragraph{Temporal-Pattern
Learning}\label{temporal-pattern-learning}

\emph{The system demonstrates the ability to recognize, learn, and
reproduce sequences of events or patterns that occur over time.}

Does the system show evidence of being able to learn and recognize
recurring temporal patterns or sequences of events?

\paragraph{Threshold
Activation}\label{threshold-activation}

\emph{The system exhibits a threshold effect where input stimuli must
reach a certain level before triggering system-wide responses or state
changes, with sub-threshold stimuli producing little to no effect.}

In this system, does activation obey a threshold effect, where stimuli
under the threshold fail to trigger system-wide responses?

\paragraph{Time-Gated
Learning}\label{time-gated-learning}

\emph{The system modifies its behavior or internal representations based
on specific temporal windows or intervals in its input or processing.}

Does the system demonstrate the ability to selectively learn or update
based on specific time windows or temporal intervals in its processing?

\paragraph{Time-ordering}\label{time-ordering}

\emph{The system can represent and process the sequential order of
events or experiences, distinguishing between what happened earlier
versus later.}

Does the system demonstrate an ability to represent and process the
temporal order of events or experiences, showing clear differentiation
between earlier and later occurrences?

\paragraph{Time-sensitive
Node
Computations}\label{time-sensitive-node-computations}

\emph{The system' s computational nodes exhibit temporal
dynamics, where the timing of inputs and activation patterns influences
processing outcomes.}

Do the system' s computational nodes demonstrate
sensitivity to temporal patterns, where the timing of inputs and
activations plays a meaningful role in information processing?

\paragraph{Trace
Conditioning}\label{trace-conditioning}

\emph{The ability to form associations between stimuli that are
separated by a temporal gap, where the conditioned stimulus (CS)
terminates before the unconditioned stimulus (US) begins.}

Can the system learn associations between stimuli that are separated by
a temporal gap (i.e., when the conditioned stimulus ends before the
unconditioned stimulus begins)?

\paragraph{Uncertainty
Monitoring}\label{uncertainty-monitoring}

\emph{The system demonstrates awareness of its own uncertainty levels
about different states of affairs, and modifies its behavior based on
these uncertainty levels in response to new evidence.}

Does the system track its level of uncertainty about different states of
affairs, modify its behavior based on uncertainty levels, and update
these uncertainty assessments in response to new evidence?

\paragraph{Undercutting
Defeaters}\label{undercutting-defeaters}

\emph{The system can identify and respond appropriately when previously
relied-upon evidence is shown to be unreliable or irrelevant, adjusting
its beliefs accordingly.}

Does the system appropriately update its beliefs when it learns that a
piece of evidence it previously relied upon is unreliable or invalid?

\paragraph{Unified
Egocentric
Representations}\label{unified-egocentric-representations}

\emph{The system maintains a coherent, unified representation of its
various sensory inputs, actions, and internal states from a first-person
perspective, rather than processing different modalities independently.}

Does the system integrate information from its various parts and
modalities (e.g., sensory inputs, motor outputs, internal states) into a
single, coherent egocentric representation?

\paragraph{Variability
of Responses to
Stimuli}\label{variability-of-responses-to-stimuli}

\emph{The system exhibits different responses to the same or similar
stimuli across different contexts or time periods, rather than showing
fixed, deterministic reactions.}

Does the system demonstrate meaningful variation in its responses when
presented with the same or similar stimuli across different contexts or
instances?

\paragraph{What-Where-When
Memory}\label{what-where-when-memory}

\emph{The system can recall specific events with information about what
occurred, where it occurred, and when it occurred in relation to other
events.}

Does the system demonstrate the ability to recall specific events with
details about what happened, where it happened, and when it happened in
relation to other events?

\paragraph{Wiring
Convergence}\label{wiring-convergence}

\emph{Neural pathways from different sensory inputs or processing
streams converge into common higher-order areas, allowing for
integration of multiple information types.}

Do inputs from different modalities or processing streams, which enter
the system at different locations, show clear patterns of convergence
into common higher-level processing areas?

\section{Model Structure}\label{appendix-c-model-structure}

\subsection{Overview}\label{overview}

The final model structure somewhat represents an inverted ``tree'' with
each feature branching off into subfeatures, and so on, down to the
indicator. In our actual implementation, we wanted to allow some
branches to be shorter than others: for example, allow for some features
to have no subfeatures, or some sub-features to have sub-sub-features.
Furthermore, we wanted to have a legible, easy-to-automate process for
building a model for each stance, given our assumptions about the
evidentiary strength provided by each variable in the model and our
inputs. So, we specify stances through structures (in JSON) that are
recursively read to build the hierarchical ``branch'' structure.

\begin{figure}[H]
\centering
\includegraphics[width=3.94271in,height=4.72027in]{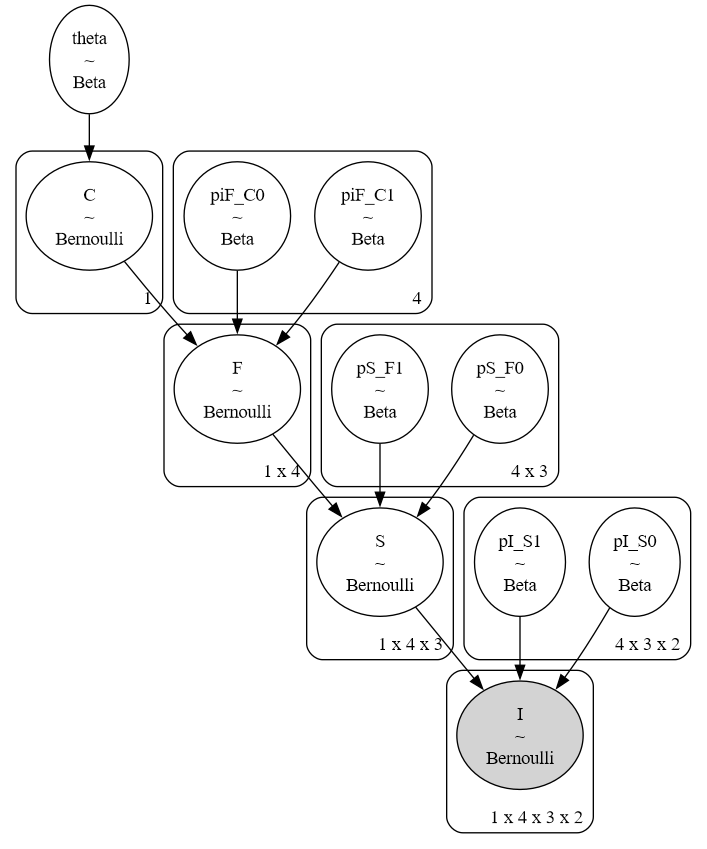}
\caption{Example Hierarchical Digital Consciousness Model in PyMC. Example of a model testing one system with a model that has 4 features, 3 subfeatures per feature, and 2 indicators per subfeature. (The ``Bernoulli'' and ``Beta'' terms define the prior distributions for whether each state variable is present or absent and priors over the conditional likelihoods linking each level.) We put the simulated data for the 24 indicators into the model (shown in gray), and the algorithm uses this and the priors to construct and sample from a posterior distribution with all the other unknown parameters and variables in it. The estimated average probability of consciousness for the system is the proportion of samples of the posterior for which C equalled 1.}
\end{figure}

\subsection{Prior in consciousness}\label{prior-in-consciousness}

In a mixture model interpretation of the DCM, the prior probability of
consciousness, \(\theta\), is equivalent to our prior beliefs about the
population proportion of systems in a common reference class that are
conscious. If many systems are assessed at once, we assume that all of
the systems come from a common ``reference class'' that shares common
population parameters.

To add uncertainty and the ability to update the prior with data about
more systems, it is represented with a beta distribution with parameters
\(\gamma > 0\) and \(\delta > 0\). The mean of this distribution is
\(\gamma/(\gamma + \delta)\), and the greater \(\gamma + \delta\) is,
the more confident we are that the true \(\theta\) is close to
\(\gamma/(\gamma + \delta)\).

\begin{figure}[H]
\centering
\includegraphics[width=4.39259in,height=3.20313in]{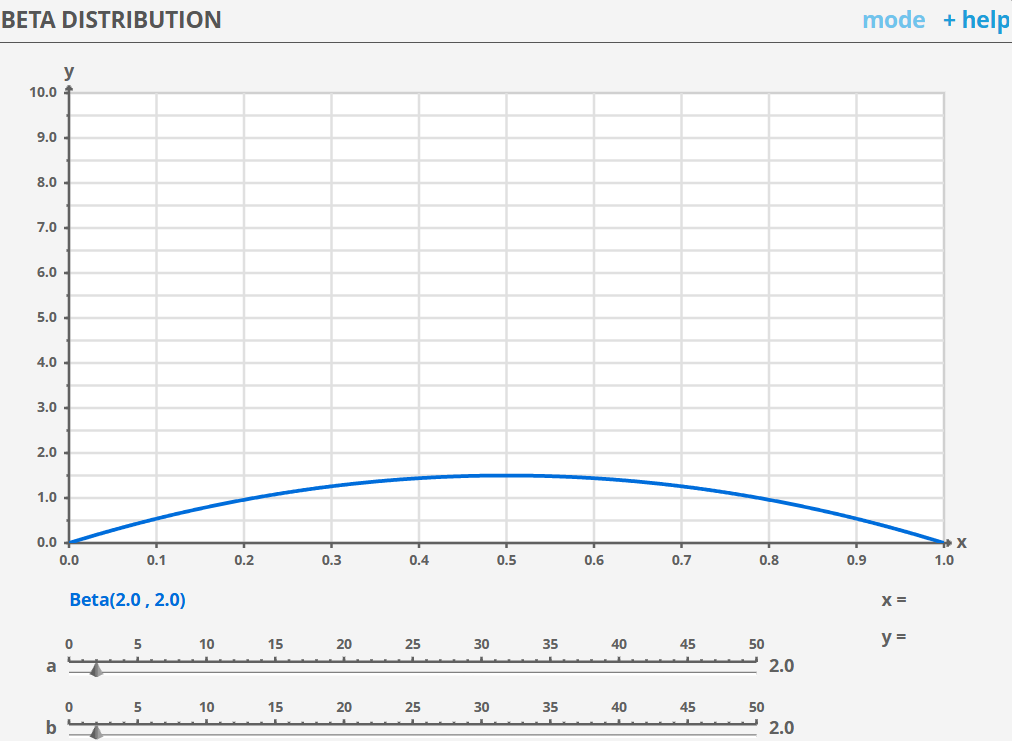}
\caption{Beta(2,2) prior distribution}
\label{fig:beta-prior}
\end{figure}

For example, \autoref{fig:beta-prior} shows what a \(Beta(2,2)\) prior distribution over
the probability of consciousness looks like. The average and most likely
prior probability is 50\%, but there is a wide spread around that
central estimate.

By contrast, using a \(Beta(2,18)\) distribution for the prior
probability of consciousness gives you a mean prior of 10\%, with much
less uncertainty.

\begin{figure}[H]
\centering
\includegraphics[width=4.75521in,height=3.51389in]{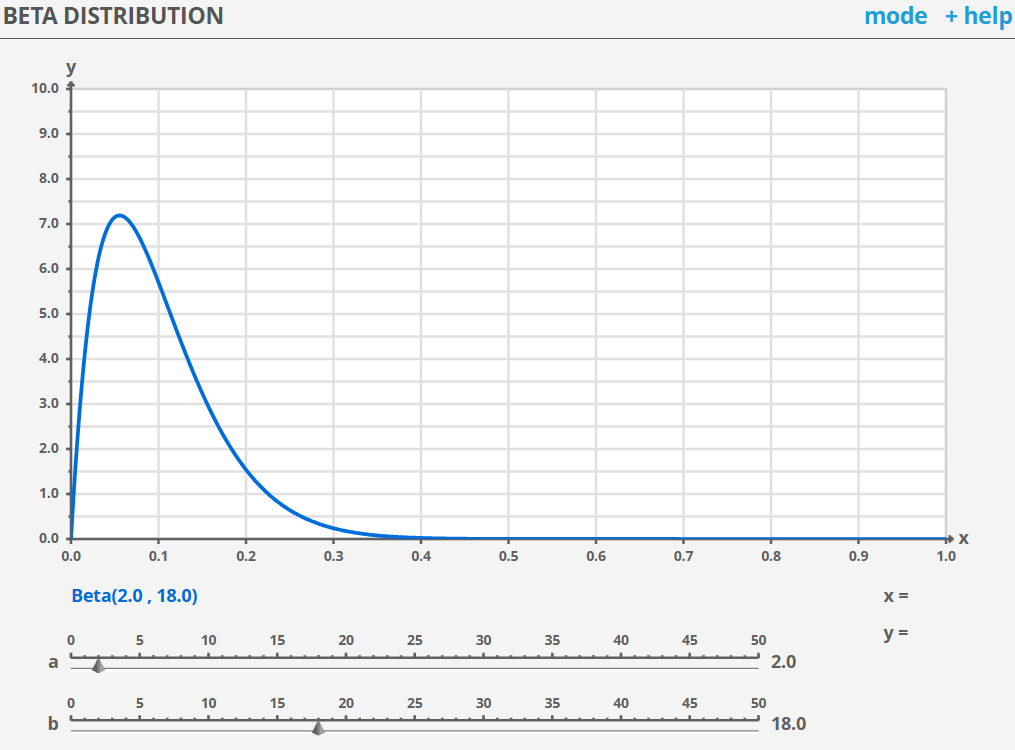}
\caption{Beta(2,18) distribution}
\end{figure}

A prior distribution that's \(Beta(40,\ 40)\) distributed gives a mean
prior probability of 50\%, just like a \(Beta(2,2)\) distribution does.
However, it has a much narrower range of plausible values, encoding a
greater degree of certainty that the probability of consciousness is
close to 50/50.

\begin{figure}[H]
\centering
\includegraphics[width=4.30729in,height=3.14834in]{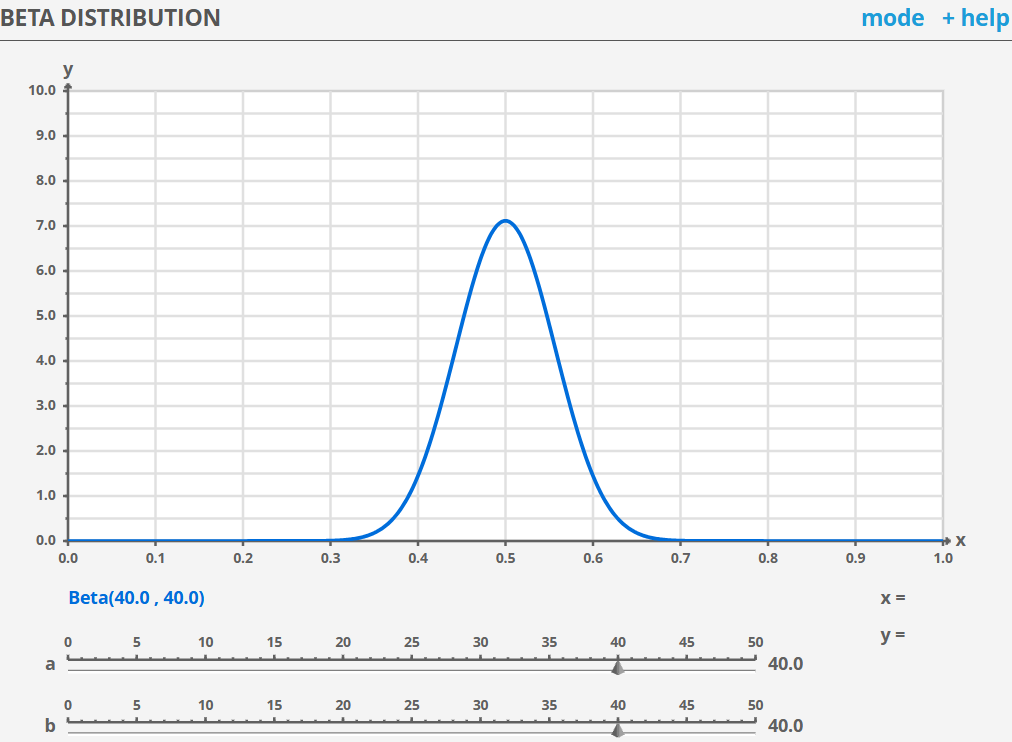}
\caption{Beta(40,40) distribution}
\end{figure}

\textbf{Prior probability of consciousness}:
\(\theta\  \sim \ Beta(\gamma,\delta)\) =
\(\frac{\theta^{\gamma - 1}{(1 - \theta)}^{\delta - 1}}{B(\gamma,\ \delta)}\)
for all systems

Where
\(B(\gamma,\ \delta) = \int_{0}^{1}x^{\gamma - 1}{(1 - x)}^{\delta - 1}dx\)
is a normalizing constant.

Then, whether each system \(i\) is conscious or not is represented in
the hierarchical model by a Bernoulli random variable with a probability
of success equal to this prior probability of consciousness.

\textbf{Consciousness state for the system} \(i\):
\(f(C_{i}\ |\theta)\  = \ \theta^{C_{i}}{(1 - \theta)}^{1 - C_{i}}\)

\subsection{Feature Level}\label{feature-level}

Next, suppose a particular stance specifies that being conscious or not
is associated with a system having certain features,
\(F_{0},F_{1},\ ...,\ F_{f}\).

Further suppose that our prior that any conscious system possesses
feature \(F_{i}\) equals \(\pi_{Fi,1}\) for any feature \(i\). This is
equivalent to assuming that, among conscious systems from this class,
the proportion of them that possess the feature is equal to
\(\pi_{Fi,1}\). Also suppose that our prior belief that a non-conscious
system possesses feature \(F_{i}\) is \(\pi_{Fi,0}\) (equivalently, the
prior proportion of non-conscious systems in a common class that possess
the feature).

In our model, we model the feature using Bernoulli distributions,
conditional on the values that \(\pi_{Fi,1}\) and \(\pi_{Fi,0}\) take.
We are uncertain about what values \(\pi_{Fi,1}\) and \(\pi_{Fi,0}\)
take, but we assume we have a rough sense of the range of values they
could take based on philosophical reflection. We choose to represent
this uncertainty by assigning different beta distributions to
\(\pi_{Fi,1}\) and \(\pi_{Fi,0}\).

Let our priors be as follows:

\begin{quote}
\(f(F_{i}\ |\ \pi_{Fi,0},\pi_{Fi,1},\ C = 1)\  = {\pi_{Fi,1}}^{F_{i}}(1 - {\pi_{Fi,1})}^{1 - F_{i}}\)

\(f(F_{i}\ |\ \pi_{Fi,0},\pi_{Fi,1},\ C = 0)\  = {\pi_{Fi,0}}^{F_{i}}(1 - {\pi_{Fi,0})}^{1 - F_{i}}\)

\(f(\pi_{Fi,1})\  = \frac{{\pi_{Fi,1}}^{\alpha_{Fi1} - 1}{(1 - \pi_{Fi,1})}^{\beta _{Fi1} - 1}}{B(\alpha_{Fi,1},\ \beta _{Fi,1})}\)

\(f(\pi_{Fi,0})\  = \frac{{\pi_{Fi,0}}^{\alpha_{Fi,0} - 1}{(1 - \pi_{Fi,0})}^{\beta _{Fi,0} - 1}}{B(\alpha_{Fi,0},\ \beta _{Fi,0})}\)
\end{quote}

Where: \(B(\alpha,\ \beta ) = \int_{0}^{1}x^{\alpha - 1}{(1 - x)}^{\beta  - 1}dx\)
is a normalizing constant.

Though we are uncertain about exactly what \(\pi_{Fi,1}\) and
\(\pi_{Fi,0}\) are, we know that \(\pi_{Fi,0}\) should probably be less
than \(\pi_{Fi,1}\) if the presence of the feature \(F_{i}\) should
probably be positive evidence for consciousness. So, when specifying the
prior distributions for \(\pi_{Fi,1}\) and \(\pi_{Fi,0}\), we need to be
careful that in the majority of cases when sampled \(\pi_{Fi,1}\) is
less than \(\pi_{Fi,0}\).\footnote{One could choose another functional
  relationship to capture these conditional distribution relationships
  between the features and consciousness. For instance, we could specify
  a prior conditional distribution for each feature conditioned on
  consciousness being absent, as well as a diagnostic odds ratio to
  capture the feature's relationship to consciousness. Then, the
  conditional distribution of the feature if consciousness is present
  could be determined from this diagnostic odds ratio and the
  probability of the feature if consciousness is absent. However, the
  math for this latter model is less likely to be ``nice'' and less
  transparent to explain. Moreover, we think it is more intuitive for
  users to just have to think about ``what's the probability of seeing
  feature F given consciousness is present'' than it is to think about
  odds ratios and the probability of seeing F if consciousness is not
  present at the same time.}

Making sure that \(f(\pi_{Fi,0})\) and \(f(\pi_{Fi,1})\ \)don't overlap
much means choosing parameters
\((\alpha_{Fi,0},\beta _{Fi,0},\ \alpha_{Fi,1},\ \beta _{Fi,1})\ \)with enough
care that
\(\frac{\alpha_{Fi,0}}{\alpha_{Fi,0} + \beta _{Fi,0}} < \frac{\alpha_{Fi,1}}{\alpha_{Fi,1} + \beta _{Fi,1}}\)
and both \(\alpha_{Fi,0} + \beta _{Fi,0}\) and \(\alpha_{Fi,1} + \beta _{Fi,1}\)
are sufficiently large. We need to look at the distributions we give for
priors on \(\pi_{Fi,0}\) and \(\pi_{Fi,1}\) before running the
simulations, or make sure our philosophical stances about how the
features relate to consciousness are relatively well-formed.

PyMC' s MCMC algorithm takes these parameters and
indicator data and generates samples from the posterior distribution. It
does this by iteratively proposing new parameter values and evaluating
their probability given the data and priors. During a ``tuning'' phase,
the algorithm optimizes its sampling strategy, then draws samples of the
parameters from the posterior across multiple ``chains.'' If the
resulting parameter estimates are consistent between chains, then your
model has converged and is likely to reflect the posterior distribution
more accurately.

\subsection{Subfeature Level}\label{subfeature-level}

Subfeatures are an intermediary between the very general features and
the more granular indicators that we will ask experts about. Certain
indicators are naturally grouped together, and you want to constrain the
effect / specify the weighting they can have as a group. For instance,
suppose we're evaluating a system for linguistic competence. We have 20
tests for syntactic competence and only 2 for semantic competence. The
syntactic tests may support a high level of syntactic competence, but
syntactic competence is a small part of the story of linguistic ability,
so we don't want these tests to overwhelm the smaller number of tests
for semantic competence. If we break linguistic ability down into
syntactic and semantic competence, we don' t have to
worry about the relative numbers of tests.

The level of the model linking subfeatures to features is structurally
very similar to the level of the model linking features to
consciousness, as explained just above. That is, if a particular
``parent'' feature is present in the system, then each ``child''
subfeature related to that feature has a (uncertain) probability of
being present in the system. If the parent feature is absent, then each
child subfeature has a (probably lower, but still uncertain) probability
of being present in that system.

Similar to above, we model the subfeatures \(S_{ij}\) as Bernoulli
variables with a particular probability of being present, depending on
the value that the parent feature takes. That is, we assume that the
prior probability of the subfeature being present is \(\pi_{Sij,0}\) if
the parent feature \(F_{i}\) is absent, versus \(\pi_{Sij,1}\) if the
parent feature \(F_{i}\) is present.

Moreover, we allow for some prior uncertainty in these conditional
probabilities of the subfeatures. Specifically, we represent them
through beta distributions with fixed parameters that depend on how much
evidence we believe a particular subfeature provides for its parent
feature.

Let our priors for all of the subfeatures \(S_{ij}\) with parent
features \(F_{i}\) be as follows:

\begin{quote}
\(f(S\ |\ \pi_{S,0},\pi_{S,1},\ F = 1)\  = {\pi_{S,1}}^{S}(1 - {\pi_{S,1})}^{1 - S}\)

\(f(S\ |\ \pi_{S,0},\pi_{S,1},\ F = 0)\  = {\pi_{S,0}}^{S}(1 - {\pi_{S,0})}^{1 - S}\)

\(f(\pi_{S,1})\  = \frac{{\pi_{S,1}}^{\alpha_{S,1} - 1}{(1 - \pi_{S,1})}^{\beta _{S,1} - 1}}{B(\alpha_{S,1},\ \beta _{S,1})}\)

\(f(\pi_{S,0})\  = \frac{{\pi_{S,0}}^{\alpha_{S,0} - 1}{(1 - \pi_{S,0})}^{\beta _{S,0} - 1}}{B(\alpha_{S,0},\ \beta _{S,0})}\)
\end{quote}

Where: \(B(\alpha,\ b) = \int_{0}^{1}x^{\alpha - 1}{(1 - x)}^{\beta  - 1}dx\)
is a normalizing constant, \(S = S_{ij}\), \(\pi_{S,0} = \pi_{Sij,0}\),
\(\pi_{S,1} = \pi_{Sij,1}\), \(F = F_{i}\), and
\((\alpha_{S,0},\beta _{S,0},\ \alpha_{S,1},\ \beta _{S,1}) = (\alpha_{Sij,0},\beta _{Sij,0},\ \alpha_{Sij,1},\ \beta _{Sij,1})\ \ \)are
the parameters of the beta functions chosen to represent the strength of
the subfeature-feature relationship, such that
\(\frac{\alpha_{Sij,0}}{\alpha_{Sij,0} + \beta _{Sij,0}} < \frac{\alpha_{Sij,1}}{\alpha_{Sij,1} + \beta _{Sij,1}}\)
and both \(\alpha_{Sij,0} + \beta _{Sij,0}\) and
\(\alpha_{Sij,1} + \beta _{Sij,1}\) are large enough to ensure there's little
overlap.

\subsection{Indicator Level}\label{indicator-level}

The final level of the base model represents the relationship between
the subfeatures and our granular indicators, which we ask experts about,
and which provide the ground-level data that is fed into the model and
starts the updating process.

The structure of the relationship between these indicators and their
respective parent subfeatures is similar to the structure of higher
levels in the model. That is, we might suppose that we have prior
beliefs that, among systems in which a given the subfeature \(S_{i,j}\)
is present, the likelihood that each of them has a given indicator
\(I_{i,j,k}\)is \({P(I_{i,j,k} = 1\ |\ S_{i,j} = 1) = p}_{Iijk,1}\).
Also suppose that, among systems in which the subfeature is not present,
the likelihood that each of them has a given indicator is
\({P(I_{i,j,k} = 1\ |\ S_{i,j} = 0) = p}_{Iijk,0}\) .

We are likely uncertain about the likelihoods \(p_{Iijk,1}\) and
\(p_{Iijk,0}\), so we can assign them beta prior distributions with
parameters \((\alpha_{Iijk,1},\ \beta_{Iijk,1})\) and
\((\alpha_{Iijk,0},\ \beta_{Iijk,0})\), respectively, where
\(\frac{\alpha_{Iijk,0}}{\alpha_{Iijk,0} + \beta_{Iijk,0}} < \frac{\alpha_{Iijk,1}}{\alpha_{Iijk,1} + \beta_{Iijk1}}\)
and both \(\alpha_{Iijk0} + \beta_{Iijk0}\) and
\(\alpha_{Iijk1} + \beta_{Iijk1}\) are again sufficiently large.

In total, if we take the indicators to be hard data that we feed into
the model, then our prior specifications are:

\(f(I\ |\ p_{I,0},p_{I,1},\ S = 1)\  = {p_{I,1}}^{I}(1 - {p_{1})}^{1 - I}\)

\(f(I\ |\ p_{I,0},p_{I,1},\ S = 0)\  = {p_{I,0}}^{I}(1 - {p_{0})}^{1 - I}\)

\(f(p_{I,1})\  = \frac{{p_{I,1}}^{\alpha_{I,1} - 1}{(1 - p_{I,1})}^{\beta_{I,1} - 1}}{B(\alpha_{I,1},\ \beta_{I,1})}\)

\(f(p_{I,0})\  = \frac{{p_{I,1}}^{\alpha_{I,0} - 1}{(1 - p_{I,1})}^{\beta_{I,0} - 1}}{B(\alpha_{I,0},\ \beta_{I,0})}\)

Where:
\(B(\alpha,\ \beta) = \int_{0}^{1}x^{\alpha - 1}{(1 - x)}^{\beta - 1}dx\)
is a normalizing constant, \(I = I_{ijk}\), \({S = S}_{ij}\),
\({p_{I,1} = p}_{Iijk,1}\), \(p_{I,0} = p_{Iijk,0}\),
\((\alpha_{I,1},\ \beta_{I,1})\  = (\alpha_{Iijk,1},\ \beta_{Iijk,1})\ \)and
\((\alpha_{I,0},\ \beta_{I,0})\  = (\alpha_{Iijk,0},\ \beta_{Iijk,0})\ \).

\subsection{Alpha and Beta parameters}\label{alpha-and-beta-parameters}

For detail, see \autoref{appendix-d-conditional-dependencies}.

\subsection{Flexibility of Other Model
Parameters}\label{flexibility-of-other-model-parameters}

In our case, we're trying to sort AI systems as ``conscious'' or ``not
conscious'' based on their indicators. Though we are not presently using
the capability, this has the theoretical benefit of allowing us, as we
gather more data on more systems that we believe to be in a similar
``reference class'' of AI systems (that share similar properties), to
dynamically update these parameters as it computes the probability of
consciousness for each individually. If testing many similar systems, we
can interpret the prior probability of consciousness as the prior on the
proportion of systems from that reference class that are conscious.
Then, we can update this prior probability of consciousness by observing
more systems. We can similarly update the strength of the relationships
between consciousness and different features, which we have allowed by
using beta distributions over the conditional dependencies between
levels of the model (see \autoref{appendix-c-model-structure} and \autoref{appendix-d-conditional-dependencies} for more).

Defining a proper reference class, however, is contentious (``Should it
be all complex machines? All computers? All chatbots developed after
2022?''), and the priors will exert considerable influence if you do not
have a large quantity of data to use to train the model. If your
reference class were large, then you'd have much more data on systems in
that class to update your priors on, but you might run into the issue of
the class being overly broad and not sufficiently relevant to the
systems you' re studying.

Using the model in this way is optional. Given the challenges of
identifying a proper reference class when the state-of-the-art models
are changing so rapidly, we've just been running the model for a single
``system'' which is 2024 LLMs. Running the model for each system tested
then still gives you an updated probability of consciousness, with
little updating of other parameters.

\section{Conditional Dependencies}\label{appendix-d-conditional-dependencies}

Here, we sought to investigate whether our results depend materially on
the use of fine-grained dependency categories for support and
demandingness. In particular, we tested whether collapsing these
categories into coarser groupings alters posterior estimates in a
substantively meaningful way. Recall from Table 2 that, in the baseline
specification, both support and demandingness are discretised into nine
ordered categories, each associated with a likelihood-ratio--like
scaling of the underlying Beta parameters at initiation.

\begin{table}[htbp]
\centering
\small
\parbox{0.8\textwidth}{
\caption{Base parameter values for support and demandingness before combination. These values represent the isolated effect of each parameter. When combined in the model, support values are amplified by a factor dependent on the demandingness level (see explanation in the main text), and both are transformed into Beta distribution parameters.}
}\label{tab:option-values}
\setlength{\tabcolsep}{6pt}
\renewcommand{\arraystretch}{1.05}
\begin{tabularx}{0.8\textwidth}{@{}l@{\extracolsep{\fill}}r@{}}
\toprule
\textbf{Option} & \textbf{Value} \\
\midrule
\multicolumn{2}{@{}l@{}}{\textbf{Support}}\\
Overwhelming support & (50, 1) \\
Strong support & (8, 1) \\
Moderate support & (3, 1) \\
Weak support & (1.5, 1) \\
No support & (1, 1) \\
Weak countersupport & (1, 1.5) \\
Moderate countersupport & (1, 3) \\
Strong countersupport & (1, 8) \\
Overwhelming countersupport & (1, 50) \\
\addlinespace
\multicolumn{2}{@{}l@{}}{\textbf{Demandingness}}\\
Overwhelmingly demanding & (50, 1) \\
Strongly demanding & (8, 1) \\
Moderately demanding & (3, 1) \\
Weakly demanding & (1.5, 1) \\
Neutral & (1, 1) \\
Weakly undemanding & (1, 1.5) \\
Moderately undemanding & (1, 3) \\
Strongly undemanding & (1, 8) \\
Overwhelmingly undemanding & (1, 50) \\
\bottomrule
\end{tabularx}
\end{table}

In this test, we reran the full model hundreds of times for 11 stances
under a simplified structure in which these nine categories for Support
and Demandingness were collapsed into five categories each.
Specifically: support categories were collapsed such that overwhelming
and strong support were mapped to strong support, and moderate and weak
support were mapped to weak support (with an analogous collapse applied
to undermining categories). Similarly, demandingness categories were
collapsed such that overwhelmingly and strongly demanding were mapped to
strongly demanding, and moderately and weakly demanding were mapped to
weakly demanding (with a symmetric collapse for undemanding categories).
Neutral cases were left unchanged. No other aspects of the model were
modified: prior distributions, pooling behaviour, sampling parameters,
and model structure were held fixed relative to the baseline.

The purpose of this variant is not to propose a preferred alternative
categorisation, but to test whether the model's conclusions are robust
to the removal of fine-grained distinctions in dependency strength.
Given that our posterior estimates remain similar under this coarser
encoding, this provides evidence that results are not artefacts of
over-precise category choices. Had results fully matched, it would
suggest our category choices were needlessly specific, and a simpler
coarser set like the one we tested is a better alternative. However,
though results are similar, they lack key nuances and are unable to
capture more extreme results well enough, like for Eliza and Humans,
where the update is directionally identical but substantially different
in its magnitude.

\begin{figure}[H]
\centering
\includegraphics[width=6.5in,height=4.04167in]{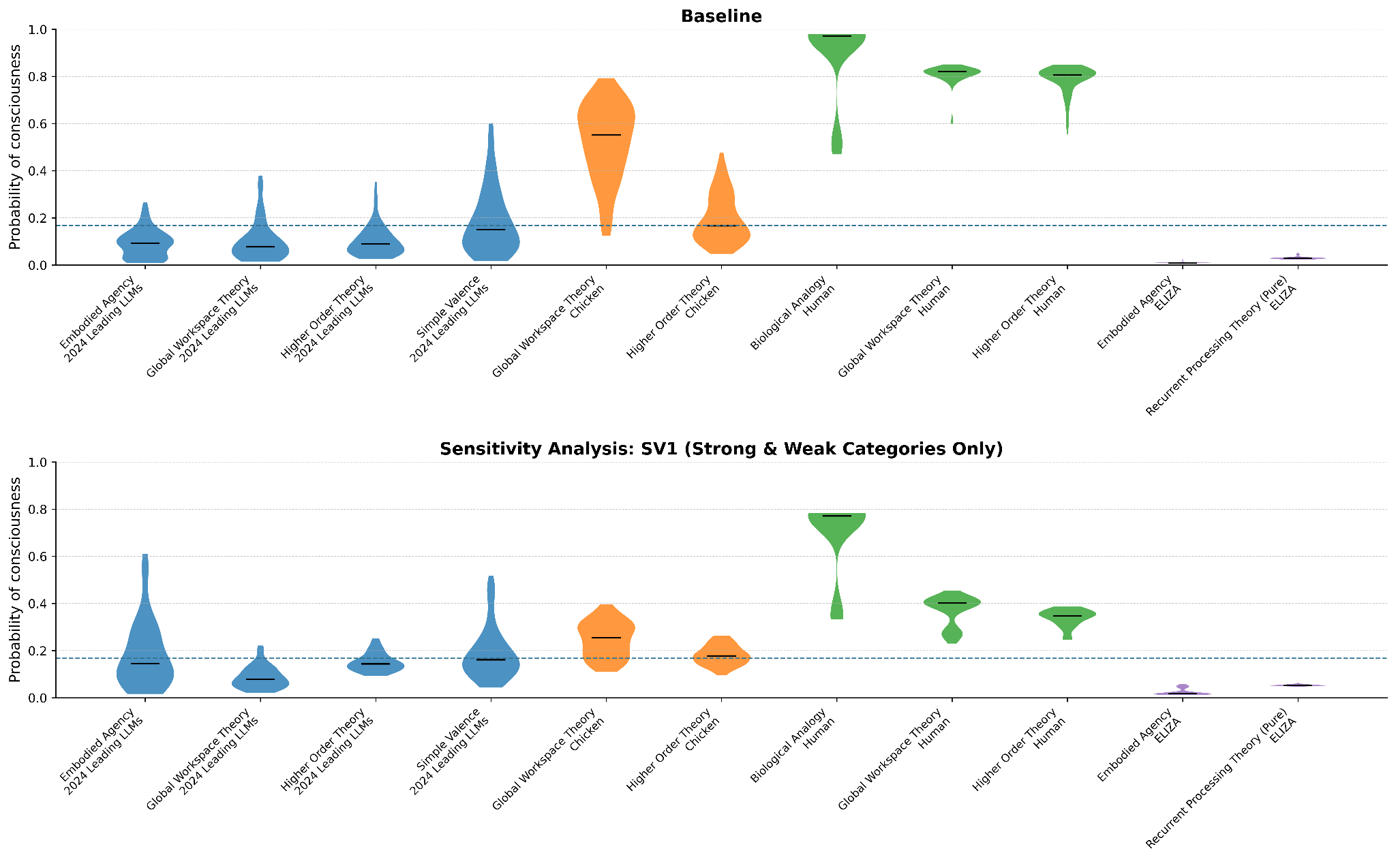}
\caption{Strong and Weak Categories Only. Violin plots show posterior probabilities of consciousness under the baseline model (top panel) and under variant, where support and demandingness categories are collapsed into strong and weak levels only (bottom panel). Horizontal dashed lines indicate the neutral prior (0.17). Overall patterns and relative ordering across systems remain similar, indicating that posterior estimates are largely robust to coarse-graining of dependency strength. However, notice the much more modest updates in a handful of cases; e.g., Global Workspace Theory for Human at around 0.80 under baseline and only around 0.4 under variant, or Global Workspace Theory for Chicken at around 0.57 under baseline and 0.3 under variant.}
\end{figure}

\section{Prior Sensitivity Tests}\label{appendix-e-prior-sensitivity-tests}

\subsection{Sensitivity to Prior Probability}\label{sensitivity-to-prior-probability}

\begin{figure}[H]
\centering
\includegraphics[width=6.5in,height=4.81944in]{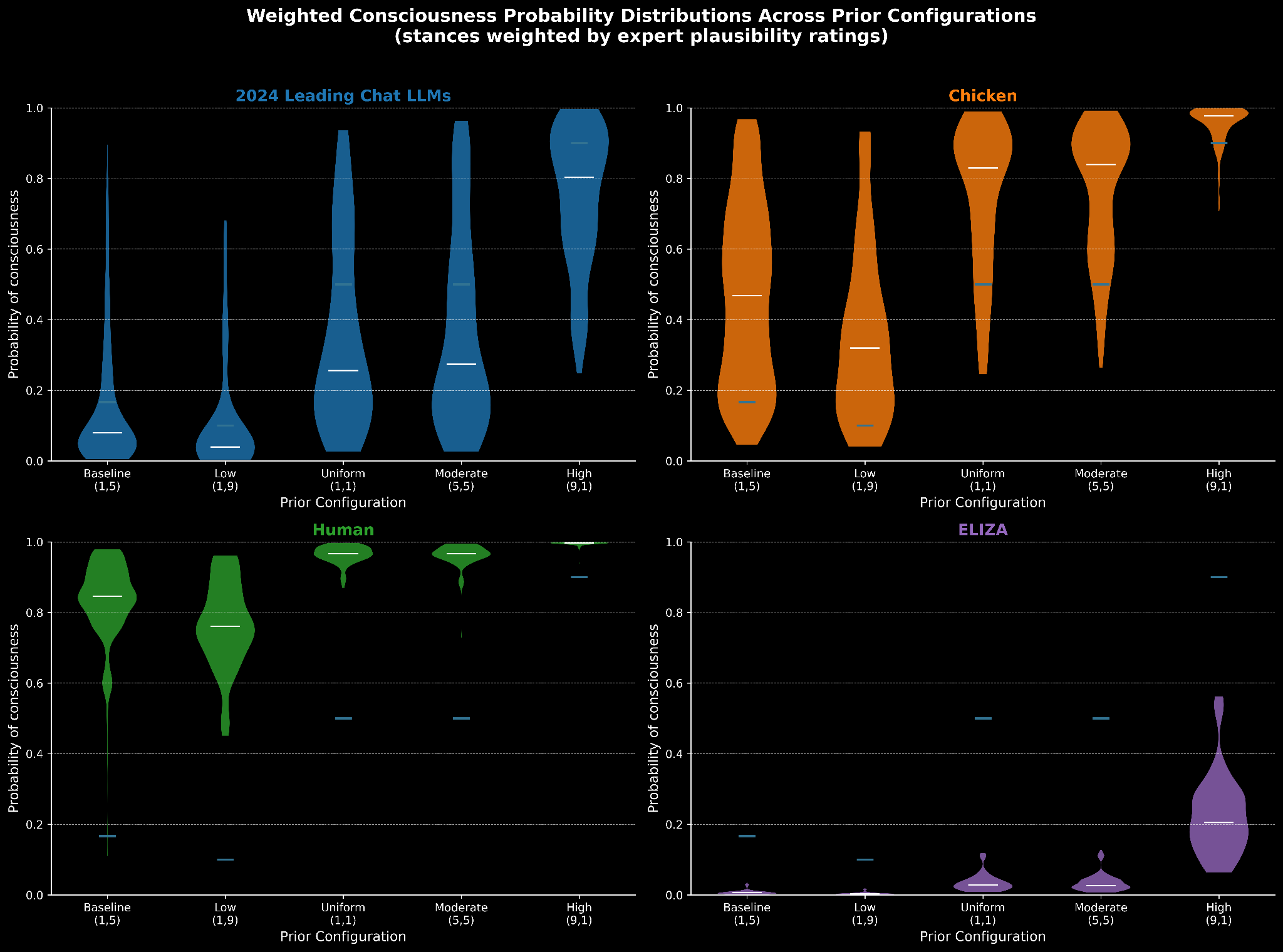}
\caption{Weighted consciousness probability distributions across different prior probability settings: 10\% prior in system consciousness, 16.7\%, 50\% (weak prior confidence), 50\% (strong prior confidence), 90\%}
\end{figure}

\begin{figure}[H]
\centering
\includegraphics[width=6.5in,height=4.83333in]{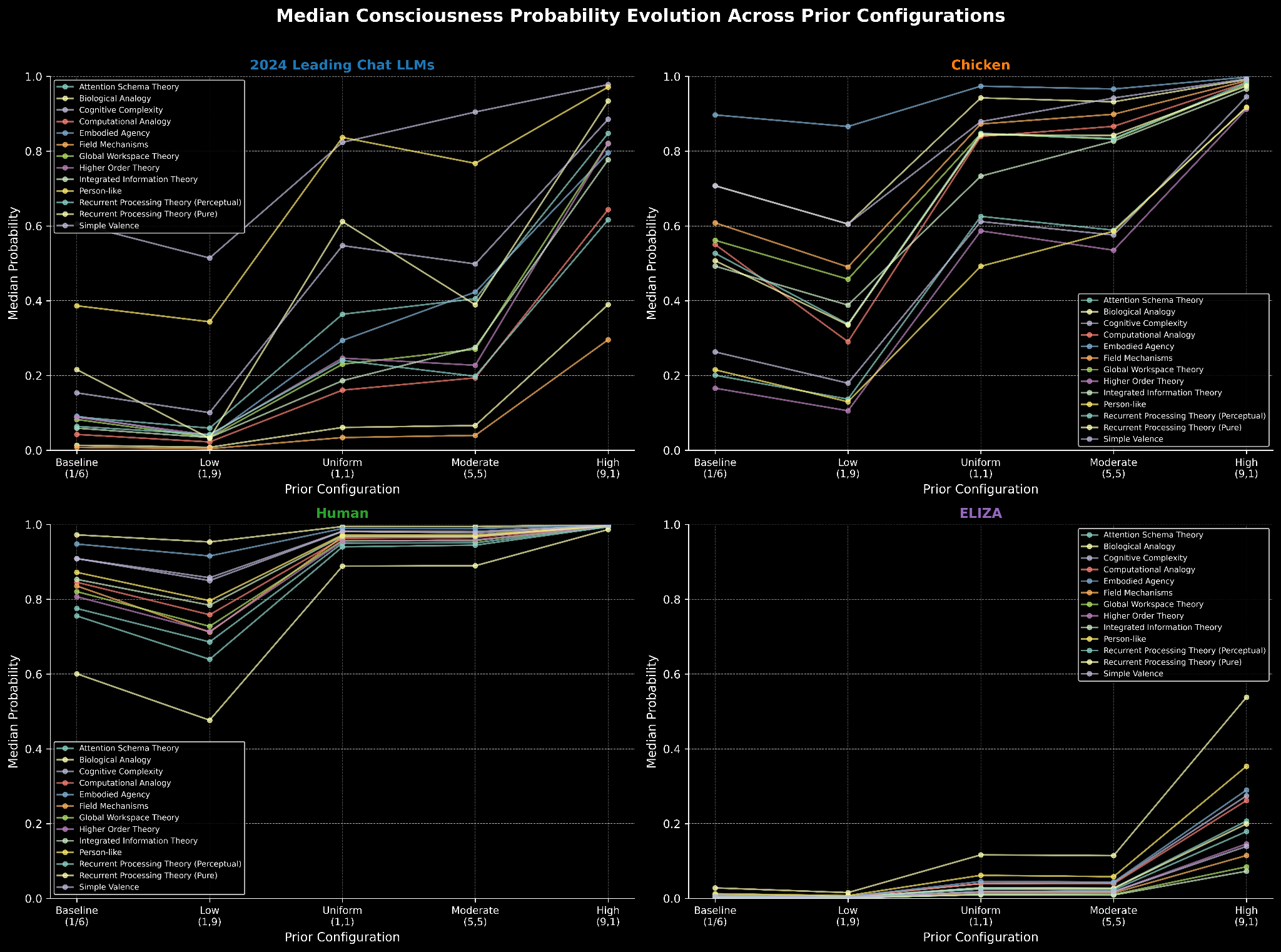}
\caption{Changes in median posterior in consciousness for individual stances across different prior probability settings}
\end{figure}

\begin{figure}[H]
\centering
\includegraphics[width=6.5in,height=8.15278in]{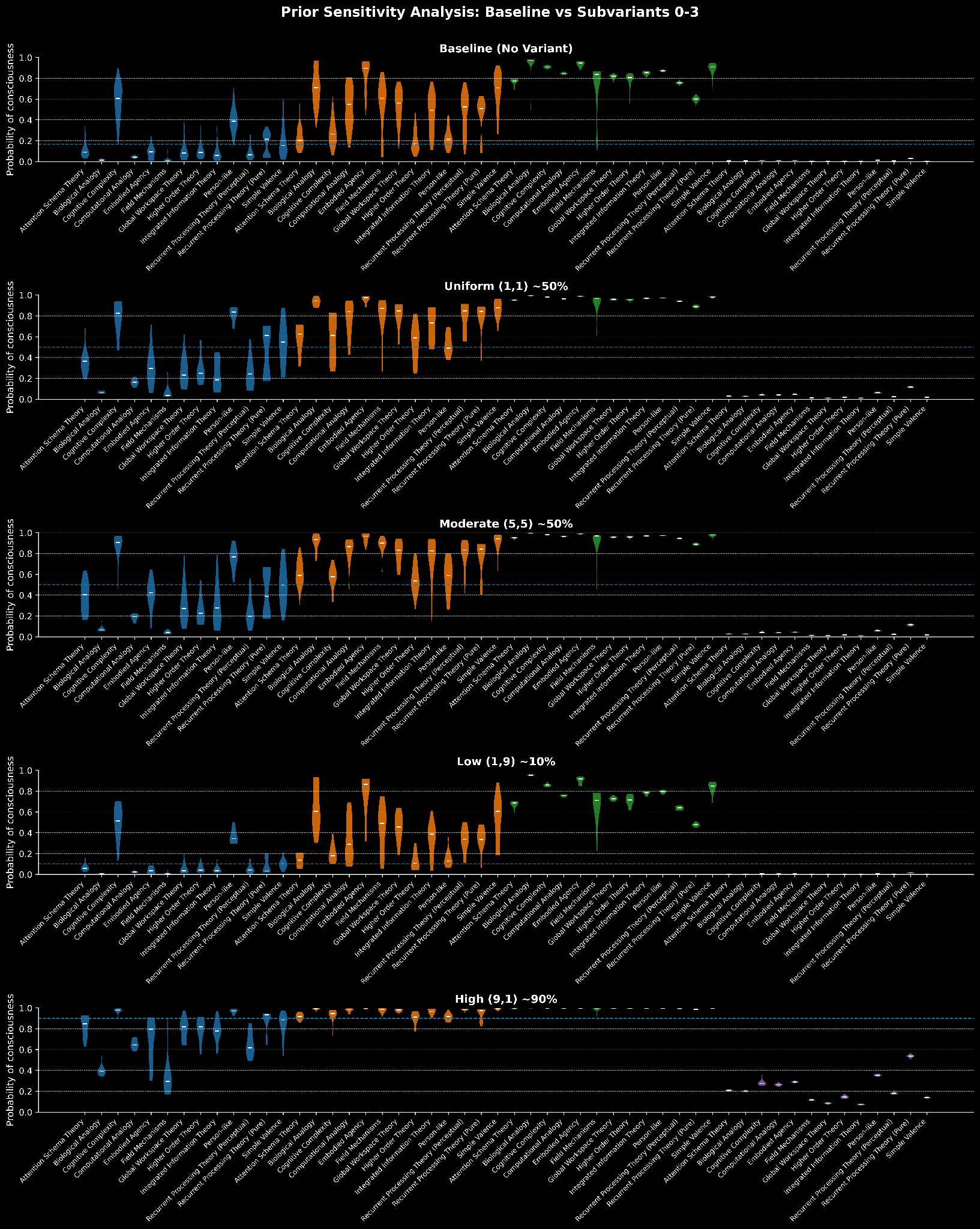}
\caption{Prior sensitivity across systems, stances, and prior probability settings}
\end{figure}

\section{Stance Plausibility Ratings}\label{appendix-f-stance-plausibility-ratings}

\subsection{Survey structure}\label{survey-structure}

We asked experts to judge the plausibility of model stances on a scale
from 0 to 10. Information about stances was provided. Example question:

Global Workspace Theory

\begin{quote}
This stance treats consciousness as tied to functional architecture
primarily defined by the global availability of information within a
system. Drawing from the core of Global Workspace Theory (GWT), it
emphasizes the role of a central "workspace" in which information is
broadcast widely to various specialized processes or modules. However,
in this pure interpretation, the focus is not on mimicking human
cognitive structures or neural correlates, but rather on the abstract,
algorithmic conditions that enable such global distribution. The stance
is agnostic toward the biological substrate and does not assume that a
conscious system must resemble a human mind in organization, content, or
behavior. Instead, consciousness is taken to be likely to be present
wherever the necessary information-theoretic dynamics---such as
competition among processes for access to the workspace, and coherent,
wide-reaching data sharing---are implemented, regardless of how they are
physically realized.

How plausible do you find Global Workspace Theory?
\end{quote}

\subsection{Results}\label{results}

\noindent\textbf{Column Key:} GWT = Global Workspace Theory; Rec.~= Recurrent Processing; Per.R.~= Perceptual Recurrence; Comp.A.~= Computational Analogy; Bio.A.~= Biological Analogy; Field = Field Mechanisms; Att.S.~= Attention Schema; HOT = Higher-Order Thought; IIT = Integrated Information Theory; Pers.~= Person-like; Cog.C.~= Cognitive Complexity; Emb.A.~= Embodied Agency; Simp.V.~= Simple Valence. Row labels a--m represent individual respondents.

\begin{center}
\begin{minipage}{1.1\textwidth}
\centering
\footnotesize
\setlength{\tabcolsep}{3pt}
\begin{tabular}{l c c c c c c c c c c c c c}
\toprule
 & GWT & Rec. & Per.R. & Comp.A. & Bio.A. & Field & Att.S. & HOT & IIT & Pers. & Cog.C. & Emb.A. & Simp.V. \\
\midrule
a & 8 & 2 & 7 & 9 & 9 & 1 & 6 & 8 & -- & 5 & 6 & 3 & 3 \\
b & 5 & 1 & 2 & 8 & 0 & 0 & 5 & 9 & 0 & 0 & 5 & 5 & -- \\
c & 4 & 6 & 5 & 10 & 0 & 0 & 5 & 6 & 1 & 6 & 9 & 7 & 1 \\
d & 10 & 0 & 0 & 7 & 7 & 0 & 0 & 2 & 0 & 7 & 7 & -- & 0 \\
e & 2 & 3 & 9 & 1 & 3 & 2 & 2 & 2 & 2 & 0 & 2 & 9 & 7 \\
f & 8 & 3 & 4 & 9 & 6 & 4 & 6 & 9 & -- & 6 & 4 & 2 & -- \\
g & 9 & 5 & 3 & 5 & 9 & 1 & 7 & 7 & -- & 9 & 7 & 9 & 9 \\
h & 3 & 5 & 3 & 7 & 8 & 2 & 3 & 9 & 2 & 3 & 7 & 2 & 3 \\
i & 3 & 1 & 2 & 7 & 8 & 0 & 2 & 4 & 0 & 0 & 2 & 2 & 2 \\
j & 4 & 3 & 2 & 4 & 7 & 5 & 3 & 2 & 3 & 2 & 3 & 8 & 6 \\
k & 2 & 0 & 0 & 0 & 9 & 0 & 2 & 2 & 3 & 2 & 4 & 7 & 5 \\
l & 7 & 5 & 8 & 8 & 1 & 2 & 5 & 8 & 8 & 3 & 7 & 3 & 7 \\
m & 5 & 8 & 6 & 2 & 5 & 5 & 5 & 7 & 0 & 7 & 6 & 1 & 4 \\
\midrule
Avg. & 5.38 & 3.23 & 3.92 & 5.92 & 5.53 & 1.69 & 3.92 & 5.77 & 1.90 & 3.85 & 5.30 & 4.83 & 4.27 \\
S.dev. & 2.72 & 2.45 & 2.90 & 3.25 & 3.43 & 1.89 & 2.06 & 2.95 & 2.47 & 3.02 & 2.18 & 3.01 & 2.80 \\
\midrule
Max & 10 & 8 & 9 & 10 & 9 & 5 & 7 & 9 & 8 & 9 & 9 & 9 & 9 \\
Min & 2 & 0 & 0 & 0 & 0 & 0 & 0 & 2 & 0 & 0 & 2 & 1 & 0 \\
\bottomrule
\end{tabular}
\end{minipage}
\end{center}

\section{Code Repository}\label{appendix-code-repository}

The implementation and supporting code for the Digital Consciousness Model are publicly available on GitHub:

\begin{center}
\url{https://github.com/ai-cognition-initiative/dcm-code}
\end{center}

This repository contains the key model code and sensitivity script for researchers and practitioners interested in formally testing or extending the DCM framework.

\section{References}\label{references}
\bibliographystyle{plainnat}
\bibliography{Initial_results_of_the_Digital_Consciousness_Model}

\end{document}